\documentclass[11pt,a4paper]{article}
\usepackage{authblk}

\usepackage[table,dvipsnames]{xcolor}
\usepackage{graphicx}
\usepackage{tikz}
\usetikzlibrary{patterns,calc,decorations.pathmorphing}

\usepackage{amsmath,amssymb,amsthm}
\usepackage{bm}

\usepackage{array,tabularx,makecell,stmaryrd}
\usepackage[export]{adjustbox}
\usepackage[a4paper,margin=3cm]{geometry}
\usepackage{float}
\usepackage{afterpage,changepage}
\usepackage{enumitem}

\usepackage[linesnumbered,ruled,vlined]{algorithm2e}
\usepackage{tcolorbox}
\usepackage{comment}
\tcbuselibrary{listings,breakable}

\usepackage{stackengine}
\usepackage{import}

\usepackage[hidelinks]{hyperref}

\usepackage[
    backend=bibtex,   
    style=numeric,    
    sorting=nyt,      
    doi=true,         
    url=false,        
    maxbibnames=99    
]{biblatex}

\newtheorem{theorem}{Theorem}[subsection]
\newtheorem{lemma}[theorem]{Lemma}
\newtheorem{corollary}[theorem]{Corollary}

\theoremstyle{definition}
\newtheorem{definition}[theorem]{Definition}
\theoremstyle{remark}
\newtheorem{remark}[theorem]{Remark}

\newcommand{\pred}{\operatorname{pred}^{(c)}}
\newcommand{\succv}{\operatorname{succ}^{(c)}}
\newcommand{\apred}{\operatorname{pred}^{(ac)}}
\newcommand{\asuccv}{\operatorname{succ}^{(ac)}}
\newcommand{\Path}[1]{P^{(c)}[#1]}
\newcommand{\APath}[1]{P^{(ac)}[#1]}
\newcommand{\SPI}{\ensuremath{\mathcal{I}=(W,R)}}

\definecolor{darkgreen}{rgb}{0.0,0.3,0.0}

\newtcolorbox{algobox}[1]{%
  colback=white,colframe=black,boxrule=0.5pt,arc=0mm,
  fonttitle=\bfseries\scshape,title=\textsc{#1},breakable}

\title{\textbf{Revisiting Chazelle's Implementation of the Bottom-Left Heuristic: A Corrected and Rigorous Analysis}}

\author{Stefan Michel\thanks{The author is supported by the Deutschlandstipendium scholarship of the University of Bonn.}}
\affil{\footnotesize Research Institute for Discrete Mathematics, University of Bonn, Germany \\
  Email: \texttt{michel@dm.uni-bonn.de}}

\date{}

\begin{document}
\maketitle
\vspace{-7ex}
\begin{abstract}
    The Strip Packing Problem is a classical optimization problem in which a given set of rectangles must be packed, without 
    overlap, into a strip of fixed width and infinite height, while minimizing the total height of the packing. 
    A straightforward and widely studied approach to this problem is the Bottom-Left Heuristic.
    It consists of iteratively placing each rectangle in the given order at the lowest feasible position in the strip and, 
    in case of ties, at the leftmost of those. Due to its simplicity and good empirical performance, this heuristic is 
    widely used in practical applications. The most efficient implementation of this heuristic was proposed by Chazelle 
    in 1983, requiring $O(n^2)$ time and $O(n)$ space to place $n$ rectangles. However, although Chazelle's original 
    description was largely correct, it omitted several formal details. Furthermore, our analysis revealed a critical 
    flaw in the original runtime analysis, which, in certain cases, results in $\Omega(n^3)$ running time. 
    Motivated by this finding, this paper provides a rigorous and corrected presentation of the implementation, addressing
    the imprecise arguments and resolving the identified flaw. The resulting analysis establishes a formally verified 
    version of Chazelle's implementation and confirms its quadratic time complexity.
\end{abstract}

\section{Introduction}

The Strip Packing Problem is a classical problem in combinatorial optimization and 
computational geometry, arising in various real-world scenarios such as scheduling problems or VLSI-design.
In this problem, we are given a rectangular strip of fixed width and infinite height, 
together with a set of rectangles $r_1, \ldots ,r_n$. 
The objective is to find a packing of the given rectangles into the strip,
such that no two rectangles overlap and the total height of the packing is minimized. 
Moreover, only \textit{orthogonal} and \textit{oriented} packings are allowed. 
A packing is orthogonal if all edges of the packed rectangles are parallel to either the horizontal or the vertical 
edges of the strip, and oriented if rectangles cannot be rotated during placement.

A straightforward reduction from the Bin Packing Problem demonstrates that the Strip Packing Problem is NP-hard~\cite{KRP}, and even 
strongly NP-hard \cite{StrongNP}. Furthermore, this reduction implies that, unless $P = NP$, no $(\tfrac{3}{2} - \varepsilon)$-approximation 
algorithm for the Strip Packing Problem can run in polynomial time. Up to this date, the best currently known approximation
algorithm was developed by Harren, Jansen, Pr\"adel and van Stee~\cite{BestSP} and achieves an approximation ratio of $\frac{5}{3} + \varepsilon$.
However, the algorithm is rather intricate and therefore seems to have limited practical applicability. 

In contrast, the Bottom-Left Heuristic (which we will abbreviate by BL-heuristic) is a fairly simple algorithm that iteratively 
places each rectangle in the given order at the lowest feasible position and, in case of ties, at the leftmost of those. 
This heuristic was first analyzed by Baker, Coffman and Rivest~\cite{BCR}, who showed that the approximation
ratio of this method heavily depends on the order in which the rectangles are given. 
More specifically, they have shown that although an arbitrary ordering can result in an unbounded approximation ratio, 
sorting the rectangles by decreasing width guarantees a (tight) approximation ratio of $3$.

In recent work, Hougardy and Zondervan proved that there cannot exist an ordering of the rectangles under which the 
BL-heuristic guarantees an approximation ratio better than $\tfrac{4}{3} - \varepsilon$~\cite{BLHZ}. Nevertheless, they also showed that 
there always exists an ordering --- constructible in time asymptotically equivalent to sorting --- under which the BL-heuristic 
achieves a $\tfrac{13}{6}$-approximation for the Strip Packing Problem~\cite{BLHZ2}. 

In addition to its simplicity and solid theoretical performance, the BL-heuristic has been observed to perform
well in practice, making it a highly attractive choice for practical applications. This observation motivates the study 
of efficient implementations of this heuristic.

To date, the most efficient implementation of this heuristic was proposed by Chazelle in 1983~\cite{BLCHZ}, requiring 
$O(n^2)$ time and $O(n)$ space to place $n$ rectangles according to the rules of the BL-heuristic. 
While the overall approach of Chazelle's implementation was largely correct, several parts of his original description 
lacked formal precision. Furthermore, we found a critical error in the accompanying runtime analysis, which,
in certain cases, results in $\Omega(n^3)$ running time. Motivated by this finding, after collecting notations and preliminaries in Section 2, 
Section 3 revisits and formalizes Chazelle's implementation, providing a corrected and fully rigorous analysis that resolves 
the inaccuracies and fixes the identified flaw. This, in turn, established a formally verified version of Chazelle's algorithm
 and confirms its quadratic time complexity.

\section{Notation and Preliminaries}

We begin by collecting the basic notations, where the notations concerning points in the plane are based on 
the notation provided by Chazelle in~\cite{BLCHZ}.

\subsection{Basic Notation}

\begin{itemize}
    \item For a $n \in \mathbb{N}$, let $[n] := \{1, \ldots ,n\}$.
    \item For a point $p = (p_1,p_2) \in \mathbb{R}^2$, let $x(p):= p_1$ and $y(p) := p_2$.
    \item For a point $p = (p_1,p_2) \in \mathbb{R}^2$ and an $l\in \mathbb{R}$, let $p +_x l := (p_1 +l,p_2)$. The symbols $-_x,+_y,-_y$ are defined analogously.
    \item For two points $p = (p_1,p_2), p' = (p_1',p_2') \in \mathbb{R}^2$ we write that $p <_x p'$ if and only if $x(p) < x(p')$. Analogous definitions apply to all other coordinate-wise comparisons.
    \item For two points $p,p' \in \mathbb{R}^2$ with $p \neq p'$, we define $\text{line}(p,p')$ as the unique line passing through both $p$ and $p'$ : 
        \begin{equation*}
            \text{line}(p,p') := \{(1-\lambda)p+ \lambda p'| \lambda \in \mathbb{R}\}
        \end{equation*}
    \item For two points $p,p' \in \mathbb{R}^2$ with $p \neq p'$, we define $\text{seg}(p,p')$ as the unique straight segment connecting them:
        \begin{equation*}
            \text{seg}(p,p') := \{(1-\lambda)p+ \lambda p'| \lambda \in [0,1]\}
        \end{equation*}
    \item Let $k\in \mathbb{N}$ and $p_1, \ldots ,p_k \in \mathbb{R}^2$. Then we define 
    \begin{equation*}
        \text{seg}(p_1, \ldots ,p_k) := \bigcup_{i=1}^{k-1} seg(p_i,p_{i+1}).
    \end{equation*}
\end{itemize}

\subsection{Preliminaries}

We proceed by formally defining the Strip Packing Problem and the Bottom-Left Heuristic. The following definitions are 
based on a paper of Hougardy and Zondervan~\cite{BLHZ}, with some modifications in structure and notation.

\begin{definition}
    A \textbf{rectangle} is a tuple $r = (w,h) \in {\mathbb{R}_{\geq0}}^2$. We call $w(r) := w$ the \textbf{width}
    of the rectangle $r$ and we call $h(r) := h$ its \textbf{height}. 

    For a $W \in \mathbb{R}_{\geq 0}$, we define the \textbf{vertical strip} with width $W$ (and infinite height) 
    as the set $S_W := [0,W] \times [0,\infty)$.
\end{definition}

\begin{definition}
    A \textbf{Strip Packing Instance} is denoted by $\mathcal{I} = (W,R)$, where 
    \begin{itemize}
        \item $W\in \mathbb{R}_{\geq 0}$ specifies the fixed width of a vertical strip and
        \item $R =\{r_1, \ldots ,r_n\}$ is a multiset of $n$ rectangles.
    \end{itemize}
    
    Each rectangle $r_i \in R $ has to fulfill $w(r_i) \leq W$, and we abbreviate $w_i := w(r_i)$ and $h_i := h(r_i)$ 
    for all $i \in [n]$.
    
    A \textbf{packing} of $R$ into $S_W$ is a function $sp : R \to S_W$, where $ (x_i,y_i) := sp(r_i)$ specifies 
    the lower left coordinate of the rectangle $r_i$ in the packing. We call a packing \textbf{feasible}, if all 
    rectangles $r_i$ satisfy the \textbf{containment constraints}, which enforce that the rectangle lies within the strip: 

    \begin{equation*}
        x_i \geq 0, x_i + w_i \leq W, y_i \geq 0
    \end{equation*}
    and all pairs of rectangles $(r_i,r_j)$ with $i \neq j$ satisfy the \textbf{non-overlap constraint}: 
    \begin{equation*}
        \big((x_i, x_i + w_i) \times (y_i, y_i + h_i)\big) \cap \big((x_j, x_j + w_j) \times (y_j, y_j + h_j)\big) = \emptyset
    \end{equation*}
    (We say that a pair of rectangles \textbf{overlaps} if it does not satisfy the non-overlap constraint).
    The \textbf{height} of a feasible packing is defined as $h(sp) :=\max \{ y_i + h_i : i \in [n] \}$.
    
\end{definition}

\medskip

With these definitions in mind, we can define the \textbf{Strip Packing Problem}

\begin{tcolorbox}[colback=white, colframe=black, boxrule=0.5pt, arc=0mm]
{\Large\textsc{Strip Packing Problem}} 
\vspace{0.1cm} \\ 
\textit{Instance:} A Strip Packing Instance $\mathcal{I} = (W,R)$. \\ 
\textit{Task:} Find a feasible packing $sp$ of $R$ into $S_W$ with minimum height

\end{tcolorbox}
We will abbreviate the Strip Packing Problem as SPP and a Strip Packing Instance by SPI.

Now we proceed to describe the Bottom-Left Heuristic. Motivated by the iterative nature of the approach, we first introduce the 
following definition. 
\newpage

\begin{definition}
    Let $\mathcal{I} = (W,R)$ be a Strip Packing Instance. A \textbf{partial packing}  of $R$ into $S_W$ is a function 
    $sp' : R \to S_W \cup \{(\infty,\infty) \}$.  We denote $R_{sp'} := \{r \in R : sp'(r) \neq (\infty,\infty)$\} and 
    we call a partial packing $sp'$ \textbf{feasible} if the restriction $sp'_{|R_{sp'}}$ is a feasible packing of 
    $R_{sp'}$ into $S_W$. The rectangles in $R_{sp'}$ are named \textbf{placed} rectangles
     and the rectangles in $R\setminus R_{sp'}$ are called \textbf{unplaced} rectangles.
    
    If $R = \{r_1, \ldots ,r_n\}$, then a \textbf{$\bm{k}$-partial packing} denotes a partial packing $sp_k$ with 
    $R_{sp_k} = \{r_1, \ldots ,r_k\}$. For $k < n$, a  \textbf{feasible extension} of a feasible $k$-partial packing~$sp_k$ 
    is a feasible $(k+1)$-partial packing $sp_{k+1}$ with  $sp_k(r_i) = sp_{k+1}(r_i) \text{ for all } i \in [k]$. 
    We call the points $p \in \mathbb{R}^2$ that imply a feasible extension for $sp_k$ (as a possible value for~$sp_{k+1}(r_{k+1}))$
    \textbf{feasible locations} for the rectangle $r_{k+1}$ (with regard to $sp_k$) and by "placing the rectangle 
    $r_{k+1}$ at $p\in \mathbb{R}^2$" (with regard to $sp_k$) we mean setting $sp_k(r_{k+1}) = p$ and thus extending the packing to a $(k+1)$-partial packing
\end{definition} 

\begin{remark}
    As every feasible partial packing is a feasible packing on a subset of the rectangle set $R$, 
    each definition concerning feasible packings extends naturally to feasible partial packings. 
    Therefore throughout this paper we will frequently use definitions and lemmas that were stated for 
    feasible packings in the context of feasible partial packings. 
\end{remark}

With this definition, we can now describe the Bottom-Left Heuristic:

\begin{tcolorbox}[colback=white, colframe=black, boxrule=0.5pt, arc=0mm]
{\Large\textsc{Bottom-Left Heuristic}}

\textit{Input:  \ } \quad A Strip Packing Instance $\mathcal{I} = (W,R)$.

\textit{Output:} \quad A feasible packing of $R$ into $S_W$ 

\begin{enumerate}[label=\textcircled{\small\arabic*}, leftmargin=2em]
    \item Let $R = \{r_1, \ldots ,r_n\} , sp(r_i) = \{(\infty, \infty) \} \ \forall i \in [n]$
    \item \textbf{For} $i = 1$  \textbf{to} $n$ \textbf{do:}
        \begin{adjustwidth}{1em}{}
          Let $(x_i,y_i)\in\mathbb{R}$ be a feasible location for $r_i$ such that $(y_i,x_i)$
          is lexicographically minimal among those \\
          $sp(r_i) \gets (x_i,y_i)$
        \end{adjustwidth}
    \item \textbf{return} $sp$

\end{enumerate}

\end{tcolorbox}
Thus, the Bottom-Left Heuristic, which we will abbreviate by BL-heuristic, starts with a feasible $0$-partial packing 
of the rectangles and extends in each iteration of the for-loop the computed feasible $i$-partial packing to a feasible 
$i+1$-partial packing by placing the next rectangle $r_{i+1}$ in order at the lowest feasible location, 
and at the leftmost of those in the case of ties. We call this position the \textbf{BL-location} of $r_{i+1}$ 
(w.r.t. the computed feasible $i$-partial packing). 

As is evident from the algorithm formulation itself, the BL-heuristic is highly sensitive to the order of the rectangles
 --- the same set of rectangles, when arranged differently, can lead to drastically different packings. 

By our formulation of the heuristic it is evident that different implementations vary only in how they compute the 
BL-location for the next rectangle in order. Therefore, specifying this computation is sufficient to describe an 
implementation. Although the partial packings throughout this iterative algorithm have the property of having been 
constructed by the BL-heuristic, Chazelle's implementation does not rely on that assumption. Instead, it only requires 
the partial packing to satisfy a more general condition, which is described in the following definition.

\newpage 

\begin{definition}
    Let $\mathcal{I}= (W,R)$ be a SPI and $sp : R \to S_W$ be a feasible packing of $R$ into~$S_W$. 
    We say that for a rectangle $r \in R$ the location $sp(r)$ is \textbf{Bottom-Left-stable (BL-stable)}, 
    if $r$ cannot be shifted further downwards or leftwards in the packing defined by $sp$ without making it infeasible.
    Formally speaking, we call $sp(r)$ BL-stable, if there exists a $\varepsilon > 0$ such that changing $sp(r)$ to 
    any point in $\big(x(sp(r))- \varepsilon, x(sp(r))\big) \times \{y(sp(r))\}$ results in the packing becoming 
    infeasible and if the analogous condition holds for the vertical direction.

    We call a packing $sp$ BL-stable, if for each rectangle $r$ the location $sp(r)$ is BL-stable.
\end{definition}
\begin{remark}
    It is trivial that a packing produced by the BL-heuristic will always be BL-stable. However, 
    one can easily construct examples of BL-stable packings that cannot be constructed by the BL-heuristic.
\end{remark}

To conclude this preliminary section, we introduce a definition to refer more conveniently to the coordinates of the four corners of the placed rectangles.

\begin{definition}
Let $\mathcal{I}= (W,R)$ be a SPI and $sp : R \to S_W$ be a feasible packing of $R$ into $S_W$. For each rectangle $r \in R$, we define :
\begin{equation*}
     x_{\min}^{sp}(r) := x(sp(r)) ,\ x_{\max}^{sp}(r) := x(sp(r)) + w(r)
\end{equation*}
\begin{equation*}
    y_{\min}^{sp}(r) := y(sp(r)),\ y_{\max}^{sp}(r) := y(sp(r)) + h(r)
\end{equation*}
\end{definition}

\begin{remark}
    For the sake of readability, whenever it is clear from the context which packing $sp$ we are referring to, 
    we will omit the superscript and refer to those coordinates simply as $x_{\min}(r),x_{\max}(r), y_{\min}(r),y_{\max}(r)$.
\end{remark}

\section{A Corrected and Rigorous Analysis of Chazelle's Implementation.}

We now turn to our rigorous and corrected presentation of Chazelle's implementation. 
In general, Chazelle's implementation will compute all feasible BL-stable locations for placing a new rectangle
in $O(k)$ time with regard to a given BL-stable partial packing, where $k$ is the number of rectangles already placed.
Consequently, placing $n$ rectangles using this approach will require $O(n^2)$ time overall. To implement the BL-heuristic accordingly,
it suffices in each iteration to retain only the feasible BL-stable location $(x_i,y_i)$ that minimizes $(y_i,x_i)$
lexicographically.   

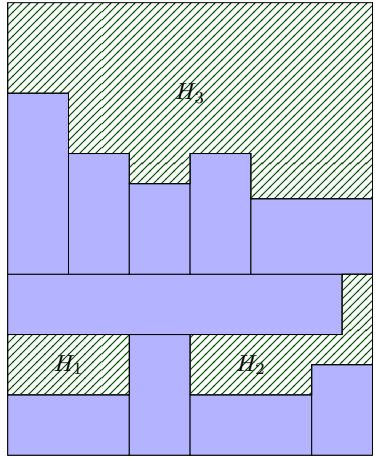
\begin{figure}[!tp]\label{chz:fig:first_holes}
    \centering
    \scalebox{0.8}{
\begin{tikzpicture}
    \pgfmathsetmacro{\stripwidth}{6}
    \pgfmathsetmacro{\stripheight}{7.5}

    \draw (0,0) -- (0,\stripheight);
    \draw (0,0) -- (\stripwidth, 0);
    \draw (\stripwidth, 0) -- (\stripwidth, \stripheight);

    \filldraw[fill=blue!30, draw = black] (0,0) rectangle (2,1);
    \filldraw[fill=blue!30, draw = black] (2,0) rectangle (3,2);
    \filldraw[fill=blue!30, draw = black] (3,0) rectangle (5,1);
    \filldraw[fill=blue!30, draw = black] (5,0) rectangle (6,1.5);

    \filldraw[fill=blue!30, draw = black] (0,2) rectangle (5.5,3) ;

    \filldraw[fill=blue!30, draw = black](0,3) rectangle (1,6) ;
    \filldraw[fill=blue!30, draw = black](1,3) rectangle (2,5) ;
    \filldraw[fill=blue!30, draw = black] (2,3) rectangle (3,4.5) ;
    \filldraw[fill=blue!30, draw = black] (3,3) rectangle (4,5);
    \filldraw[fill=blue!30, draw = black] (4,3) rectangle (6,4.25);

    \filldraw[pattern = north east lines, pattern color = darkgreen]
    (0,1) -- (2,1) -- (2,2) -- (0,2) -- cycle;
    \filldraw[pattern = north east lines, pattern color = darkgreen]
    (3,1) -- (5,1) -- (5,1.5) -- (6,1.5) -- (6,3) -- (5.5,3) -- (5.5,2) -- (3,2) -- cycle ;

    \filldraw[pattern = north east lines, pattern color = darkgreen]
    (0,\stripheight) -- (0, 6) -- (1,6) -- (1,5) -- (2,5) -- (2,4.5) -- (3,4.5) -- (3,5) -- (4,5) -- (4,4.25) -- (6,4.25) -- (6,\stripheight) -- cycle ;

   \node at (1,1.5) {$H_1$};

   \node at (4,1.5) {$H_2$};

   \node at (3,6) {$H_3$};
    
\end{tikzpicture}}
    \caption{Holes of a packing produced by BL}
    \label{fig:chazelle:intro:holes}
\end{figure}

With Chazelle's implementation, we determine these feasible BL-stable locations separately within each \textit{hole} of the partial packing, where a hole of 
a partial packing is defined as a connected region of the strip that remains after removing the area occupied by the 
already placed rectangles.
(A packing that can be produced by the BL-heuristic with its holes is shown in Figure~\ref{fig:chazelle:intro:holes}. The placed rectangles are filled in blue and the holes are labeled $H_1,H_2,H_3$)

Before examining how to compute the feasible BL-stable locations for a given hole, we will first formally define the concept of a hole in Section~3.1, 
investigate its basic properties and show that these holes have a more specific structure when they arise from a BL-stable packing. 
We will refer to this restricted subclass as \textit{BLS-holes}. Afterwards, in Section~3.2, we will show 
how to compute the feasible BL-stable locations for an even more restricted subclass of holes, referred to as \textit{nice holes}. 
This will enable us to also compute all feasible locations for a BLS-hole, as we will show in Section~3.3 that each such BLS-hole can be partitioned into a set of nice holes 
(the significant error in Chazelle's original description occurs in this partitioning step).

Simultaneously, we will show that all the functions involved operate in linear time with regard to the size of the data structures
used to store these holes. Finally, Section~3.4 will show how we manage this data structure
throughout the algorithm in linear space and thus preserve the overall linear runtime for placing a new single rectangle.

Throughout this whole section, whenever a definition or a lemma is taken from or inspired by Chazelle's description,
it will be explicitly referenced. In absence of such a reference, 
the respective result should be understood as an original contribution of this paper,
which aims to formalize the description of the implementation and to address the flaw in the original analysis.

\subsection{Properties of Holes and BLS-Holes}

\subsubsection{Basic Definitions}

We begin by properly defining the term \textit{hole}. 
In Chazelle's description, holes are defined as polygons with horizontal and vertical edges that may arise
in a BL-stable packing~\cite{BLCHZ}. We will provide a similar definition using Jordan curves. 
However, we will slightly adapt the terminology, as we will use the term hole in a more general sense to describe
regions that could not just occur in BL-stable packings but also in arbitrary ones. 
Since the BL-stability imposes more constraints on the structure of holes, we will use the term \textit{BLS-holes} to 
refer to a more restricted class of holes that excludes certain configurations that could not occur in BL-stable packings.

Thus, we begin by defining Jordan curves, closely following the presentation by Korte and Vygen in~\cite[Chapter~2.5]{KVY}.
\begin{definition}
    A \textbf{simple Jordan curve} is the image of a continuous injective function $\varphi : [0,1] \to \mathbb{R}^2$; 
    its \textbf{endpoints} are $\varphi(0)$ and $\varphi(1)$. A \textbf{closed Jordan curve} is the image of a 
    continuous function $\varphi : [0,1] \to \mathbb{R}^2$ with $\varphi(0) = \varphi(1)$ and 
    $\varphi( \tau) \neq \varphi(\tau')$ for~$0 \leq \tau < \tau' < 1$.
    
    A \textbf{rectilinear arc} (respectively \textbf{closed rectilinear arc}) is a simple 
    (respectively closed) Jordan curve $J$ that is the union of finitely many axis-aligned straight line segments. We 
    call a set $\{p,p'\} \subset \mathbb{R}^2$ an \textbf{edge} of $J$, if $\text{seg}(p,p')$ is an inclusion-wise maximal
    straight line segment of $J$. A point $p\in \mathbb{R}^2$ is called a \textbf{vertex} of $J$ if $p \in J$
    and $p$ is the intersection of two edges of $J$.

    Let $R = \mathbb{R}^2\setminus J$, where $J$ is the union of finitely many axis-aligned straight line segments. 
    We define the $\textbf{connected regions}$ of $R$ as equivalence classes where two points in $R$ are equivalent 
    if they can be joined by a simple Jordan curve within $R$.
\end{definition}

To define and analyze properties the properties of holes, we will frequently rely on the following theorem, 
which is a special case of the famous Jordan curve theorem.
\begin{theorem}
    If $J$ is a closed rectilinear arc, then $\mathbb{R}^2\setminus J$ splits into exactly two connected regions, 
    each of which has $J$ as its boundary. Exactly one of the regions, called the \textbf{outer region} of $J$, is unbounded.
    We call the other region, which thus is bounded, the \textbf{inner region} of $J$.
\end{theorem} 
\noindent
\textit{Proof: } For a proof, we again refer to~\cite[Chapter~2.5]{KVY}. \qedsymbol
\medskip

With this theorem in mind, we can now define holes: 

\begin{definition}
    We call a set $H \subsetneq \mathbb{R}^2$ a \textbf{hole}, if there exists a closed rectilinear arc $J$ 
    such that $H = J \cup I$, where $I$ is the inner region of $J$. We call $J$ the \textbf{boundary} of $H$. 
    The \textbf{edges and vertices} of $H$ are the edges and vertices of $J$.
\end{definition}

\begin{remark}
    Since the rectilinear arc $J$ in the definition is uniquely determined if it exists, the definition is well-defined.
\end{remark}

It is now easy to see, that if $sp$ is a feasible BL-stable partial packing of a SPI \SPI, then \begin{equation*}
    R = S_W \setminus \bigcup_{r\in R_{sp}} [x_{\min}(r),x_{\max}(r)] \times [y_{\min}(r),y_{\max}(r)]
\end{equation*} splits into open connected regions where exactly one is unbounded while all the others are 
bounded and qualify as holes according to our definition. To ensure uniform treatment of all such connected regions, 
we introduce an artificial upper boundary for the unbounded region at $y = \sum_{r \in R} h(r)$, 
as this boundary does not exclude any BL-stable packings (nor any other arbitrary packings of interest).
Since we aim to treat these connected regions as holes, which are defined as closed regions, 
we therefore consider the \textit{topological closure} of these regions, where the topological closure 
is understood in the classical sense, as presented in standard topology references such as \cite[Chapter~2.17]{TPLGY}.
In summary, this yields the following definition:

\begin{definition}
    Let $sp$ be a partial packing of a SPI \SPI. Then we define the $\textbf{holes}$ of $sp$ as the topological closures of the connected regions of
    \begin{equation*}
        R = ([0,W] \times [0,\sum_{r \in R} h(r)])  \setminus \bigcup_{r\in R_{sp}} [x_{\min}(r),x_{\max}(r)] \times [y_{\min}(r),y_{\max}(r)]
    \end{equation*}
\end{definition}

This completes the formal definitions of the geometric objects, denoted as holes, that we want to examine. 
Here, in Section~3.1, our focus will lie on properties that arise from the structure of their boundary. 
For proving those properties, we will often traverse the boundary in a clockwise direction, and occasionally 
in a counterclockwise direction. This motivates the following definition: 

\begin{definition}\label{chz:def_edge_direction}
    Let $H$ be a hole and $e$ an edge of $H$. Then we say $e$ is a \textbf{(clockwise) leftward} edge, 
    if $e$ is a horizontal edge and by traversing the hole in clockwise order we traverse $e$ from right to left.
    We say that $e$ is a \textbf{anticlockwise-leftward} edge if $e$ is a (clockwise) rightward edge. 
    Clockwise (or respectively anticlockwise) rightward, upward and downward edges are defined analogously.
\end{definition}

\begin{remark}
    Thus, when we refer to leftward (or respectively rightward, downward or upward) edges without explicitly 
    specifying whether we mean clockwise leftward or anticlockwise leftward, it should be understood as clockwise.
\end{remark}

The following very intuitive Lemma will be essential in proving these boundary-related properties mentioned above.

\begin{lemma}\label{chz::horizontal_interior}
    Let $H$ be a hole and $e$ a horizontal edge of $H$. If $e$ is a leftward edge,
    then the interior of $H$ lies above $e$. If $e$ is a rightward edge, then the interior of $H$ lies below $e$.
\end{lemma}
\begin{proof} 
This lemma is a direct consequence of the geometric properties of clockwise traversal. 
Since the result is geometrically evident and a formal proof would be quite technical without offering much additional insight, 
we omit the details here. \qedhere \end{proof}

Since many of the properties discussed above involve local properties of the boundary, 
we will now introduce the notions of \textit{successors} and \textit{predecessors} to refer precisely to neighboring elements.
 Additionally we will define \textit{path notations} to formally describe segments of the boundary between two points or edges.

\begin{definition}
    Let $H$ be a hole, $v$ a vertex of $H$ and $e$ an edge of $H$. We define the \textbf{clockwise predecessor} $\pred(v)$ 
    of $v$ as the vertex that comes right before $v$ in clockwise order and the \textbf{clockwise successor} $\succv(v)$
    of $v$ as the vertex that comes right after $v$ in clockwise order. Analogously, we define $\pred(e), \succv(e)$
    as the (clockwise) preceding/succeeding edge of $e$ respectively.

    Similarly, we define $\apred(v), \asuccv(v), \apred(e), \asuccv(e)$ as the \textbf{anticlockwise predecessor/successor}.
\end{definition}

\begin{definition}
    Let $H$ be a hole and $p_1, p_2 \in \mathbb{R}^2$ two points on the boundary of $H$. 
    Then we define the \textbf{clockwise} $p_1$-$p_2$-\textbf{path} $\Path{p_1,p_2}$ as the rectilinear arc
    which corresponds to the clockwise traversal of the boundary of $H$ starting at $p_1$ and ending at $p_2$. 
    Analogously, we define the \textbf{anticlockwise} $p_1$-$p_2$-\textbf{path} $\APath{p_1,p_2}$ as the 
    rectilinear arc which corresponds to the anticlockwise traversal of the boundary of $H$ starting at $p_1$ and ending at $p_2$.
    
    Now let $e = \{p, \succv(p)\}, e' = \{p', \succv(p')\}$ be edges of $H$. Then we define 
    \begin{equation*}
        \Path{e,e'} := \Path{\succv(p), p'} \text{ and } \APath{e,e'} = \APath{p, \succv(p')}
    \end{equation*}
\end{definition}

And finally, we introduce notation to describe the geometric positions of the edges of a hole. 

\begin{definition}
    For an edge $e = \{p_1,p_2\} \subset \mathbb{R}^2$ of a hole $H$ we define $x_{\max}(e) := \max\{x(p_1),x(p_2)\}$
     and analogously we define $x_{\min}(e), y_{\min}(e)$ and $y_{\min}(e)$. If $e$ is a horizontal edge, then we write  
    \begin{equation*}
        y(e) := y_{\max}(e) = y_{\min}(e)
    \end{equation*}
    and if $e$ is a vertical edge we write 
    \begin{equation*}
        x(e) := x_{\max}(e) = x_{\min}(e)
    \end{equation*}
    
    For a rectilinear arc $J$ we write $x_{\max}(J) := \max\{x(p) | p\in J\}$ and analogously we define 
    $x_{\min}(J),y_{\min}(J),y_{\max}(J)$.
\end{definition}

\subsubsection{Properties of General Holes}

With these definitions in place, we can now begin analyzing the structure of a hole. 
We start by examining specific types of edges (and one special type of vertex) that exhibit special geometric properties.
These include for example \textit{leftmost edges} and \textit{left notches}. 
All of these special edges were also defined in Chazelle's description \cite{BLCHZ} with a slight difference in terminology. 
For example, Chazelle uses the terms \textit{upper notches} and \textit{lower notches}, whereas we will refer to them 
as \textit{top notches} and \textit{bottom notches} to ensure uniqueness in our abbreviations --- as we will abbreviate
a left notch as LN.

The following definition closely follows the definition in Chazelle's description \cite{BLCHZ}.

\begin{definition}\label{chz:special_edges}
    Let $H$ be an arbitrary hole. A \textbf{notch} of $H$ is an edge that displays a reflex angle at each of its endpoints.
    It is said to be a \textbf{left} (respectively \textbf{right, top, bottom}) notch if its preceding edge and its 
    succeeding edge both lie to its left (respectively to its right, above it, below it). 

    An edge of $H$ is said to be a \textbf{leftmost} (respectively \textbf{rightmost, topmost, botmost}) if it displays
    a convex angle at each of its endpoints and if its preceding edge and its succeeding edge both lie entirely to its 
    right (respectively to its left, below it, above it).

    And lastly, a vertex that has a preceding downward edge and a succeeding rightward edge is called a \textbf{falling corner}.

\end{definition}

The hole which can be seen in the figure below provides a clear illustration of all previously defined terms.

\begin{figure}[H]
    \centering
    \begin{tikzpicture}
    \coordinate (A) at (1,0);
    \coordinate (B) at (1,2);
    \coordinate (C) at (2,2);
    \coordinate (D) at (2,3);
    \coordinate (E) at (1,3);
    \coordinate (F) at (1,4);
    \coordinate (G) at (0,4);
    \coordinate (H) at (0,5);
    \coordinate (I) at (3,5);
    \coordinate (J) at (3,6);
    \coordinate (K) at (5,6);
    \coordinate (L) at (5,4);
    \coordinate (M) at (7,4);
    \coordinate (N) at (7,6);
    \coordinate (O) at (9,6);
    \coordinate (P) at (9,5);
    \coordinate (Q) at (11,5);
    \coordinate (R) at (11,4);
    \coordinate (S) at (9,4);
    \coordinate (T) at (9,3);
    \coordinate (U) at (11,3);
    \coordinate (V) at (11,1);
    \coordinate (W) at (9,1);
    \coordinate (X) at (9,0);
    \coordinate (Y) at (8,0);
    \coordinate (Z) at (8,1);
    \coordinate (Z') at (7,1);
    \coordinate (Y') at (7,0);
    \coordinate (X') at (6,0);
    \coordinate (W') at (6,1);
    \coordinate (V') at (4,1);
    \coordinate (U') at (4,0);

    \draw[thick] (A) -- (B) node[midway, right = 0.5cm] (lmelabel) {{\boldmath\scriptsize\textsf{leftmost edge}}};
    \draw[->] (lmelabel) -- ($($(A)!0.5!(B)$) + (1mm,0)$);
    \draw[thick] (B) -- (C);
    \draw[thick] (C) -- (D) node[midway, left = 0.5cm] (lnlabel) {{\boldmath\scriptsize\textsf{left notch}}};
    \draw[->] (lnlabel) -- ($($(C)!0.5!(D)$) + (-1mm,0)$) ;
    \draw[thick] (D) -- (E);
    \draw[thick] (E) -- (F);
    \draw[thick] (F) -- (G);
    \draw[thick] (G) -- (H);
    \draw[thick] (H) -- (I);
    \draw[thick] (I) -- (J);
    \draw[thick] (J) -- (K) node[midway, below = 0.5cm] (umlabel) {{\boldmath\scriptsize\textsf{topmost edge}}};
    \draw[->] (umlabel) -- ($($(J)!0.5!(K)$) + (0,-1mm)$) ;
    
    \draw[thick] (K) -- (L);
    \draw[thick] (L) -- (M) node[midway, below = 0.5cm] (unlabel) {{\boldmath\scriptsize\textsf{top notch}}};
    \draw[->] (unlabel) -- ($($(L)!0.5!(M)$) + (0,-1mm)$) ;
    \draw[thick] (M) -- (N);
    \draw[thick] (N) -- (O);
    \draw[thick] (O) -- (P) node[midway, right = 0.5cm] (fclabel) {{\boldmath\scriptsize\textsf{falling corner}}};
    \draw[->] (fclabel) -- ($ (P) + (1mm,1mm)$) ;
    \draw[thick] (P) -- (Q);
    \draw[thick] (Q) -- (R) node[midway, left = 0.5cm] (rmlabel) {{\boldmath\scriptsize\textsf{rightmost edge}}};
    \draw[->] (rmlabel) -- ($($(Q)!0.5!(R)$) + (-1mm,0)$) ;
    \draw[thick] (R) -- (S);
    \draw[thick] (S) -- (T) node[midway, right = 0.5cm] (rnlabel) {{\boldmath\scriptsize\textsf{right notch}}};
    \draw[->] (rnlabel) -- ($($(S)!0.5!(T)$) + (1mm,0)$) ;
    \draw[thick] (T) -- (U);
    \draw[thick] (U) -- (V);
    \draw[thick] (V) -- (W);
    \draw[thick] (W) -- (X);
    \draw[thick] (X) -- (Y) node[midway, above = 1cm] (bmlabel) {{\boldmath\scriptsize\textsf{bottommost edge}}};
    \draw[->] (bmlabel) -- ($($(X)!0.5!(Y)$) + (0,1mm)$) ;
    \draw[thick] (Y) -- (Z);
    \draw[thick] (Z) -- (Z');
    \draw[thick] (Z') -- (Y');
    \draw[thick] (Y') -- (X');
    \draw[thick] (X') -- (W');
    \draw[thick] (W') -- (V') node[midway, above = 0.5cm] (bnlabel) {{\boldmath\scriptsize\textsf{bottom notch}}};
    \draw[->] (bnlabel) -- ($($(W')!0.5!(V')$) + (0,1mm)$) ;
    \draw[thick] (V') -- (U');
    \draw[thick] (U') -- (A);

\end{tikzpicture}
    \caption{Special edges of a hole (Figure adapted from~\cite{BLCHZ})}
    \label{fig:chz:special_edges_hole}
\end{figure}
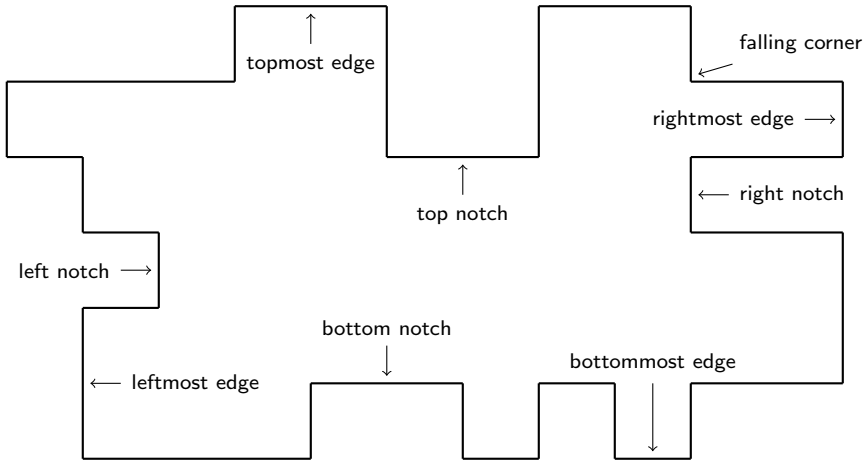

With the formulation of Definition~\ref{chz:def_edge_direction} we can now present an equivalent formulation of
 Definition~\ref{chz:special_edges}  

\begin{lemma}\label{chz:special_edges_equiv}
    Let $H$ be an arbitrary hole and let $e$ be an edge of $H$. Then it holds:
    \begin{itemize}
        \item $e$ is a leftmost edge if and only if $e$ is an upward edge, $\pred(e)$ is a leftward edge and
         $\succv(e)$ is a rightward edge.
        
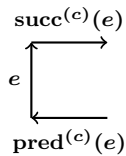
\begin{figure}[H]
            \centering
            \begin{tikzpicture}
  \coordinate (A) at (0,0);
  \coordinate (B) at (-1,0);
  \coordinate (C) at (-1,1);
  \coordinate (D) at (0,1);

  \draw[thick, ->] (A) -- (B) node[midway, below]{{\boldmath\scriptsize\textsf{$\pred(e)$}}} ;
  \draw[thick, ->] (B) -- (C)node[midway, left]{{\boldmath\scriptsize\textsf{$e$}}};
  \draw[thick, ->] (C) -- (D)node[midway, above]{{\boldmath\scriptsize\textsf{$\succv(e)$}}} ;
\end{tikzpicture}
        \end{figure}

        \item $e$ is a left notch if and only if $e$ is an upward edge, $\pred(e)$ is a rightward edge
        and $\succv(e)$ is a leftward edge.
        
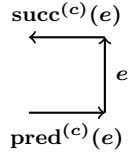
\begin{figure}[H]
            \centering
            \begin{tikzpicture}
  \coordinate (A) at (0,0);
  \coordinate (B) at (1,0);
  \coordinate (C) at (1,1);
  \coordinate (D) at (0,1);

  \draw[thick, ->] (A) -- (B) node[midway, below]{{\boldmath\scriptsize\textsf{$\pred(e)$}}} ;
  \draw[thick, ->] (B) -- (C)node[midway, right]{{\boldmath\scriptsize\textsf{$e$}}};
  \draw[thick, ->] (C) -- (D)node[midway, above]{{\boldmath\scriptsize\textsf{$\succv(e)$}}} ;
\end{tikzpicture}
        \end{figure}
    \end{itemize}
    And equivalently we have (now the statement is only stated by the drawing) : 
\begin{itemize}
    \item \raisebox{-0.1\height}{
    \begin{minipage}[t]{\linewidth}
      \begin{tabularx}{\linewidth}{@{}p{0.26\linewidth}@{\hspace{0.5em}}p{0.18\linewidth}@{\hspace{0em}}p{0.26\linewidth}@{\hspace{0.5em}}p{0.18\linewidth}@{}}
        $e$ is a RME $\Leftrightarrow$ &
        \begin{tikzpicture}[trim left=0pt, trim right=0pt, baseline=(current bounding box.center)]
  \coordinate (A) at (0,0);
  \coordinate (B) at (0.75,0);
  \coordinate (C) at (0.75,0.75);
  \coordinate (D) at (0,0.75);

  \draw[thick, ->] (D) -- (C) node[midway, above]{{\boldmath\scriptsize\textsf{$p^{(c)}(e)$}}};
  \draw[thick, ->] (C) -- (B)node[midway, right]{{\boldmath\scriptsize\textsf{$e$}}};
  \draw[thick, ->] (B) -- (A)node[midway, below]{{\boldmath\scriptsize\textsf{$s^{(c)}(e)$}}} ;
\end{tikzpicture}&
        ; $e$ is a RN $\Leftrightarrow$ &
        \begin{tikzpicture}[trim left=0pt, trim right=0pt, baseline=(current bounding box.center)]
  \coordinate (A) at (0,0);
  \coordinate (B) at (0.75,0);
  \coordinate (C) at (0.75,0.75);
  \coordinate (D) at (0,0.75);

  \draw[thick, ->] (C) -- (D) node[midway, above]{{\boldmath\scriptsize\textsf{$p^{(c)}(e)$}}};
  \draw[thick, ->] (D) -- (A)node[midway, left]{{\boldmath\scriptsize\textsf{$e$}}};
  \draw[thick, ->] (A) -- (B)node[midway, below]{{\boldmath\scriptsize\textsf{$s^{(c)}(e)$}}} ;
\end{tikzpicture}
      \end{tabularx}
    \end{minipage}
    }
    
    \item \raisebox{0.1\height}{
    \begin{minipage}[t]{\linewidth}
      \begin{tabularx}{\linewidth}{@{}p{0.26\linewidth}@{\hspace{0.5em}}p{0.18\linewidth}@{\hspace{0em}}p{0.26\linewidth}@{\hspace{0.5em}}p{0.18\linewidth}@{}}
        $e$ is a TME $\Leftrightarrow$ &
        \raisebox{3mm}{\begin{tikzpicture}[trim left=0pt, trim right=0pt, baseline=(current bounding box.center)]
  \coordinate (A) at (0,0);
  \coordinate (B) at (0.75,0);
  \coordinate (C) at (0.75,0.75);
  \coordinate (D) at (0,0.75);

  \draw[thick, ->] (A) -- (D) node[midway, left]{{\boldmath\scriptsize\textsf{$p^{(c)}(e)$}}};
  \draw[thick, ->] (D) -- (C)node[midway, above]{{\boldmath\scriptsize\textsf{$e$}}};
  \draw[thick, ->] (C) -- (B)node[midway, right]{{\boldmath\scriptsize\textsf{$s^{(c)}(e)$}}};
\end{tikzpicture}}&
        ; $e$ is a TN $\Leftrightarrow$ &
        \begin{tikzpicture}[trim left=0pt, trim right=0pt, baseline=(current bounding box.center)]
  \coordinate (A) at (0,0);
  \coordinate (B) at (0.75,0);
  \coordinate (C) at (0.75,0.75);
  \coordinate (D) at (0,0.75);

  \draw[thick, ->] (D) -- (A) node[midway, left]{{\boldmath\scriptsize\textsf{$p^{(c)}(e)$}}};
  \draw[thick, ->] (A) -- (B) node[midway, below]{{\boldmath\scriptsize\textsf{$e$}}};
  \draw[thick, ->] (B) -- (C)node[midway, right]{{\boldmath\scriptsize\textsf{$s^{(c)}(e)$}}};
\end{tikzpicture}
      \end{tabularx}
    \end{minipage}
  }

    \item \raisebox{0.1\height}{
    \begin{minipage}[t]{\linewidth}
      \begin{tabularx}{\linewidth}{@{}p{0.26\linewidth}@{\hspace{0.5em}}p{0.18\linewidth}@{\hspace{0em}}p{0.26\linewidth}@{\hspace{0.5em}}p{0.18\linewidth}@{}}
        $e$ is a BME $\Leftrightarrow$ &
        \begin{tikzpicture}[trim left=0pt, trim right=0pt, baseline=(current bounding box.center)]
  \coordinate (A) at (0,0);
  \coordinate (B) at (0.75,0);
  \coordinate (C) at (0.75,0.75);
  \coordinate (D) at (0,0.75);

  \draw[thick, ->] (C) -- (B) node[midway, right]{{\boldmath\scriptsize\textsf{$p^{(c)}(e)$}}};
  \draw[thick, ->] (B) -- (A) node[midway, below]{{\boldmath\scriptsize\textsf{$e$}}};
  \draw[thick, ->] (A) -- (D)node[midway, left]{{\boldmath\scriptsize\textsf{$s^{(c)}(e)$}}} ;
\end{tikzpicture}&
        ; $e$ is a BN $\Leftrightarrow$ &
        \raisebox{3mm}{\begin{tikzpicture}[trim left=0pt, trim right=0pt, baseline=(current bounding box.center)]
  \coordinate (A) at (0,0);
  \coordinate (B) at (0.75,0);
  \coordinate (C) at (0.75,0.75);
  \coordinate (D) at (0,0.75);

  \draw[thick, ->] (B) -- (C) node[midway, right]{{\boldmath\scriptsize\textsf{$p^{(c)}(e)$}}};
  \draw[thick, ->] (C) -- (D) node[midway, above]{{\boldmath\scriptsize\textsf{$e$}}};
  \draw[thick, ->] (D) -- (A)node[midway, left]{{\boldmath\scriptsize\textsf{$s^{(c)}(e)$}}} ;
\end{tikzpicture}}
      \end{tabularx}
    \end{minipage}
  }
  
\end{itemize}
where $s^{(c)}(e):= \succv(e), p^{(c)}(e) := \pred(e)$. Moreover, RME stands for rightmost edge, RN for right notch
 and the remaining abbreviations are defined analogously.
\end{lemma}
\noindent
\begin{proof} This statement is a direct consequence of Definition~\ref{chz:special_edges} and Lemma~\ref{chz::horizontal_interior}

\end{proof}

With this equivalent definition of the special edges in mind, we can now develop an intuitive criterion for 
determining when a path $\Path{p_1,p_2}$ contains such a special edge. 

\newcommand{\alignedimport}[2]{%
  \begin{tikzpicture}
    \begin{scope}
      \input{#1/#2}
    \end{scope}
    \path let \p1 = (#1/#2.alignpoint) in
      node at ($(-\x1,0)$) {};
  \end{tikzpicture}%
}

\newcommand{\alignedimporrt}[2]{%
  \begin{tikzpicture}[y=1cm, x=1cm, yscale=1, xscale=-1, 
  every node/.append style={scale=1}, 
  inner sep=0pt, outer sep=0pt]
    \begin{scope}
      \input{#1/#2}
    \end{scope}
    \path let \p1 = (alignpoint) in node[shift={(-\x1+,0)}] {};
  \end{tikzpicture}%
}

\begin{lemma}\label{chz::special_existence}
    Let $p_1,p_2$ be two points on the boundary of a hole $H$ with $x(p_1) = x(p_2)$ and $y(p_1) < y(p_2)$. Then it holds :
    \begin{enumerate}[label=(\roman*)]
        \item If $x_{\min}(\Path{p_1,p_2}) < x(p_1)$ and $x_{\max}{(\Path{p_1,p_2})} \leq x(p_1)$, 
        then the path $\Path{p_1,p_2}$ contains a LME. 
        
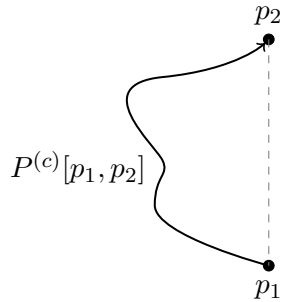
\begin{figure}[H]
          \centering
          \begin{tikzpicture}[y=1cm, x=1cm, yscale=1, xscale=1, 
  every node/.append style={scale=1}, 
  inner sep=0pt, outer sep=0pt]
\node (alignpoint) at (3.0, 0) {};
\begin{scope}[xshift = -2cm]
  \path[draw=black, ->, line cap=butt, line join=miter, line width=0.0265cm] 
    (3.0, 1.0)
    .. controls (3.0, 1.0) and (1.5, 1.4) .. (1.5, 1.8)
    .. controls (1.5, 2.2) and (1.7, 2.2) .. (1.6, 2.4)
    .. controls (1.5, 2.6) and (0.6, 3.4) .. (1.6, 3.5)
    .. controls (2.6, 3.6) and (2.95, 3.95) .. (2.95, 3.95);

  \filldraw[black] (3.0, 1.0) circle (2pt) node[below = 0.2cm] {$p_1$};
  \filldraw[black] (3.0, 4.0) circle (2pt) node[above = 0.2cm] {$p_2$};

  \node at (0.49,2.3) {$\Path{p_1,p_2}$} ;
  \draw[dashed, gray] (3.0, 4.0) -- (3.0, 1.0);
\end{scope}
\end{tikzpicture}
        \end{figure}
        \item If $x_{\max}(\Path{p_1,p_2}) > x(p_1)$ and $x_{\min}{(\Path{p_1,p_2})} \geq x(p_1)$, 
        then the path $\Path{p_1,p_2}$ contains a LN.
        
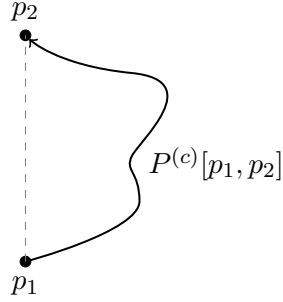
\begin{figure}[H]
          \centering
          \begin{tikzpicture}[y=1cm, x=1cm, yscale=1, xscale=-1, 
  every node/.append style={scale=1}, 
  inner sep=0pt, outer sep=0pt]
\node (alignpoint) at (3.0, 0) {};
\begin{scope}[xshift = -3.5cm]
  \path[draw=black, ->, line cap=butt, line join=miter, line width=0.0265cm] 
    (3.0, 1.0)
    .. controls (3.0, 1.0) and (1.5, 1.4) .. (1.5, 1.8)
    .. controls (1.5, 2.2) and (1.7, 2.2) .. (1.6, 2.4)
    .. controls (1.5, 2.6) and (0.6, 3.4) .. (1.6, 3.5)
    .. controls (2.6, 3.6) and (2.95, 3.95) .. (2.95, 3.95);
  \filldraw[black] (3.0, 1.0) circle (2pt) node[below = 0.2cm] {$p_1$};
  \filldraw[black] (3.0, 4.0) circle (2pt) node[above = 0.2cm] {$p_2$};

  \node at (0.49,2.3) {$\Path{p_1,p_2}$} ;
  \draw[dashed, gray] (3.0, 1.0) -- (3.0, 4.0);
\end{scope}
\end{tikzpicture}
        \end{figure}
    \end{enumerate}
\end{lemma}

\begin{figure}[H]
    \centering
    \scalebox{0.85}{\begin{tikzpicture}
    \filldraw[black] (5.0, 1.0) circle (2pt) node[right = 0.2cm] {$p_1$};
    \filldraw[black] (5.0, 4.0) circle (2pt) node[right = 0.2cm] {$p_2$};
    \filldraw[black] (0.9, 2.2) circle (1.5pt) node[above = 0.2cm] {$v_0$};
    \filldraw[black] (0.9, 1.1) circle (1.5pt) node [left = 0.2cm] {$v_1$};
    \filldraw[black] (1.7, 1.1) circle (1.5pt) node [right = 0.2cm] {$v_2$};

    \draw[thick, ->]
      (5.0, 1.0)
        .. controls (4.0, 1.8) and (4.5, 2.4) .. (3.8, 2.7)
        .. controls (3.0, 3.0) and (2.0, 2.8) .. (1.0, 2.2);

    \draw[thick , ->] (0.9,2.2) -- (0.9,1.2);
    \draw[thick , ->] (0.9,1.1) -- (1.6,1.1);

    \draw[thick,blue] (0.9,1.1) -- (0.9,-3);
    \draw[thick, blue] (0.9,-3) -- (5.0, -3);
    \draw[thick, blue] (5.0,-3) -- (5.0,1.0);

    \node at (3, 3.3) {$\Path{p_1,v_1}$} ;
    \node at (0.7,1.7) {$e$}; 
    \node[anchor=east, black] at (0.8, -3) {\footnotesize $y_{\min}(P)-1$};
    \draw[very thick] (0.7,-3) -- (1.1,-3); 

    \path[pattern=north east lines, pattern color=gray]
      (5.0, 1.0)
        .. controls (4.0, 1.8) and (4.5, 2.4) .. (3.8, 2.7)
        .. controls (3.0, 3.0) and (2.0, 2.8) .. (1.0, 2.2)
        -- (0.9,2.2)
        -- (0.9,-3)
        -- (5.0,-3)
        -- (5.0,1.0)
        -- cycle;

    \node at ({(5+0.9)/2}, {(1.1-3)/2}) {$I$};

\end{tikzpicture}}
    \caption{Illustration of Proof \ref{chz::special_existence}, here $I$ is the inner region of $J$}
    \label{chz:fig:path_proof}
\end{figure}
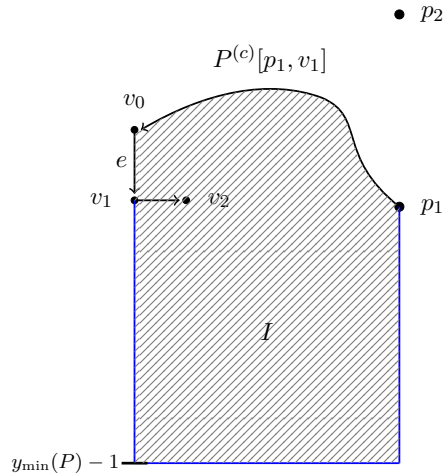

\noindent
\begin{proof} We will only prove $(i)$, as the proof of $(ii)$ is completely analogous. 
  So assume that we are in the setting of $(i)$. We may assume without loss of generality, 
  that there is no point $p' \in \Path{p_1,p_2}$ with $p' \notin \{p_1,p_2\}$ but $x(p') = x(p_1)$. 
  Thus, the first edge of $\Path{p_1,p_2}$ has to be a leftward edge and the last edge has to be a rightward edge. 
  Now let $e$ be a vertical edge of $\Path{p_1,p_2}$ with $x(e)$ minimal. Since a vertical edge can neither be the last 
  nor the first edge on $\Path{p_1,p_2}$, we have that both $\pred(e)$ and $\succv(e)$ must lie on $\Path{p_1,p_2}$ and 
  both of them have to lie to the right of $e$. If $e$ is an upward edge, then the statement follows directly from 
  Lemma~\ref{chz:special_edges_equiv}. So assume $e$ is a downward edge. Then $\succv(e)$ is a rightward edge and we 
  denote its endpoints by $v_1$ and $v_2$, where $v_1$ is the left vertex and $v_2$ the right. Now consider the closed 
  rectilinear arc (see Figure~\ref{chz:fig:path_proof} for an illustration)
\begin{equation*}
    J := \Path{p_1, v_1} \cup \text{seg}\big(v_1, (x(v_1), y_{\min}(P)-1),  (x(p_1), y_{\min}(P)-1), p_1\big)
\end{equation*}
We have that $x_{\max}(J) \leq x(p_2)$ and thus $p^* = p_2 +_{x} 1$ lies in the outer region of $J$. But this also 
implies that $p_2$ lies in the outer region of $J$ since $p^*$ and $p_2$ can trivially be joined by a rectilinear arc 
within $\mathbb{R}^2\setminus J$. However, it is also easy to see that $v_2$ lies in the inner region of $J$. This now 
leads to a contradiction, because this implies that $\Path{v_2,p_2}$ has to cross the boundary of $J$, which is disallowed
 by definition of $J$ since both $\Path{v_2,p_2}$ and $\Path{p_1,v_1}$ are sub-paths of $P$. This proves the statement. \qedhere
\end{proof}

\begin{remark}
    Using Lemma \ref{chz:special_edges_equiv} one can determine analogous criteria for the existence of right/top/bottom-most edges
    (and respectively notches). These criteria can also be derived directly from Lemma \ref{chz::special_existence} by rotating
    the hole, as for example rotating a hole by 180 degrees transforms rightmost edges into leftmost edges and right notches
    into left notches.
\end{remark}

With this criterion for the existence of special edges in mind, we will now derive a lemma which examines the 
relations among those special edges. We will again formulate this statement only for leftmost edges and left notches, 
and the analogous statement for the other special edges can again be derived via rotations of the hole.

\begin{lemma}\label{chz:most_notch_relation} Let $H$ be an arbitrary hole. Then the following statements hold: 
    \begin{enumerate}[label=(\roman*), labelwidth=1.5em, labelsep=0.5em, leftmargin=*, align=left, widest=ii]
        \item Let $L_1,L_2$ be two (different) leftmost edges of $H$. If the path $\Path{L_1,L_2}$ 
        does not contain a left notch, then the path $\Path{L_2,L_1}$ does. Moreover, in this case the path
        $\Path{L_1,L_2}$ contains a rightmost edge.
        \item Let $N_1,N_2$ be two (different) left notches of $H$. If the path $\Path{N_1,N_2}$ does not contain a 
        leftmost edge, then the path $\Path{N_2,N_1}$ does. Moreover, in this case the path $\Path{N_1,N_2}$ contains a right notch.
    \end{enumerate} 
\end{lemma}
\begin{proof} We will again only prove $(i)$, as the proof for $(ii)$ is again completely analogous. 
So assume we are in the scenario of $(i)$ and let
\begin{equation*}
    \pred(L_1) = \{v_1,v_2\}, L_1 = \{v_2,v_3\}, \succv(L_1) = \{v_3,v_4\}
\end{equation*} and 
\begin{equation*}
    \pred(L_2) = \{w_1,w_2\}, L_2 = \{w_2,w_3\}, \succv(L_2) = \{w_3,w_4\}
\end{equation*}
 Without loss of generality we can assume that $x(v_3) \leq x(w_3) < x(w_4)$. Now let $p$ be the last point on the path 
 $\Path{v_3,w_2}$ such that $x(p) = x(w_2)$ and $p \neq w_2$. Observe that we have $x_{\min}(\Path{p,w_2}) \geq x(w_2)$ 
 and $x_{\max}(\Path{p,w_2}) \geq x(w_1) > x(w_2)$, due to the minimality of $p$ and the fact that $w_1$ lies on that path. 
 Thus, if $y(p) < y(w_2)$, Lemma~\ref{chz::special_existence} implies that the sub-path $\Path{p,w_2}$ of $\Path{L_1,L_2}$ 
 contains a left notch. 

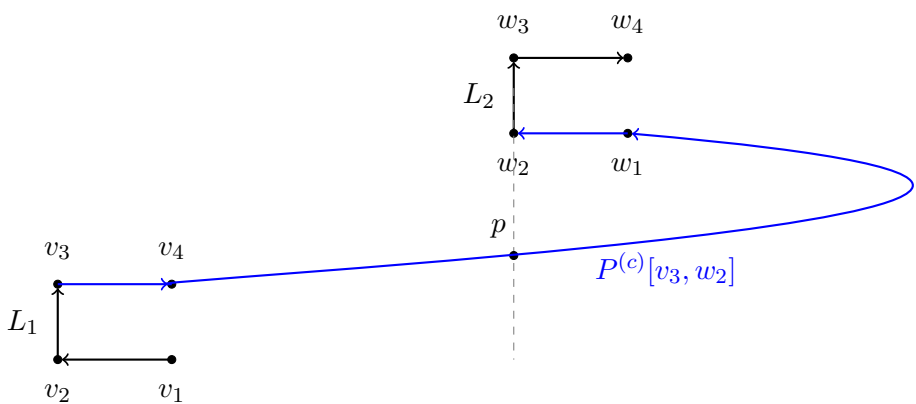
\begin{figure}[H]
    \centering
    \begin{tikzpicture}
\coordinate (v1) at (1.5, 0);
\coordinate (v2) at (0, 0);
\coordinate (v3) at (0, 1);
\coordinate (v4) at (1.5, 1);

\coordinate (w1) at (7.5, 3);
\coordinate (w2) at (6, 3);
\coordinate (w3) at (6, 4);
\coordinate (w4) at (7.5, 4);

\filldraw[black] (v1) circle (1.5pt) node[below = 0.2cm] {$v_1$};
\filldraw[black] (v2) circle (1.5pt) node[below = 0.2cm] {$v_2$};
\filldraw[black] (v3) circle (1.5pt) node[above = 0.2cm] {$v_3$};
\filldraw[black] (v4) circle (1.5pt) node[above = 0.2cm] {$v_4$};

\filldraw[black] (w1) circle (1.5pt) node[below = 0.2cm] {$w_1$};
\filldraw[black] (w2) circle (1.5pt) node[below = 0.2cm] {$w_2$};
\filldraw[black] (w3) circle (1.5pt) node[above = 0.2cm] {$w_3$};
\filldraw[black] (w4) circle (1.5pt) node[above = 0.2cm] {$w_4$};

\draw[thick, ->, shorten >=1.5pt] (v1) -- (v2);
\draw[thick, ->, shorten >=1.5pt] (v2) -- (v3) node[pos = 0.5, left = 0.1cm]{$L_1$};
\draw[thick, ->, shorten >=1.5pt, blue] (v3) -- (v4);

\draw[thick, ->, shorten >=1.5pt, blue] (w1) -- (w2);
\draw[thick, ->, shorten >=1.5pt] (w2) -- (w3) node[pos = 0.5, left = 0.1cm]{$L_2$};
\draw[thick, ->, shorten >=1.5pt] (w3) -- (w4);

\draw[dashed,gray] ($(w2) + (0,0.6)$) -- (6,0);

\filldraw[black] (6,1.38) circle (1.5pt) node[left = 0.2cm, above = 0.1cm] {$p$};

\node at (8, 1.2) {$\textcolor{blue}{\Path{v_3,w_2}}$};

\draw[thick, ->, shorten >=1.5pt, blue]
  (v4) .. controls (0, 1) and (19, 2) .. (w1);

\end{tikzpicture}
    \caption{Proof of Lemma~\ref{chz:most_notch_relation}}
\end{figure}

So we have in our setting that $y(p) > y(w_2)$ and thus by Lemma \ref{chz::special_existence} we conclude that 
$\Path{p,w_2}$ contains a rightmost edge which shows the second part of the statement. Now consider the closed 
rectilinear arc $J = \Path{p,w_2} \cup \text{seg}(w_2,p)$. It is easy to verify that $w_4$ lies in the inner region of 
$J$. As $x(v_2) \leq x(w_2) = x_{\min}(J)$, we deduce that $v_2$ lies either on the boundary of $J$ or in the outer 
region of $J$. However, in both cases, the path $\Path{w_3,v_2} = \Path{L_2,L_1}$ must have a point on the boundary of 
$J$ that is not $w_3$. Let $p^*$ be the first such point. Since the path $\Path{w_3,v_2}$ can intersect $J$ only at the 
segment $\text{seg}(w_3,p)$, we have that $x(p^*) = x(w_3)$ and $y(p^*) > y(w_3)$. But since $p^*$ is the first such point 
and since the path starts with the edge $\{w_3,w_4\}$, we have $x_{\min}(\Path{w_3,p^*}) \geq x(w_3)$ and 
$x_{\max}(\Path{w_3,p^*}) >  x(w_3)$. Thus, by Lemma~\ref{chz::special_existence} we conclude that the sub-path 
$\Path{w_3,p^*}$ of the path $\Path{L_2,L_1}$ contains a left notch (an illustration of that scenario can be seen in 
Figure~\ref{chz:fig:proof:most_notch_relation}). This proves the statement

\vspace*{-7\baselineskip}
\begin{figure}[H]
    \centering
    \definecolor{myred}{rgb}{1, 0, 0}
\definecolor{mydarkgreen}{rgb}{0,0.25,0}
\begin{tikzpicture}
\coordinate (v1) at (1.5, 0);
\coordinate (v2) at (0, 0);
\coordinate (v3) at (0, 1);
\coordinate (v4) at (1.5, 1);

\coordinate (w1) at (7.5, 3);
\coordinate (w2) at (6, 3);
\coordinate (w3) at (6, 4);
\coordinate (w4) at (7.5, 4);

\filldraw[black] (v1) circle (1.5pt) node[below = 0.2cm] {$v_1$};
\filldraw[black] (v2) circle (1.5pt) node[below = 0.2cm] {$v_2$};
\filldraw[black] (v3) circle (1.5pt) node[above = 0.2cm] {$v_3$};
\filldraw[black] (v4) circle (1.5pt) node[above = 0.2cm, xshift = -0.25cm] {$v_4$};

\filldraw[black] (w1) circle (1.5pt) node[below = 0.2cm] {$w_1$};
\filldraw[black] (w2) circle (1.5pt) node[below = 0.2cm] {$w_2$};
\filldraw[black] (w3) circle (1.5pt) node[above = 0.2cm, xshift = 0.25cm] {$w_3$};
\filldraw[black] (w4) circle (1.5pt) node[above = 0.2cm] {$w_4$};

\draw[thick, ->, shorten >=1.5pt] (v1) -- (v2);
\draw[thick, ->, shorten >=1.5pt] (v2) -- (v3) node[pos = 0.5, left = 0.1cm]{$L_1$};
\draw[thick, ->, shorten >=1.5pt, blue] (v3) -- (v4);

\draw[thick, ->, shorten >=1.5pt, blue] (w1) -- (w2);
\draw[thick, ->, shorten >=1.5pt, red] (w2) -- (w3) node[pos = 0.5, left = 0.1cm]{$\textcolor{black}{L_2}$};
\draw[thick, ->, shorten >=1.5pt, mydarkgreen] (w3) -- (w4);

\node at (4, 5.8) {$\textcolor{blue}{\Path{v_3,w_2}}$};

\draw[thick, ->, shorten >=1.5pt, blue]
  (v4) .. controls (4, 10) and (16, 4) .. (w1);

\draw[thick, myred, shorten <= 1.5pt]
    (w3) -- ($(w2)+(0,2.72)$);

\node at (5.2,4.5) {\footnotesize $\textcolor{myred}{\text{seg}(w_2,p)}$} ;

\filldraw[black] ($(w2)+(0,2.72)$) circle (1.5pt) node[above = 0.2cm] {$p$};

\draw[thick, ->, shorten >=1.5pt, mydarkgreen]
  (w4) .. controls ($(w4)+(4,0.2)$) and ($(w2)+(1,2.23)$) .. ($(w2)+(0,2.23)$);

\filldraw[mydarkgreen] ($(w2)+(0,2.23)$) circle (1.5pt) node[left = 0.1cm] {\footnotesize$\textcolor{mydarkgreen}{p^*}$} ;

\node at (9,3.8) {\footnotesize $\textcolor{mydarkgreen}{\Path{w_3,p^*}}$};

\end{tikzpicture}
    \caption{Proof of Lemma~\ref{chz:most_notch_relation}}
    \label{chz:fig:proof:most_notch_relation}
\end{figure}
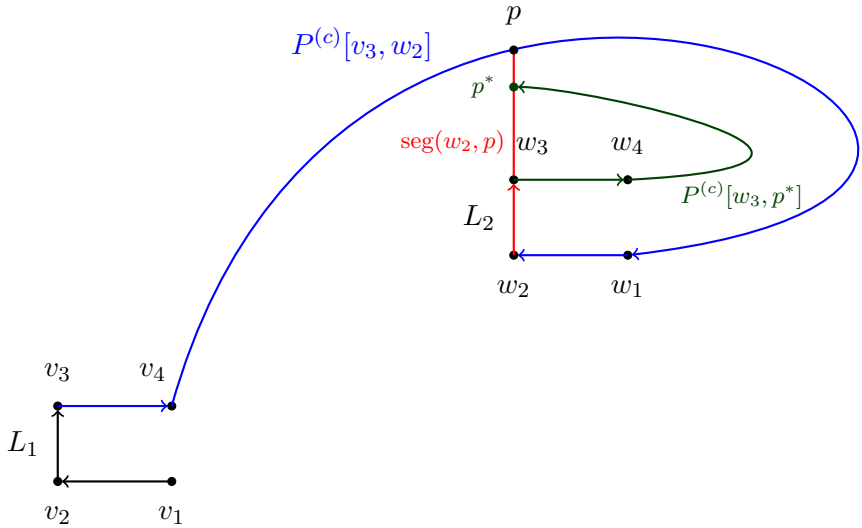

\qedhere
\end{proof}

We now conclude our examination of general hole properties with an intuitive lemma about the presence of leftmost and rightmost edges.

\begin{lemma}\label{chz:geq_one_most_edge}
    Every hole $H$ contains at least one leftmost edge and at least one rightmost edge
\end{lemma}
\begin{proof} Consider the vertical edge $e$ of $H$ which minimizes $x(e)$. Thus both $\pred(e)$ and $\succv(e)$ lie to 
  the right of e. From Lemma~\ref{chz:special_edges_equiv} we deduce, that $e$ is either a leftmost edge or a right notch. 
  However as the interior of the hole is on the left side of a right notch, the latter is impossible. This proves the leftmost 
  case, the rightmost case follows again by rotation of the hole. \qedhere 
\end{proof}

\subsubsection{Properties of BLS-Holes}

We will now start to examine the properties of a restricted class of holes, which exclude certain configurations that 
could not possibly occur as holes of feasible BL-stable packings. This restriction is fundamentally based on the 
following two lemmas, where both of them were also described in Chazelle's paper~\cite{BLCHZ}. The proof of the first 
lemma is taken directly from that work. However, for the second lemma, Chazelle's description contained a flaw in the 
case analysis. Thus in our version, the Case 1 analysis follows Chazelle's argument, while Case 2 is proven using an 
alternative approach to correct the flaw.

\begin{lemma}\label{chz:BLS_lemma}
    Let $sp$ be a feasible BL-stable partial packing of a SPI \SPI \ and let $H$ be a hole of $sp$. 
    Then $H$ contains neither right notches nor top notches.
\end{lemma}
\begin{proof} Assume $H$ would have a right notch $e$. Thus by the definition of a hole of a partial packing there is a 
    placed rectangle $r \in R_{sp}$ with $x_{\min}(r) = x(e)$ and $(y_{\min}(r), y_{\max}(r)) \subseteq (y_{\min}(e),y_{\max}(e))$. 
    This however implies, that $r$ could be shifted more to the left without breaking the feasibility of $sp$, 
    which is a contradiction to the BL-stability. 
\begin{figure}[H]
    \centering
    \begin{tikzpicture}
    \filldraw[fill = blue!30, draw = black] (0,0) rectangle (1.9,1);
    \draw[thick, red] (0,0) -- (1,0);
    \draw[thick, red] (0,0) -- (0,1) node[midway, left] {$e$};
    \draw[thick, red] (0,1) -- (1,1);
    \draw[thick, red] (1,1) -- (1,2);
    \draw[thick, red] (1,0) -- (1,-1);

    \fill[pattern = north east lines, pattern color = darkgreen] (1,-1) -- (-1,-1) -- (-1,2) 
    -- (1,2) -- (1,1) -- (0,1) -- (0,0) -- (1,0) -- cycle;
    \node at ({(1.9)/2}, {(1)/2}) {$r$} ;

    \draw[->, thick, black] (2.8,0.5) -- (2, 0.5);
\end{tikzpicture}
    \caption{The right notch scenario (interior of $H$ is hatched green)}
\end{figure}
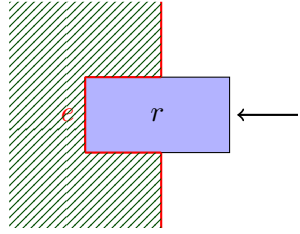

The top notch case is completely analogous, as $r$ could be shifted more downward in this case. \qedhere 
\end{proof}
\begin{lemma}\label{chz:falling_corner_lemma}
   Let $sp$ be a feasible BL-stable partial packing of a SPI \SPI \ and let $H$ be a hole of $sp$. Then $H$ contains at 
   most one falling corner.
\end{lemma}
\begin{proof} Assume for contradiction that $H$ contains two distinct falling corners $c_1,c_2$. We can assume without 
    loss of generality that the path $\Path{c_1,c_2}$ does not contain more edges than the path $\Path{c_2,c_1}$. \\ 
\textit{Case 1: } $\Path{c_1,c_2}$ contains at most 2 edges. \\  
It is easy to verify that $\Path{c_1,c_2}$ cannot contain only one edge by definition of a falling corner, so in this 
case we have that $\Path{c_1,c_2}$ contains exactly 2 edges. Now let 
\begin{equation*} 
    e_1 := \{\pred (c_1),c_1\}, e_2 := \{c_1,\succv(c_1)\}
\end{equation*}
and 
\begin{equation*}
    e_3 = \{\pred(c_2),c_2\}, e_4 = \{c_2, \succv(c_2)\}
\end{equation*} By definition of a hole of a feasible packing, we get that there is a placed rectangle $r_1$ with 
$(x_{\min}(r_1),y_{\min}(r_1)) = c_1$. It now has to hold that $x_{\max}(r_1) > x_{\max}(e_2)$ as otherwise the 
rectangle $r_1$ could be shifted more downwards without breaking the feasibility of $sp$.
Analogously one can verify, that there has to be a rectangle $r_2$ with $(x_{\min}(r_2), y_{\min}(r_2)) = c_2$. 
But now we must have $y_{\max}(r_2) < y_{\min}(r_1) = y_{\max}(e_3)$, as otherwise those two rectangles would overlap. 
This however implies that the rectangle $r_2$ could be shifted more to the left without breaking the feasibility of $sp$, 
again contradicting the BL-stability~(illustrated in Figure 7). 
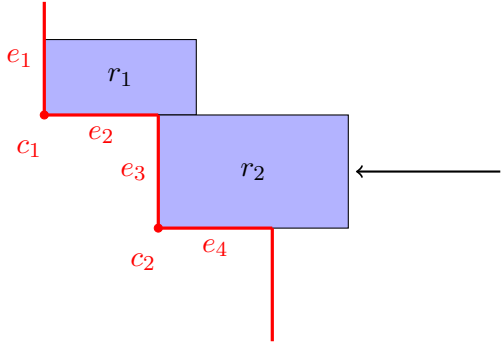
\begin{figure}[!tp]
    \centering
    \scalebox{1}{\begin{tikzpicture}
    \coordinate (pc1) at (0,1.5);
    \coordinate (c1) at (0,0);
    \coordinate (sc1) at (1.5,0);
    \coordinate (c2) at (1.5,-1.5);
    \coordinate (sc2) at (3,-1.5);

    \filldraw[fill = blue!30, draw = black] (c1) rectangle (2,1);
    \filldraw[fill = blue!30, draw = black] (c2) rectangle (4,0);

    \draw[very thick, red] (pc1) -- (c1) node[midway, left] {$e_1$};
    \draw[very thick, red] (c1) -- (sc1) node[midway, below] {$e_2$};
    \draw[very thick, red] (sc1) -- (c2) node[midway, left] {$e_3$};
    \draw[very thick, red] (c2) -- (sc2) node[midway, below] {$e_4$};
    \draw[very thick,red] (sc2) -- (3,-3);

    \filldraw[red] (c1) circle (1.5pt) node[left = 0.2cm, below = 0.2cm] {$c_1$};
    \node at (1,0.5) {$r_1$};
    \node at ({(4+1.5)/2},{(-1.5)/2}) {$r_2$};

    \filldraw[red] (c2) circle (1.5pt) node[left = 0.2cm, below = 0.2cm] {$c_2$};
    \draw[->, thick] (6,-0.75) -- (4.1,-0.75);

\end{tikzpicture}}
    \caption{Case 1 of Lemma~\ref{chz:falling_corner_lemma}}
\end{figure}

\noindent 
\textit{Case 2:} $\Path{c_1,c_2}$ contains more than 2 edges. \\ 
Let $e_1,e_2,e_3$ be the first 3 edges of the path $\Path{c_1,c_2}$ in this order. By definition of a falling corner, 
$e_1$ is a rightward edge. Therefore, $e_2$ is a vertical edge, but it cannot be an upward edge as this would imply the 
existence of a top notch by Lemma~\ref{chz:special_edges_equiv}. This implies that $e_2$ is a downward edge. Now $e_3$ 
is again a horizontal edge which cannot be a rightward edge, as this would imply the existence of a falling corner just 
two edges away from $c_1$, which is prohibited by the Case 1 analysis. Thus, we have that $e_3$ is a leftward edge and 
this implies that by Lemma~\ref{chz:special_edges_equiv}, $e_2$ is a rightmost edge. Analogously, one can show that the 
path $\Path{c_2,c_1}$ also contains a rightmost edge, since this path also contains more than two edges by assumption. 
Thus, the boundary of $H$ contains 2 distinct rightmost edges $R_1,R_2$. However, since Lemma~\ref{chz:most_notch_relation} 
now implies that the path $\Path{R_1,R_2}$ or the path $\Path{R_2,R_1}$ must contain a right notch, this is a contradiction 
to Lemma~\ref{chz:BLS_lemma} . \qedhere    
\end{proof}

Having now identified these specific types of boundary structures that cannot occur in BL-stable packings, we can now 
define the term of a BLS-hole, which corresponds to Chazelle's definition of a \textit{hole} \cite{BLCHZ}.
\begin{definition}
    A hole $H$ is a \textbf{BLS-hole} if it contains neither right notches nor top notches and contains at most one 
    falling corner. 
\end{definition}

\begin{remark}
    By Lemma~\ref{chz:BLS_lemma} and~\ref{chz:falling_corner_lemma}, every hole of a feasible BL-stable partial packing 
    is already a BLS-hole.
\end{remark}

This restriction to BLS-holes provides a significantly more constrained structure of the boundary of a hole. With this 
added regularity, we will be able to establish several structural properties of BLS-holes that rely fundamentally on the 
absence of right and top notches. 

\begin{corollary}\label{chz:exactly:one:rme:and:tme}
    Let $H$ be a BLS-hole. Then $H$ contains exactly one rightmost edge and exactly one topmost edge.
\end{corollary}
\begin{proof} By Lemma~\ref{chz:geq_one_most_edge} we have that $H$ contains at least one rightmost edge and in the 
    proof of Lemma~\ref{chz:falling_corner_lemma} it was already shown that $H$ cannot contain two distinct rightmost 
    edges. The topmost case is now completely analogous. \qedhere
\end{proof}

\begin{corollary}\label{chz:left_notch_paths_contain_leftmosts}
    Let $N_1,N_2$ be two different left notches of a BLS-hole $H$. Then both $\Path{N_1,N_2}$ and $\Path{N_2,N_1}$ 
    contain a leftmost edge.
\end{corollary}
\begin{proof}
    Follows directly from Lemma~\ref{chz:most_notch_relation}. \qedhere 
\end{proof}

\begin{corollary}\label{chz:notch_rightmost_contains_lefmost}
    Let $H$ be a BLS-hole, $R$ the unique rightmost edge of $H$ and $N$ a left notch of $H$. Then both paths 
    $\Path{R,N}$ and $\Path{N,R}$ contain a leftmost edge. 
\end{corollary}
\begin{proof}
Let $L$ be the vertical edge of the path $\Path{R,N}$ with minimal $x$-value. If $L$ 
 is neither the first nor the last vertical edge on this path, then it follows directly from the minimality that 
 both adjacent horizontal edges lie to its right. However, since the first edge of the path $\Path{R,N}$ is a leftward 
 edge and the last edge of this path is a rightward edge, it follows that this is also the case if $L$ is the first or 
 the last vertical edge on the path. Thus, by Lemma~\ref{chz:special_edges_equiv}, $L$ is either a leftmost edge or a 
 right notch, and thus by the BL-stability of $H$ we have that $L$ is a leftmost edge. \qedhere
\end{proof}

 \begin{corollary}\label{chz:one_less_notch_than_most}
     Let $H$ be a BLS-hole with exactly $k$ distinct leftmost edges. Then $H$ contains exactly $k-1$ distinct left notches.
 \end{corollary}
 \begin{proof} 
    Let $L_1, \ldots ,L_{k}$ be the leftmost edges of $H$ enumerated in clockwise order and let $R$ be the unique 
    rightmost edge of $H$. We can assume without loss of generality that $R$ lies on $\Path{L_k,L_1}$ and that no other 
    leftmost edge lies on that path. This however implies, that $\Path{L_k,L_1}$ contains no left notch. This is because
    otherwise, if $N$ would be a left notch on this path, then either $\Path{N,R}$ or $\Path{R,N}$ would not contain a leftmost edge, 
    contradicting Corollary~\ref{chz:notch_rightmost_contains_lefmost} (note that this argument also holds if $k=1$). 
    Now consider the paths $\Path{L_i,L_{i+1}}$ for $i \in [k-1]$. Each of these paths does not contain $R$ by assumption. 
    Therefore, each such path must contain a left notch by Lemma~\ref{chz:most_notch_relation}. However, since no such path 
    contains a leftmost edge, it follows that no such path can contain two distinct left notches. This is because otherwise, 
    one of the two paths between these two left notches would not contain a leftmost edge, contradicting Corollary~\ref{chz:left_notch_paths_contain_leftmosts}. 
    This means that all such paths $\Path{L_i,L_{i+1}}$ for $i\in [k-1]$ contain exactly one left notch. Since the path $\Path{L_k,L_1}$ 
    contains no left notch at all, this completes the proof of the statement. \qedhere 
 \end{proof}

 Now, with all those corollaries in place, we can state the main structural statement concerning BLS-holes. 
 \begin{theorem}\label{chz:bls_theorem}
     Let $H$ be a BLS-hole and $R$ its unique rightmost edge. Furthermore, let $L_1, \ldots ,L_k$ be the leftmost edges of 
     $H$ listed in clockwise order, with $L_1$ being the first such edge following $R$ in clockwise order. Similarly, 
     let $N_2, \ldots ,N_{k}$ be the left notches of $H$ in clockwise order, where $N_2$ is the first left notch following 
     $R$ in clockwise order. Then we have that in a clockwise traversal starting at $R$, we traverse these edges in the following order 
     \begin{equation*}
    R,L_1,N_2,L_2,N_3, \ldots ,N_{k},L_k,R
     \end{equation*}
     Additionally, if $H$ contains a falling corner $c_f$, then this vertex is traversed between $L_k$ and $R$.
 \end{theorem}
\begin{proof} The first part of the statement follows directly from the proof of Corollary~\ref{chz:one_less_notch_than_most}. 
    The second part follows from the proof of Lemma~\ref{chz:falling_corner_lemma}, as we have shown there that the first 
    vertical edge encountered after a falling corner in clockwise order must be a rightmost edge. \qedhere
\end{proof}

\begin{definition}\label{chz:canonical:ordering}
    Let $H$ be a BLS-hole with exactly $k$ leftmost edges. Then we call \begin{equation*}
         L_1(H),N_2(H),L_2(H), \ldots ,N_k(H),R(H)
    \end{equation*} the \textbf{canonical ordering of the vertical special edges} of H, where $L_i(H),N_i(H)$ and $R(H)$ 
    are defined as in Theorem~\ref{chz:bls_theorem}. Furthermore, let $c_f(H)$ denote the unique falling corner of $H$ if such exists.
\end{definition}

\begin{remark}
    Although it may seem odd at first that the indexing for the left notches starts at $2$, the reason will become 
    apparent in Section~3.3. To abbreviate, we will call the canonical ordering of the vertical special edges 
    just the canonical ordering of a BLS-hole. Furthermore, one could analogously define a canonical ordering of the 
    horizontal special edges of $H$ as the absence of top notches yields analogous properties. 
\end{remark}

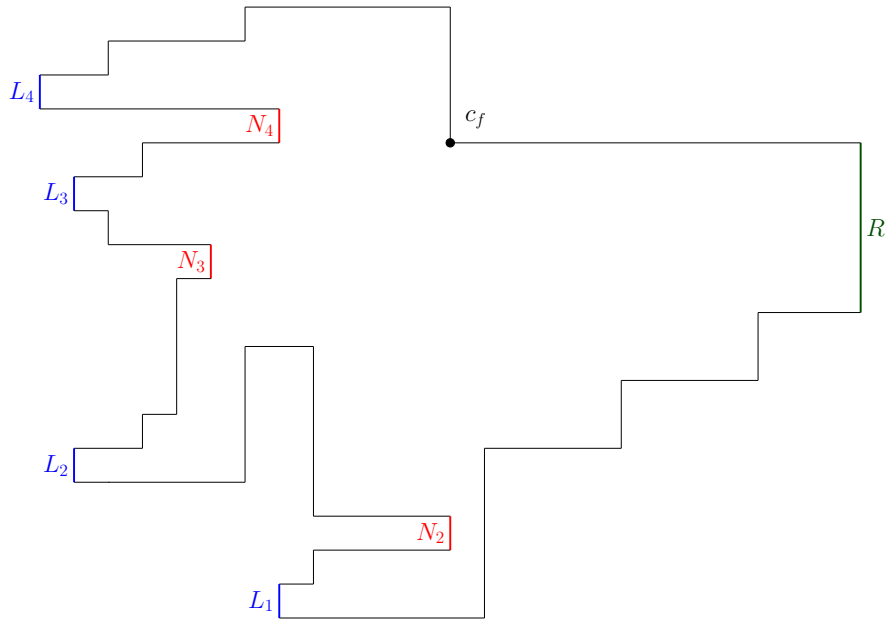
\begin{figure}[H]
    \centering
    \scalebox{0.45}{\begin{tikzpicture}
    \coordinate (A) at (7,1);
    \coordinate (B) at (13,1);
    \coordinate (C) at (13,6);
    \coordinate (D) at (17,6);
    \coordinate (E) at (17,8);
    \coordinate (F) at (21,8);
    \coordinate (G) at (21,10);
    \coordinate (H) at (24,10);
    \coordinate (I) at (24,15);
    \coordinate (J) at (12,15);
    \coordinate (K) at (12,19);
    \coordinate (L) at (6,19);
    \coordinate (M) at (6,18);
    \coordinate (N) at (2,18);
    \coordinate (O) at (2,17);
    \coordinate (P) at (0,17);
    \coordinate (Q) at (0,16);
    \coordinate (R) at (7,16);
    \coordinate (S) at (7,15);
    \coordinate (T) at (3,15);
    \coordinate (U) at (3,14);
    \coordinate (V) at (1,14);
    \coordinate (W) at (1,13);
    \coordinate (X) at (2,13);
    \coordinate (Y) at (2,12);
    \coordinate (Z) at (5,12);
    \coordinate (Z') at (5,11);
    \coordinate (Y') at (4,11);
    \coordinate (X') at (4,7);
    \coordinate (W') at (3,7);
    \coordinate (V') at (3,6);
    \coordinate (U') at (1,6);
    \coordinate (T') at (1,5);
    \coordinate (S') at (2,5);
    \coordinate (R') at (2,5);
    \coordinate (Q') at (6,5);
    \coordinate (P') at (6,9);
    \coordinate (O') at (8,9);
    \coordinate (N') at (8,4);
    \coordinate (M') at (12,4);
    \coordinate (L') at (12,3);
    \coordinate (K') at (8,3);
    \coordinate (J') at (8,2);
    \coordinate (I') at (7,2);

    \draw[thick] (A) -- (B);
    \draw[thick] (B) -- (C);
    \draw[thick] (C) -- (D);
    \draw[thick] (D) -- (E);
    \draw[thick] (E) -- (F);
    \draw[thick](F) -- (G);
    \draw[thick] (G) -- (H);
    \draw[ultra thick, darkgreen] (H) -- (I) node[midway, right] {\huge $R$};
    \draw[thick] (I) -- (J);
    \draw[thick] (J) -- (K);
    \draw[thick] (K) -- (L);
    \draw[thick] (L) -- (M);
    \draw[thick] (M) -- (N);
    \draw[thick](N) -- (O);
    \draw[thick] (O) -- (P);
    \draw[ultra thick, blue] (P) -- (Q) node[midway, left] {\huge $L_4$};
    \draw[thick] (Q) -- (R);
     \draw[ultra thick, red] (R) -- (S) node[midway, left] {\huge $N_4$};
    \draw[thick] (S) -- (T);
    \draw[thick] (T) -- (U);
    \draw[thick] (U) -- (V);
    \draw[ultra thick, blue] (V) -- (W) node[midway, left] {\huge $L_3$};
    \draw[thick] (W) -- (X);
    \draw[thick] (X) -- (Y);
    \draw[thick] (Y) -- (Z);
    \draw[ultra thick, red] (Z) -- (Z') node[midway, left] {\huge $N_3$};
    \draw[thick] (Z') -- (Y');
    \draw[thick] (Y') -- (X');
    \draw[thick] (X') -- (W');
    \draw[thick] (W') -- (V');
    \draw[thick] (V') -- (U');
    \draw[ultra thick, blue] (U') -- (T') node[midway, left] {\huge $L_2$};
    \draw[thick] (T') -- (S');
    \draw[thick] (S') -- (R');
    \draw[thick] (R') -- (Q');
    \draw[thick] (Q') -- (P');
    \draw[thick] (P') -- (O');
    \draw[thick] (O') -- (N');
    \draw[thick] (N') -- (M');
    \draw[ultra thick, red] (M') -- (L') node[midway, left] {\huge $N_2$};
    \draw[thick] (L') -- (K');
    \draw[thick] (K') -- (J');
    \draw[thick] (J') -- (I');
    \draw[ultra thick, blue] (I') -- (A) node[midway, left] {\huge $L_1$};

    \filldraw[black] (J) circle (3.5pt) node[above right = 0.3cm ] {\huge $c_f$};

\end{tikzpicture}}
    \caption{BLS-hole}
    \label{fig:weird:BLSHole}
\end{figure}

In Figure \ref{fig:weird:BLSHole} an illustration of a BLS-hole is shown with the canonical ordering labeled 
(this figure is adapted from \cite{BLCHZ}, with minor modifications to ensure that it could occur in a BL-packing).

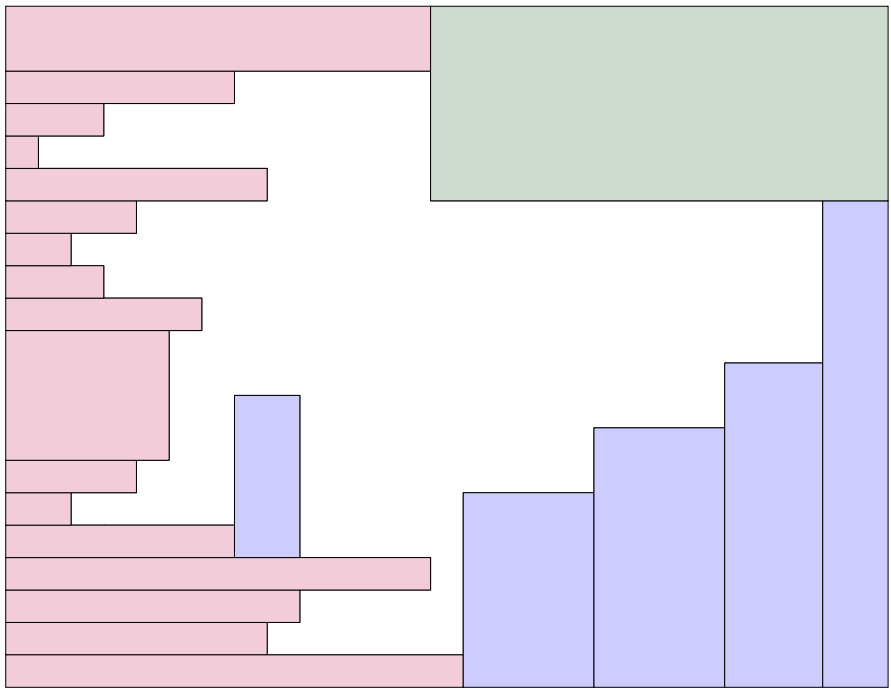
\begin{figure}[H]
    \centering
    \scalebox{0.43}{\begin{tikzpicture}
    \coordinate (A) at (7,1);
    \coordinate (B) at (13,1);
    \coordinate (C) at (13,6);
    \coordinate (D) at (17,6);
    \coordinate (E) at (17,8);
    \coordinate (F) at (21,8);
    \coordinate (G) at (21,10);
    \coordinate (H) at (24,10);
    \coordinate (I) at (24,15);
    \coordinate (J) at (12,15);
    \coordinate (K) at (12,19);
    \coordinate (L) at (6,19);
    \coordinate (M) at (6,18);
    \coordinate (N) at (2,18);
    \coordinate (O) at (2,17);
    \coordinate (P) at (0,17);
    \coordinate (Q) at (0,16);
    \coordinate (R) at (7,16);
    \coordinate (S) at (7,15);
    \coordinate (T) at (3,15);
    \coordinate (U) at (3,14);
    \coordinate (V) at (1,14);
    \coordinate (W) at (1,13);
    \coordinate (X) at (2,13);
    \coordinate (Y) at (2,12);
    \coordinate (Z) at (5,12);
    \coordinate (Z') at (5,11);
    \coordinate (Y') at (4,11);
    \coordinate (X') at (4,7);
    \coordinate (W') at (3,7);
    \coordinate (V') at (3,6);
    \coordinate (U') at (1,6);
    \coordinate (T') at (1,5);
    \coordinate (S') at (2,5);
    \coordinate (R') at (2,5);
    \coordinate (Q') at (6,5);
    \coordinate (P') at (6,9);
    \coordinate (O') at (8,9);
    \coordinate (N') at (8,4);
    \coordinate (M') at (12,4);
    \coordinate (L') at (12,3);
    \coordinate (K') at (8,3);
    \coordinate (J') at (8,2);
    \coordinate (I') at (7,2);

    \draw[thick] (A) -- (B);
    \draw[thick] (B) -- (C);
    \draw[thick] (C) -- (D);
    \draw[thick] (D) -- (E);
    \draw[thick] (E) -- (F);
    \draw[thick](F) -- (G);
    \draw[thick] (G) -- (H);
    \draw[thick] (H) -- (I);
    \draw[thick] (I) -- (J);
    \draw[thick] (J) -- (K);
    \draw[thick] (K) -- (L);
    \draw[thick] (L) -- (M);
    \draw[thick] (M) -- (N);
    \draw[thick](N) -- (O);
    \draw[thick] (O) -- (P);
    \draw[thick] (P) -- (Q);
    \draw[thick] (Q) -- (R);
    \draw[thick] (R) -- (S);
    \draw[thick] (S) -- (T);
    \draw[thick] (T) -- (U);
    \draw[thick] (U) -- (V);
    \draw[thick] (V) -- (W);
    \draw[thick] (W) -- (X);
    \draw[thick] (X) -- (Y);
    \draw[thick] (Y) -- (Z);
    \draw[thick] (Z) -- (Z');
    \draw[thick] (Z') -- (Y');
    \draw[thick] (Y') -- (X');
    \draw[thick] (X') -- (W');
    \draw[thick] (W') -- (V');
    \draw[thick] (V') -- (U');
    \draw[thick] (U') -- (T');
    \draw[thick] (T') -- (S');
    \draw[thick] (S') -- (R');
    \draw[thick] (R') -- (Q');
    \draw[thick] (Q') -- (P');
    \draw[thick] (P') -- (O');
    \draw[thick] (O') -- (N');
    \draw[thick] (N') -- (M');
    \draw[thick] (M') -- (L');
    \draw[thick] (L') -- (K');
    \draw[thick] (K') -- (J');
    \draw[thick] (J') -- (I');
    \draw[thick] (I') -- (A);

    \filldraw[draw = black, fill = purple!20] (-1,1) rectangle (I');
    \filldraw[draw = black, fill = purple!20] (-1,2) rectangle (K');
    \filldraw[draw = black, fill = purple!20] (-1,3) rectangle (M');
    \filldraw[draw = black, fill = purple!20] (-1,4) rectangle (Q');
    \filldraw[draw = black, fill = purple!20] (-1,5) rectangle (U');
    \filldraw[draw = black, fill = purple!20] (-1,6) rectangle (W');
    \filldraw[draw = black, fill = purple!20] (-1,7) rectangle (Y');
    \filldraw[draw = black, fill = purple!20] (-1,11) rectangle (Z);
    \filldraw[draw = black, fill = purple!20] (-1,12) rectangle (X);
    \filldraw[draw = black, fill = purple!20] (-1,13) rectangle (V);
    \filldraw[draw = black, fill = purple!20] (-1,14) rectangle (T);
    \filldraw[draw = black, fill = purple!20] (-1,15) rectangle (R);
    \filldraw[draw = black, fill = purple!20] (-1,16) rectangle (P);
    \filldraw[draw = black, fill = purple!20] (-1,17) rectangle (N);
    \filldraw[draw = black, fill = purple!20] (-1,18) rectangle (L);
    \filldraw[draw = black, fill = purple!20] (-1,19) rectangle ($(K) + (0,2)$);
    \filldraw[draw= black, fill = purple!20] (-1,0) rectangle (B);

    \filldraw[draw = black, fill = blue!20] (6,4) rectangle (O');
    \filldraw[draw = black, fill = blue!20] (13,0) rectangle (D);
    \filldraw[draw = black, fill = blue!20] (17,0) rectangle (F);
    \filldraw[draw = black, fill = blue!20] (21,0) rectangle (H);
    \filldraw[draw = black, fill = blue!20] (24,0) rectangle ($(I) + (2,0)$);
    \filldraw[draw = black, fill = darkgreen!20] (J) rectangle ($(I) + (2,6)$);

\end{tikzpicture}}
    \caption{BL-packing corresponding to the hole}
    \label{fig:weird:BLpackingHole}
\end{figure}

At first glance, it may seem rather implausible that such a seemingly complicated hole could arise in a packing produced 
by the BL-heuristic. However, Figure \ref{fig:weird:BLpackingHole} demonstrates that exactly this hole can actually occur 
in such a packing. Here, the colored shapes in the figure should correspond to the already placed rectangle. 

While the packing, as drawn, clearly cannot be constructed by the BL-heuristic in its illustrated form 
(since many of the red rectangles could obviously be placed in a lower position), one can imagine a scenario where the 
left side of each red rectangle is extended sufficiently far to the left. Here, sufficiently far means that no such two 
adapted red rectangles could fit horizontally next to each other in the strip. Assuming these rectangles are now first 
to place for the BL-heuristic and are given in the bottom-to-top order in which they are illustrated, the BL-heuristic 
would be forced to position them as modeled in the illustration. Assume now in a similar manner, that the upper side of 
the green rectangle is extended far enough upwards as such it could no longer be moved to the interior of the hole. With this final assumption one can now verify easily that if the BL-heuristic receives the remaining 
rectangles ordered lexicographically by $(y_{\min}(r), x_{\min}(r))$, where the coordinates are based on their illustrated 
positions, that this would then produce exactly the packing shown. 

\vspace{0.5cm}

With the canonical ordering defined, we can now describe the \textbf{data structure} that will be used in the following 
algorithms to store a single BLS-hole, following the description from \cite{BLCHZ}.

This data structure consists primarily of a doubly-linked list of the vertices of the hole, in the order in which they 
appear on a traversal of the boundary of the hole. Additionally, it stores special pointers to all edges involved in 
the canonical ordering of the hole, as well as a pointer to the unique falling corner, if one exists. Therefore, the 
size of this data structure is linear in the number of vertices of the hole, which motivates the following definition.

\begin{definition}
    Let $H$ be a hole. We denote by $nv(H)$ the number of vertices of $H$.
\end{definition} 

It is clear that this data structure allows traversal of the boundary of $H$ in clockwise order as well as in 
anticlockwise order in $O(nv(H))$ time. This property will be useful for the algorithms presented in Section~3.2 and~3.3.

\subsection{Computing BL-stable Locations for Nice Holes}

Having established these structural properties of (BLS)-holes, we will now turn to the problem of computing all 
BL-stable locations contained in such a BLS-hole for placing a new rectangle. However, as stated in the introduction, 
we will first examine this problem for the more restricted subclass of \textit{nice holes}.  From this point on, we assume
that the dimensions of the new rectangle to place and the coordinates of all vertices of our holes are integral. 
This assumption is made purely for readability, as the entire algorithm and all proofs could be extended easily for rational values.

\subsubsection{Nice Holes}

We first define the term \textit{nice hole}, which is based on Chazelle's definition from \cite{BLCHZ}. 
\begin{definition}
    A \textbf{nice hole} is a BLS-hole that does not contain a left notch. 
\end{definition}

This restriction now imposes significantly stronger structural constraints, which allows us to compute the BL-stable 
locations quite more easily. A key factor for that is that due to the absence of left notches, a nice hole contains 
exactly one leftmost edge. The reason why this simplifies the computation will first be explained with a visual 
interpretation, which is taken from Chazelle's description \cite{BLCHZ}.

We can view the rectangle $r_{new}$ that we want to place next as a mechanical device consisting of two horizontal bars 
of length $w(r_{new})$, held together by a spring pushing outward. Now we can search for such feasible BL-stable 
locations by sliding this spring device from left to right in the nice hole, starting at the leftmost edge. While 
sliding, we observe the variation in height of this mechanical device and wait for it to exceed $h(r_{new})$. 
Whenever this happens, we can report the current location of the bottom-left corner of this spring as a feasible candidate. 

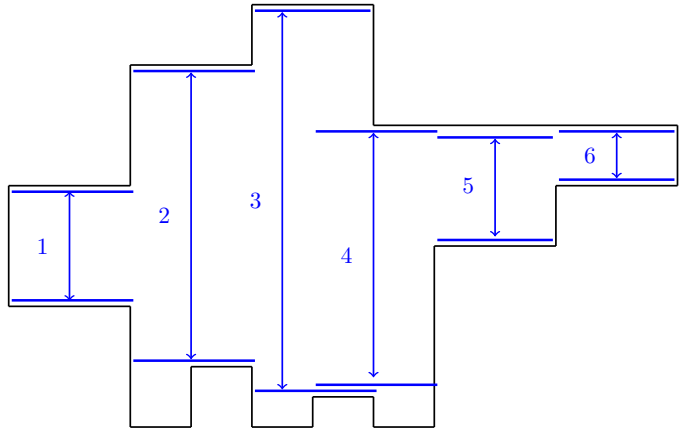
\begin{figure}[H]
    \centering
    \scalebox{0.8}{\begin{tikzpicture}
    \coordinate (A) at (0,0);
    \coordinate (B) at (2,0);
    \coordinate (C) at (2,-2);
    \coordinate (D) at (3,-2);
    \coordinate (E) at (3,-1);
    \coordinate (F) at (4,-1);
    \coordinate (G) at (4,-2);
    \coordinate (H) at (5,-2);
    \coordinate (I) at (5,-1.5);
    \coordinate (J) at (6,-1.5);
    \coordinate (K) at (6,-2);
    \coordinate (L) at (7,-2);
    \coordinate (M) at (7,1);
    \coordinate (N) at (9,1);
    \coordinate (O) at (9,2);
    \coordinate (P) at (11,2);
    \coordinate (Q) at (11,3);
    \coordinate (R) at (6,3);
    \coordinate (S) at (6,5);
    \coordinate (T) at (4,5);
    \coordinate (U) at (4,4);
    \coordinate (V) at (2,4);
    \coordinate (W) at (2,2);
    \coordinate (X) at (0,2);

    \draw[thick] (A) -- (B);
    \draw[very thick,blue] ($(A) + (0.05,0.1)$) -- ($(B) + (0.05,0.1)$);
    \draw[very thick,blue] ($(A) + (0.05,1.9)$) -- ($(B) + (0.05,1.9)$);
    \draw[<->, thick, blue, shorten >=2.7pt, shorten <=2.7pt] (1,0) -- (1,2) node [midway, left = 0.2cm] {$1$};
    \draw[thick] (B) -- (C);
    \draw[thick] (C) -- (D);
    \draw[thick] (D) -- (E);
    \draw[thick] (E) -- (F);
    \draw[very thick,blue] ($(C) + (0.05,1.1)$) -- ($(F) + (0.05,0.1)$);
    \draw[very thick,blue] ($(C) + (0.05,5.9)$) -- ($(F) + (0.05,4.9)$);
    \draw[<->, thick, blue, shorten >=3.5pt, shorten <=3.5pt] (3,-1) -- (3,4) node [midway, left = 0.2cm] {$2$};
    \draw[thick] (F) -- (G);
    \draw[thick] (G) -- (H);
    \draw[thick] (H) -- (I);
    \draw[thick] (I) -- (J);
    \draw[thick] (J) -- (K);
    \draw[thick] (K) -- (L);
    \draw[very thick,blue] ($(G) + (0.05,0.6)$) -- ($(J) + (0.05,0.1)$);
    \draw[very thick,blue] ($(G) + (0.05,6.9)$) -- ($(J) + (-0.05,6.4)$);
    \draw[<->, thick, blue, shorten >=3.5pt, shorten <=3.5pt] (4.5,-1.5) -- (4.5,5) node [midway, left = 0.2cm] {$3$};
    \draw[thick] (L) -- (M);
    \draw[thick] (M) -- (N);
    \draw[thick] (N) -- (O);
    \draw[very thick,blue] ($(I) + (0.05,0.2)$) -- ($(L) + (0.05,0.7)$);
    \draw[very thick,blue] ($(I) + (0.05,4.4)$) -- ($(L) + (0.05,4.9)$);
    \draw[<->, thick, blue, shorten >=3.5pt, shorten <=3.5pt] (6,-1.3) -- (6,3) node [midway, left = 0.2cm] {$4$};
    \draw[thick] (O) -- (P);
    \draw[thick] (P) -- (Q);
    \draw[thick] (Q) -- (R);
    \draw[thick] (R) -- (S);
    \draw[thick] (S) -- (T);
    \draw[thick] (T) -- (U);
    \draw[very thick,blue] ($(M) + (0.05,0.1)$) -- ($(N) + (-0.05,0.1)$);
    \draw[very thick,blue] ($(M) + (0.05,1.8)$) -- ($(N) + (-0.05,1.8)$);
    \draw[<->, thick, blue, shorten >=6.3pt, shorten <=4.5pt] (8,1) -- (8,3) node [midway, left = 0.2cm] {$5$};
    \draw[thick] (U) -- (V);
    \draw[thick] (V) -- (W);
    \draw[thick] (W) -- (X);
    \draw[thick] (X) -- (A);
    \draw[very thick,blue] ($(O) + (0.05,0.1)$) -- ($(P) + (- 0.05,0.1)$);
    \draw[very thick,blue] ($(O) + (0.05,0.9)$) -- ($(P) + (-0.05,0.9)$);
    \draw[<->, thick, blue, shorten >=3.5pt, shorten <=3.5pt] (10,2) -- (10,3) node [midway, left = 0.2cm] {$6$};
    
\end{tikzpicture}}
    \caption{Illustration of sliding the spring device through a nice hole (adapted from \cite{BLCHZ})}
    \label{fig:chz:spring_device}
\end{figure}

The goal of Section~4.3 is to formalize the implementation of this visual interpretation. To do so, we partition the described 
process into three parts. Rather than sliding a mechanical spring, we will simulate the motion of those horizontal bars 
separately, first across the bottom boundary of the nice hole and then across its top boundary. Finally, we combine these 
two results to obtain the feasible BL-stable locations. This motivates the following definition, which is inspired by \cite{BLCHZ}.
\begin{definition}
    Let $S$ be a nice hole. We denote by $\bm{L(S)}$ the unique leftmost edge of $S$ and by $\bm{R(S)}$ its unique 
    rightmost edge. Furthermore, we define the \textbf{lower boundary} of $S$ as the rectilinear arc $FB(S):= \APath{L(S),R(S)}$ 
    and the \textbf{upper boundary} of $S$ as the rectilinear arc $ FT(S) :=\Path{L(S),R(S)}$. 
\end{definition}

The following lemma now provides the foundation for efficiently performing the left-to-right sweep across the two 
boundaries. While this result was implicitly used in \cite{BLCHZ}, it was neither explicitly stated nor proved.
\begin{lemma}\label{chz:nice_hole_x_sorting}
    Let $S$ be a nice hole. If one traverses the edges of FB(S) or FT(S), starting at the vertex belonging to $L(S)$, 
    then these edges $e$ are visited in non-decreasing order by the sorting keys $x_{\min}(e)$.
\end{lemma}
\begin{proof} Consider this traversal of $FB(S)$. Note that this would correspond to an anti-clockwise traversal of 
    this segment in $S$. Thus, the first horizontal edge of $FB(S)$ is an anticlockwise-rightward edge by Lemma 
    \ref{chz:special_edges_equiv}. If this holds for all horizontal edges, then the claim follows immediately. 
    Otherwise, consider the first anticlockwise-leftward edge $e$ visited on this traversal. By the minimality of 
    $e$, the prior horizontal edge is an anticlockwise rightward-edge, so the vertical edge between these two edges 
    would be by Lemma \ref{chz:special_edges_equiv} either a rightmost edge or a left notch of $S$. However, neither 
    of these edge types can belong to $FB(S)$ $\lightning$. The proof for $FT(S)$ is completely analogous.\qedhere
\end{proof}

\subsubsection{The Bottom-Function}

We start by describing the motion of the horizontal bar across the bottom boundary of the nice hole. 
The goal here will be to compute all positions where such a bar can be placed in the nice hole without crossing the 
bottom boundary. The corresponding pseudocode is taken from Chazelle's description \cite{BLCHZ}, 
with only minor modifications for improved readability and correctness, such as adjusted boundary conditions 
($<$ vs. $\leq$) and a fix for an edge case. However, Chazelle's proof of the correctness of that pseudocode lacked 
formal precision as the return value of that function was not properly defined. Therefore, Section~3.2.2 provides 
rigorous definitions and a rigorous correctness proof.

Throughout Section~3.2.2, we fix a width $w\in \mathbb{N}_{>0}$ for the horizontal bar. 
All subsequent definitions are now implicitly parameterized by this width, but we omit explicit indexing by $w$ for improved 
readability. We also assume that our nice hole $S$ is sufficiently wide for the horizontal bar, which means that $w\leq x(R(S)) - x(L(S))$.

We begin by introducing a derived definition of the bottom boundary, which will simplify reasoning about whether the 
horizontal bar crosses this lower boundary.
\begin{definition}
    Let $S$ be a nice hole. Then we define $S_{bot}$ as the nice hole given by the boundary \begin{equation*}
        FB(S)\cup \text{seg}\Big( l_u, (x(L(S), y_{\max}(S)), (x(R(S)), y_{\max}(S)), r_u \Big)
    \end{equation*} where $l_u,r_u$ are the lower vertices of $L(S),R(S)$ respectively.
\end{definition}
\begin{remark}
    It is trivial that our goal can be equivalently formulated as finding all positions where such a bar can be placed within $S_{bot}$.
\end{remark}

The following two definitions now aim to formalize what the function is expected to return.

\begin{definition}\label{chz:bot:f}
    Let $S$ be a nice hole and let $p_1 \in \{x(L(S)),x(L(S))+1, \ldots ,x(R(S)) -w\}$ be arbitrary. We define 
    \begin{equation*}
        f(p_1) := \min\{p_2 \in \mathbb{R} \mid \text{seg}\big((p_1,p_2),(p_1 + w , p_2)\big) \subseteq S_{bot} \}
    \end{equation*}
\end{definition}

So, our goal is to compute $f(p_1) $ for all $p_1 \in \{x(L(S)), \ldots ,x(R(S)) -w\}$. We cannot of course store the whole 
function as this could possibly break the linear runtime. However, since $f$ is (almost) a staircase function, it suffices 
to store only the points at which $f$ changes in value. This motivates the following definition.

\begin{definition}\label{chz:setc}
    Let $S$ be a nice hole. Then we define $C$ as the set which contains for every $p_1 \in \{x(L(S)),  \ldots , x(R(S)) - w\}$ exactly the following points
    \begin{enumerate}[label=(\roman*)]
        \item $(p_1,f(p_1))$ for $p_1 \in \{x(L(S)), x(R(S)) - w\}$
        \item $\{(p_1, f(p_1)), (p_1,z)\}$ if $f(p_1 + 1) = z > f(p_1), p_1 < x(R(S)) - w$ 
        \item $\{(p_1,z), (p_1,f(p_1))\}$ if $f(p_1 - 1) = z > f(p_1), p_1 > x(L(S))$
    \end{enumerate}
\end{definition}

One can verify that for such a $p_1 \in \{x(L(S)), \ldots , x(R(S))-w\}$ there can be $3$ distinct points $p$ in $C$ with $x(p) = p_1$. 
This is exactly the edge case missing in Chazelle's pseudocode, as it did not include the point $(p_1,f(p_1))$ in that 
case but only the other two, which may result in missing the BL-location. In this scenario, we introduce the convention 
that $C$ contains $4$ such points $p$, where the point $(p_1,f(p_1))$ appears twice in that set. We take this approach 
because we require the set $C$ to remain ordered at all times, specifically in non-decreasing order of $x$-coordinates. 
For points with equal $x$-coordinates, the ordering follows the order in the sets described in case $(ii)$ and case $(iii)$. 
If an $x$-coordinate appears four times, this corresponds to one case $(ii)$ pair and to one case $(iii)$ pair, 
in which case we have the convention that the case $(ii)$ pair precedes the case $(iii)$ pair in our ordering.
Similarly, the case $(i)$ points should be treated as distinct and in the event of a tie w.r.t. the $x$-coordinate, 
they are ordered before a case $(ii)$ pair/after a case $(iii)$ pair.
With this convention of our ordering, we have that $|C|$ is always an even number and for $C = \{c_0, \ldots ,c_{2k-1}\}$ it 
holds that $x(c_{2i-1}) = x(c_{2i})$ and $y(c_{2i}) = y(c_{2i+1})$ for every $i\in \{1, \ldots ,k-1\}$, as well as $y(c_0) = y(c_1), y(c_{2k-2}) = y(c_{2k-1})$

\begin{figure}[H]
    \centering
    \scalebox{1}{\begin{tikzpicture}
    \draw[thick] (0,0) -- (2,0);
    \draw[thick] (2,0) -- (2,-1);
    \draw[thick] (2,-1) -- (4,-1) node[midway, below = 12pt] {$\underbrace{\hspace{2cm}}_{= w}$};
    \draw[thick] (4,-1) -- (4,2);
    \draw[thick] (4,2) -- (6,2);

    \filldraw[blue] (2,0) circle (1.5pt) node[above = 0.2cm] {$c_1$};
    \filldraw[blue] (2,-1) circle (1.5pt) node[left = 0.2cm] {$c_2,c_3$};
    \filldraw[blue] (2,2) circle (1.5pt) node[left = 0.2cm] {$c_4$};
\end{tikzpicture}}
    \caption{The edge case scenario}
    \label{fig:chz:4_c_points}
\end{figure}
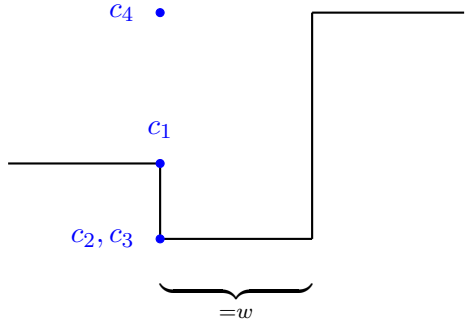

This set $C$ is now exactly what we want to compute with our function. Before presenting the pseudocode, we will first 
explain the idea behind the computation, where the intuitive description and the corresponding figures are inspired by Chazelle's description~\cite{BLCHZ}

The core intuition, as stated at the beginning, is now to slide a horizontal bar \linebreak[4] $\text{seg}(b_1,b_1+_x w)$ from left to 
right across the nice hole $S_{bot}$. Now the $x$-coordinates of the points in $C$ (except possibly for the first and last 
point of $C$) correspond exactly to the positions where the y-coordinate of the bar changes --- meaning when it either has to 
rise to remain within the boundaries of the hole or when it has to fall as there is no horizontal edge anymore which is supporting it.
To simplify computing the first point in $C$, we assume that the first edge of the bottom boundary of $S_{bot}$ 
(which is the anti-clockwise successor of $L(S_{bot})$) has a width of at least $w$. This assumption can be made without 
loss of generality, since if the edge has smaller width we may instead just consider the nice hole which arises from 
$S_{bot}$ by shifting the leftmost edge $w$ units to the left. Once the set $C$ is now computed for this modified nice hole, 
it is trivial to deduce the set $C$ for our original hole $S_{bot}$. This will now make the initialization step trivial: 
in the modified setting the first point of $C$ is simply the lower vertex $l_u$ of the (possibly shifted) edge $L(S_{bot})$. 
Therefore, we can then initialize the bar simply at $b_1 = l_u$ and then start sliding it across the bottom boundary.

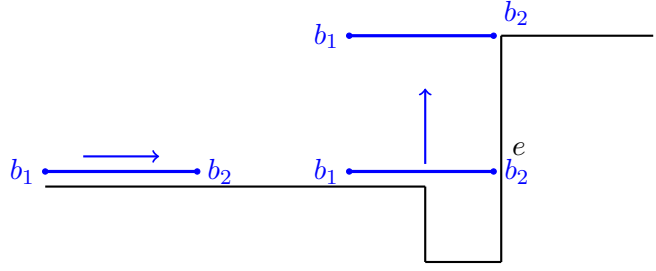
\begin{figure}[!tp]
    \centering
    \scalebox{1}{\begin{tikzpicture}
    \draw[thick] (0,0) -- (5,0);
    \draw[thick] (5,0) -- (5,-1);
    \draw[thick] (5,-1) -- (6,-1);
    \draw[thick] (6,-1) -- (6,2) node[midway, right] {$e$};
    \draw[thick] (6,2) -- (8,2);

    \draw[very thick, blue] (0,0.2) -- (2,0.2); 
    \draw[->,thick, blue] (0.5,0.4) -- (1.5,0.4);
    \filldraw[blue] (0,0.2) circle (1pt) node[left] {$b_1$};
    \filldraw[blue] (2,0.2) circle (1pt) node[right] {$b_2$};

    \draw[very thick, blue] (4,0.2) -- (5.9,0.2);
    \draw[->, thick, blue] (5, 0.3) -- (5,1.3);
    \draw[very thick, blue] (4,2) -- (5.9,2);

    \filldraw[blue] (4,0.2) circle (1pt) node[left] {$b_1$};
    \filldraw[blue] (4,2) circle (1pt) node[left] {$b_1$};

    \filldraw[blue] (5.9,0.2) circle (1pt) node[right = 0.2] {$b_2$};
    \filldraw[blue] (5.9,2) circle (1pt) node[above right = 0.001cm] {$b_2$};
\end{tikzpicture}}
    \caption{Move-up scenario}
    \label{chz:bottom:move_up}
\end{figure}

Now, as long as the bar neither has to move up nor down, no point should be added to $C$. So, assume that the bar has to 
move up at first. This occurs because we encounter a vertical edge $e$ with $y_{\max}(e) > y(b_1)$. In that case, the 
point pair $\{(x(e)-w,y(b_1)), (x(e)-w, y_{\max}(e))\}$ should be added to $C$, which is a case $(ii)$ point in 
Definition~\ref{chz:setc}, and $b_1$ should be updated to the second point of that point pair. 
(Figure \ref{chz:bottom:move_up} shows an illustration of this scenario, the two right $b_1$ points are the points that have to be added to $C$) 

\begin{figure}[H]
    \centering
    \scalebox{1}{\begin{tikzpicture}
    \draw[thick] (-1,0) -- (3,0);
    \draw[thick] (3,0) -- (3,-3) node[midway, left] {$e_1$};
    \draw[thick] (3,-3) -- (3.5,-3);
    \draw[thick] (3.5,-3) -- (3.5,-2);
    \draw[thick] (3.5,-2) -- (4.5,-2);
    \draw[thick] (4.5,-2) -- (4.5,-3);
    \draw[thick] (4.5,-3) -- (5.5,-3);
    \draw[thick] (5.5,-3) -- (5.5,-1);
    \draw[thick] (5.5,-1) -- (7,-1) node[midway, below] {$e'$};
    \draw[thick] (7,-1) -- (7,-3) node[midway, right] {$e_2$};
    \draw[thick] (7,-3) -- (9,-3);

    \draw[very thick, blue] (-1,0.2) -- (2,0.2);
    \draw[->, thick, blue] (0, 0.4) -- (1,0.4);

    \filldraw[blue] (-1,0.2) circle (1pt) node[left]{$b_1$};
    \filldraw[blue] (2,0.2) circle (1pt) node[right]{$b_2$};

    \draw[very thick, blue] (3.1,0.2) -- (6,0.2);
    \draw[->, thick, blue] (4.5, 0) -- (4.5,-0.5);
    \filldraw[blue] (3.1,0.2) circle (1pt) node[left]{$b_1$};
    \filldraw[blue] (6,0.2) circle (1pt) node[right]{$b_2$};

    \draw[very thick, blue] (3.1,-0.8) -- (6,-0.8);
    \filldraw[blue] (3.1,-0.8) circle (1pt) node[left]{$b_1$};
    \filldraw[blue] (6,-0.8) circle (1pt) node[right]{$b_2$};

    \path (3,-3) -- (6,-3)
        node[midway, below=10pt]
        {\(\underbrace{\hspace{3cm}}_{\Large I}\)};

\end{tikzpicture}}
    \caption{Move-down scenario}
    \label{chz:bottom:move_down}
\end{figure}
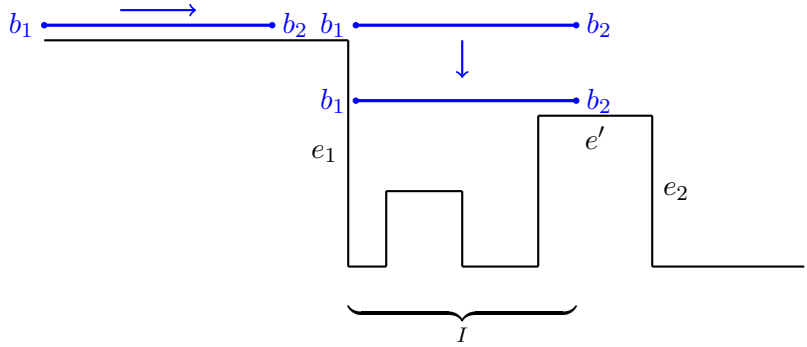

This concludes that case, so assume now instead that the bar has to fall down first. This now occurs because we find an 
interval $I = [I_{\min}, I_{\min}+w)$ such that for each horizontal edge $e$ of $FB(S_{bot}) = FB(S)$ with $x_{\min}(e) \in I$ we have $y(e) < y(b_1)$.
In this case, the point pair $\{(I_{\min}, y(b_1)),(I_{\min}, y(e'))\}$ should be added to $C$, 
where $e'$ is a horizontal edge of $FB(S_{bot})$ with  $y(e')$ maximal among those with $x_{\min}(e') \in I$. This is 
exactly a case $(iii)$ point pair in definition \ref{chz:setc}, and $b_1$ should again be updated to the second point of 
that point pair (Figure \ref{chz:bottom:move_down} shows again an illustration).

This intuitive explanation now motivates the following definition. 

\begin{definition}
    Let $S$ be a nice hole and $v$ be a vertex on the bottom boundary of $S$. Then we define 
    \begin{equation*}
        \textbf{rise}(v) := \text{argmin}\{x(e)|e \in V(FB(S)) , x(e) \geq x(v) , y_{\max}(e) > y(v)\} 
    \end{equation*}
    where $V(FB(S))$ is the set of all vertical edges on $FB(S)$. Similarly, with $H(FB(S))$ denoting the set of all horizontal edges on $FB(S)$ we define
        $\textbf{rift}(v) := \text{argmin}\{ \min(I) | I \in I_v\}$
    where 
    \begin{align*}
        I_{v} := \{I = [I_{\min},I_{\min}+w)|& I_{\min} \geq x(v),
        y(e) < y(v) \\ & \forall e \in H(FB(S)) \text{ with } x_{\min}(e) \in I\}
    \end{align*}
    Here we define additionally $\textbf{rift-start}(v) := e$ where $e$ is the unique vertical edge of $FB(S)$ with $x(e) = I_{\min}$ 
    for the interval $I= (I_{\min}, I_{\min} + w)$ attaining the minimum above. Similarly, we define $\textbf{rift-end}(v) := e'$, 
    where $e'$ is the first vertical edge (in anticlockwise order) of $FB(S)$ with $x(e) \geq I_{\min}+w$. 
\end{definition}

\begin{remark}
    In Figure \ref{chz:bottom:move_down} $e_1$ would be the rift-start and $e_2$ would be the rift-end for the mentioned $I$.
\end{remark}

So our implementation will work as follows: after adding the first point $c_0$ to $C$, which is a trivial matter by the 
prior assumption, we traverse the nice hole $S_{bot}$ from left to right to determine whether $\text{rift}(c_0)$ or 
$\text{rise}(c_0)$ comes first. This traversal from left to right can be performed by considering the edges of 
$FB(S_{bot}) = FB(S)$ in anticlockwise order, as shown in Lemma \ref{chz:nice_hole_x_sorting}. If we encounter the rise 
first, then it is clear how to add the next point pair to $C$. If we, however, encounter the rift $I = [I_{\min},I_{\min}+w)$ 
first, then we have to compute the maximal $y$-value among the horizontal edges $e$ with $x_{\min}(e) \in I$. This 
computation now cannot be done by searching the whole interval every time, as such rifts could intersect, which could 
thus lead to considering the same edges multiple times, breaking the linear runtime. 

So for that matter, we will first introduce two helper functions carrying out that procedure. These helper functions, 
which were also described in \cite{BLCHZ}, use the deque data structure to store the horizontal edges, which is a 
double-ended queue that allows insertion and removal at both ends in constant time. A detailed description of deques 
can be found in \cite[Chapter~2.2.1]{KNUTH}. For such a deque $Q$, we denote by $\{a\} \cup Q$ the insertion of $a$ at 
the front of $Q$ and by $Q \cup \{a\}$ the insertion at the back. We denote by $FRONT(Q)$ and $BACK(Q)$ the first and 
last elements of $Q$, respectively. The operations $POPFRONT(Q)$ and $POPBACK(Q)$ remove the first and last element 
respectively. Before describing the functions, we will first introduce the following notational convention, also used in \cite{BLCHZ}:

For a nice hole $S$, let $\left\{ \{h_1,d_1 \}, \{h_2,d_2\} ,  \ldots  , \{h_{m-1},d_{m-1}\}\right\}$ denote the vertical 
edges of $FB(S)$ in anticlockwise order, such that $h_i >_y d_i \ \forall i \in \{1, \ldots ,m-1\}$. In addition, we define 
$\{h_0,d_0\} := L(S)$ and $\{h_m,d_m\} := R(S)$. Similarly, let $\left\{ \{l_1,r_1\}, \ldots ,\{l_m,r_m\} \right\}$ be the 
horizontal edges of $FB(S)$, again in anticlockwise order with $l_i <_x r_i \ \forall i \in [m]$. 

 With this in mind, we can now describe both helper functions. Their respective pseudocode is based on Chazelle's 
 description in \cite{BLCHZ}, but the accompanying lemmas --- which state their goal and prove their correctness --- 
 differ from Chazelle's argument in order to provide a more rigorous presentation.

\begin{algorithm}[H]
\SetKwProg{Fn}{Function}{}{end}
\SetKwProg{Pn}{Procedure}{}{end}
\SetKwFunction{FSetup}{SETUP}
\SetKwFunction{FMerge}{MERGE}

\Fn{\FSetup{$\mathit{i},\mathit{j}$}}{
    $Q \gets \emptyset$ \;
    Let $k\in \mathbb{N}$ such that $\{l_k,r_k\} = \apred(\{h_j,d_j\})$ \;
    \While{$\{l_k,r_k\} \neq \apred(\{h_i,d_i\})$}{
        \If{$Q = \emptyset$ \textnormal{or} $FRONT(Q) <_y \{l_k,r_k\}$}{
            $Q \gets \{l_k,r_k\} \cup Q$ \;
        }
        $k \gets k -1 $\;
    }
    \Return{$Q$}
}

\Fn{\FMerge{$Q,Q'$}}{
    \If{$Q \neq \emptyset$ \textnormal{and} $Q' \neq \emptyset$}{
        \While{$BACK(Q) \leq_y FRONT(Q')$}{
            $POPBACK(Q)$ \;
        }
    }
    \Return $Q \cup Q'$
}

\caption{Helper Functions}
\end{algorithm}

\begin{lemma}\label{chz:bottom:setup}
    Let $S$ be a nice hole and $i,j \in \mathbb{N}$ with $i < j$. Then \textbf{setup(i,j)} computes a deque $Q$ which 
    stores horizontal edges of the path $\APath{\{h_{i},d_{i}\},\{h_{j}, d_{j}\}}$. This deque additionally fulfills the 
    following conditions
    \begin{enumerate}[label=(Q\arabic*)]
        \item $Q$ is ordered in the sense that for any two edges $e,e' \in Q$ with $e$ preceding $e'$ we have $x_{\max}(e) \leq x_{\min}(e'), y(e) > y(e')$.
        \item $Q$ contains the horizontal edge $e$ of the path $\APath{\{h_{i},d_{i}\},\{h_{j}, d_{j}\}}$ which maximizes $(y(e), x_{\min}(e))$ lexicographically 
        \item Let $e$ be a horizontal edge of $FB(S)$ that lies on the path $\APath{\{h_{i},d_{i}\},\{h_{j}, d_{j}\}}$ 
        and is not the last such edge . Then $Q$ contains the horizontal edge $e'$ of $FB(S)$ which maximizes 
        $(y(e'), x_{\min}(e'))$ lexicographically among those that lie on $\APath{e,\{h_{j}, d_{j}\}}$(Note that this path notation excludes $e$).
    \end{enumerate}
    
\end{lemma}
\begin{proof} $(Q1)$ and $(Q2)$ follow directly from the algorithm formulation as we traverse the edges of 
    $\APath{\{h_{i},d_{i}\},\{h_{j}, d_{j}\}}$ from right to left and add a new edge to the front of $Q$ if and only if 
    that would be the largest with regard to $y$-coordinate yet. Now for $(Q3)$ let $e$ be a horizontal edge on the path 
    which is not the last and let $\{h_{i'},d_{i'}\} = \asuccv(e)$. Note, that when $e$ is considered in the while-loop 
    of $setup(i,j)$, the state of $Q$ is identical to the output of $setup(i',j)$. Since no edges are ever removed from 
    $Q$, $(Q3)$ now follows directly from $(Q2)$ applied to $setup(i',j)$ . \qedhere 
\end{proof}

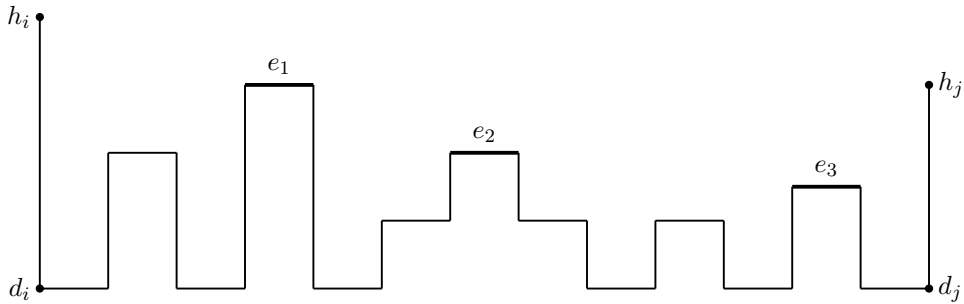
\begin{figure}[H]
    \centering
    \scalebox{0.9}{\begin{tikzpicture}
    \coordinate (A) at (0,4);
    \coordinate (B) at (0,0);
    \coordinate (C) at (1,0);
    \coordinate (D) at (1,2);
    \coordinate (E) at (2,2);
    \coordinate (F) at (2,0);
    \coordinate (G) at (3,0);
    \coordinate (H) at (3,3);
    \coordinate (I) at (4,3);
    \coordinate (J) at (4,0);
    \coordinate (K) at (5,0);
    \coordinate (L) at (5,1);
    \coordinate (M) at (6,1);
    \coordinate (N) at (6,2);
    \coordinate (O) at (7,2);
    \coordinate (P) at (7,1);
    \coordinate (Q) at (8,1);
    \coordinate (R) at (8,0);
    \coordinate (S) at (9,0);
    \coordinate (T) at (9,1);
    \coordinate (U) at (10,1);
    \coordinate (V) at (10,0);
    \coordinate (W) at (11,0);
    \coordinate (X) at (11,1.5);
    \coordinate (Y) at (12,1.5);
    \coordinate (Z) at (12,0);
    \coordinate (Z') at (13,0);
    \coordinate (Z'') at (13,3);

    \filldraw[black] (A) circle (1.5pt) node[left] {$h_i$};
    \filldraw[black] (B) circle (1.5pt) node[left] {$d_i$};
    \draw[thick] (A) -- (B);
    \draw[thick] (B) -- (C);
    \draw[thick] (C) -- (D);
    \draw[thick] (D) -- (E);
    \draw[thick] (E) -- (F);
    \draw[thick] (F) -- (G);
    \draw[thick] (G) -- (H);
    \draw[ultra thick] (H) -- (I) node[midway, above] {$e_1$}; 
    \draw[thick] (I) -- (J);
    \draw[thick] (J) -- (K);
    \draw[thick] (K) -- (L);
    \draw[thick] (L) -- (M);
    \draw[thick] (M) -- (N);
    \draw[ultra thick] (N) -- (O) node[midway, above] {$e_2$};
    \draw[thick] (O) -- (P);
    \draw[thick] (P) -- (Q);
    \draw[thick] (Q) -- (R);
    \draw[thick] (R) -- (S);
    \draw[thick] (S) -- (T);
    \draw[thick] (T) -- (U);
    \draw[thick] (U) -- (V);
    \draw[thick] (V) -- (W);
    \draw[thick] (W) -- (X);
    \draw[ultra thick] (X) -- (Y) node[midway, above] {$e_3$};
    \draw[thick] (Y) -- (Z);
    \draw[thick] (Z) -- (Z');
    \draw[thick] (Z') -- (Z'');
    \filldraw[black] (Z') circle (1.5pt) node[right] {$d_j$};
    \filldraw[black] (Z'') circle (1.5pt) node[right] {$h_j$};
\end{tikzpicture}}
    \caption{Deque $Q= \{e_1,e_2,e_3$\}}
    \label{fig:enter-label}
\end{figure} 

\begin{lemma}\label{chz:bottom:merge}
    Let $S$ be a nice hole and $i,j,k \in \mathbb{N}$ with $ i < j < k$. Then the deque returned by $setup(i,k)$ is 
    equal to the deque returned by $merge(Q_1, Q_2)$, where $Q_1$ and $Q_2$ are the deques returned by $setup(i,j)$ and $setup(j,k)$ respectively.
\end{lemma}
\begin{proof}
    This lemma is directly evident from the formulation of the algorithms. \qedhere
\end{proof}

With these helper functions in mind, we now describe the algorithm for computing the set $C$, using again the previously 
established notation for the edges of $FB(S)$. As stated earlier, the pseudocode follows again Chazelle's description~\cite{BLCHZ}, 
but we provide an alternative proof of correctness, as Chazelle's argument lacked formal precision and did not explicitly define the set $C$.

\SetKwInOut{Input}{Input}
\SetKwInOut{Output}{Output}
\SetKw{Continue}{continue}
\SetKw{Break}{break}
\begin{algorithm}[!t]
\caption{Bottom-Function}
\Input{Nice hole $S$, $w\in \mathbb{R}_{>0}$ with $x_{\max}(e) - x_{\min}(e) \geq w $ for the first edge $e$ of $FB(S)$}
\Output{Set $C$ as described in Definition \ref{chz:setc}}
\vspace{0.5cm}
$(b_1,b_2) \gets (d_0, d_0 +_x w)$ \;
$C_{\text{alg}} \gets \{b_1\}, Q \gets \emptyset$ \;
Let us denote by $\text{back}(C_{\text{alg}})$ the last element of $C_{\text{alg}}$ \;
support $\gets \{h_1,d_1\} \cap \text{line}(b_1,b_2)$ \;
start $\gets 1$, $r \gets 1$ \;
\While{True \label{chz:bot:first_while}}{
    \While{$\{h_r,d_r\} <_x \textnormal{support} + _x w$ \label{chz:bot:second_while}}{
        \If{$r = m$ \label{chz:bot:ifret}}{
            $C_{\text{alg}} \gets C_{\text{alg}} \cup \{(x(R(S))- w, y(\text{back}(C_{\text{alg}}))\}$ \; \label{chz:bot:lastel}
            \Return {$C$} \; \label{chz:bot:return}
        }
        \If{$y(h_r) >_y \textnormal{support}$ \label{chz:bot:ifrise}}{
            $u \gets (x(h_r)-w, y(\text{back}(C_{\text{alg}})))$ \;
            $(b_1,b_2) \gets (h_r -_x w , h_r)$ \;
            $C_{\text{alg}} \gets C_{\text{alg}} \cup \{u,b_1\}, Q \gets \emptyset $ \; \label{chz:bot:rise_point_add}
            $\text{support} \gets \{h_{r+1},d_{r+1}\} \cap \text{line}(b_1,b_2)$ \;
            $\text{start} \gets r+1, r \gets r+1$ \; \label{chz:bot:rise_end}
        }
        \Else{$r \gets r+1$ \; \label{chz:bot:else}}
    }
    $b_1 \gets \text{support}, b_2 \gets b_1 +_x w$ \; \label{chz:bot:firstrift}
    $Q' \gets SETUP(\text{start}, \text{r})$ \; \label{chz:bot:qprime}
    $Q \gets MERGE(Q,Q')$ \; \label{chz:bot:qprimeqmerge}
    $a \gets FRONT(Q)$ \; 
    $POPFRONT(Q)$ \;
    $(b_1, b_2) \gets (b_1,b_2) -_y [y(b_1) - y(a)]$ \;
    \If{$y(b_1) \neq y(\textnormal{support})$}{
        $C_{\text{alg}} \gets C_{\text{alg}} \cup \{\text{support}, b_1\}$ \; \label{chz:bot:riftadd}
    }
    Let $r_k$ be the right vertex of $a$ \; 
    $\text{start} \gets r, \text{support} \gets r_k$ \ ; \label{chz:bot:lastrift}

}

\Return $C_{\text{alg}}$
\end{algorithm}

\begin{theorem}\label{chz:bottom:correct}
    The algorithm \textit{Bottom-function} computes the set $C = \{c_0,c_1, \ldots ,c_l\}$ for the specified input correctly 
    and operates in $O(nv(S))$ time.
\end{theorem}
\begin{proof}
Throughout this proof, we will denote by $\text{supp}$ the index of the vertical edge of $FB(S)$ such that 
$\text{support} \in \{h_{\text{supp}}, d_{\text{supp}}\}$. For visual intuition, one can imagine support being the 
right vertex of the horizontal edge currently supporting $\text{seg}(b_1,b_2)$. We will prove the correctness by first 
showing that the following invariants hold at the start of each iteration in both while-loops:
\begin{enumerate}[label=(\arabic*)]
    \item $\exists k \in \mathbb{N}$ such that $C_{\text{alg}} = \{c_0, \ldots ,c_k\} \subseteq C$, meaning that 
        $C_{\text{alg}}$ is a subset of $C$ preserving the order, which has not skipped a point
    \item Let $c_k$ be the last point of $C_{alg}$. Then $y(c_k) = y(\text{support}), x(c_k) \leq x(\text{support}), \\ 
        x(\text{support}) \leq \min\{x(\text{rise}(c_k)), x(\text{rift-start}(c_k))\}$. In addition, \text{support} is always the right vertex of a horizontal edge of $FB(S)$.
    \item Neither $\text{rise}(c_k)$ nor $\text{rift-end}(c_k)$ lies on $\APath{\{h_{\text{supp}}, d_{\text{supp}} \} , 
        \{h_r,d_r\}}$ (Note, that this notation again excludes both end-edges)
    \item $x(\{h_{r-1},d_{r-1}\}) < x (\{h_{\text{supp}}, d_{\text{supp}}\})) + w$ 
    \item $Q$ fulfills $(Q1) - (Q3)$ from Lemma \ref{chz:bottom:setup} for $i = \text{supp}, j = \text{start}$
\end{enumerate}
We prove this by showing that if these invariants hold at the beginning of a while-loop iteration, then they also hold 
at the end of that iteration. The statement then follows inductively, since it is trivial that these invariants hold 
right before the first iteration of the outer while-loop. \\
\textit{Case 1: }The invariants hold at the beginning of the inner while-loop (line \Ref{chz:bot:second_while}).  \\ 
Assume that the function does not return in line \ref{chz:bot:return} and we instead proceed in the else branch in line 
\ref{chz:bot:else}. This implies that $\{h_r,d_r\} < \text{support} +_x w$ and $y(\text{support}) \geq y(h_r)$. From 
this we can now follow with invariant $(2)$ that $\{h_r,d_r\}$ can neither be $\text{rise}(c_k)$ nor $\text{rift-end}(c_k)$. 
Consequently, both $(3)$ and $(4)$ still hold after incrementing $r$ by 1. Since no other variables are changed, we have that 
$(1),(2)$ and $(5)$ remain valid. This concludes the proof of the preservation of the invariants when the else-branch is taken.

So now consider the case in which the if-branch at line \ref{chz:bot:ifrise} is taken. From invariants $(2)$ and $(3)$ 
we can follow that $\{h_r,d_r\} = \text{rise}(c_k)$ in this case. Moreover, $\text{rift-start}(c_k)$ cannot precede 
$\{h_r,d_r\}$ in anticlockwise order. This is because by invariant $(2)$, $\text{rift-start}(c_k)$ cannot lie between 
$c_k$ and support. And since we also have in this  case that $y(c_k) = y(\text{support})$, $\{h_r,d_r\} <_x \text{support} +_x w$ 
and $\text{rise}(c_k) = \{h_r,d_r\}$, it can also not lie between support and $\{h_r,d_r\}$. Therefore, the next points 
for $C_{alg}$ are case $(ii)$ points from Definition \ref{chz:setc}, and thus in line \Ref{chz:bot:rise_point_add} they 
are added correctly to $C_{\text{alg}}$ which thus preserves $(1)$ after this update. And since the adjustments in line 
\Ref{chz:bot:rise_end} cause $\{h_{\text{supp}},d_{\text{supp}}\} = \{h_{\text{start}}, d_{\text{start}}\} = \{h_r,d_r\}$ 
and $Q = \emptyset $, one can verify easily that $(2) - (5)$ also hold after this adjustment. This concludes the proof of the statement for Case 1.

\textit{Case 2: } The invariants hold at the beginning of the outer while-loop (line \Ref{chz:bot:first_while}). \\ 
As we have proven the statement for Case 1 already, we can assume that the invariants hold at line \Ref{chz:bot:firstrift}, 
regardless of whether we entered the inner while-loop or not. To show that invariants $(1) -(3)$ hold at the end of the 
iteration, we will perform a sub-case distinction based on whether new points are added to $C_{\text{alg}}$ at line 
\Ref{chz:bot:riftadd}. However, proving the preservation of $(4)$ and $(5)$ can be done for both cases together, so we start with that.

The preservation of $(4)$ is now trivial in this case, as the $x-$coordinate of $\{h_{\text{supp}}, d_{\text{supp}}\}$ 
in increased in line \Ref{chz:bot:lastrift} while $r$ remains unchanged. For the preservation of $(5)$, observe that in 
line \Ref{chz:bot:qprime} we compute a set $Q'$ that satisfies $(Q1) - (Q3)$ for $i= \text{start}, j = r$, as it follows 
from Lemma \ref{chz:bottom:setup}. After merging this set with the current set $Q$ in line \Ref{chz:bot:qprimeqmerge}, 
the adapted $Q$ satisfies $(Q1) - (Q3)$ for $i = \text{supp}, j = r$ by Lemma \ref{chz:bottom:merge}. Next, we remove 
the front edge in $Q$ and update support to the right vertex of that edge. As property $(Q3)$ held before that adaption 
of $Q$, one can now easily verify that the adapted set $Q$ satisfies $(Q1) - (Q3)$ for $i = \text{(the adapted) supp}, j = r$. 
Since start is updated to $r$ in that same line, this completes the argument for the preservation of invariant $(5)$. 
So now we proceed with the subcase distinction.\vspace{0.3cm} \\ 
\textit{Case 2.1: }This iteration adds new points to $C_{\text{alg}}$ in line \Ref{chz:bot:riftadd}. \\
Note, that this implies that the horizontal edge $a$ which maximizes $y(a)$ among those with 
$x_{\min}(a) \in [x(h_{\text{supp}}), x(h_r))$, satisfies $y(a) < y(\text{support}) = y(c_k)$. Now by invariant $(4)$, 
we can even follow that $x_{\min}(a) \in [x(h_{\text{supp}}), x(h_{\text{supp}}) + w)$. In combination with invariants 
$(2)$ and $(3)$ this now implies that 
\begin{align*}
    & \text{rift}(c_k) = [x(h_{\text{supp}}), x(h_{\text{supp}}) + w) , \text{rift-start}(c_k) = \{h_{\text{supp}},d_{\text{supp}}\} , \\ & \text{rift-end}(c_k) = \{h_r,d_r\}, \text{rise}(c_k) \geq_x h_r
\end{align*}
Thus, the next points to be added to $C_{\text{alg}}$ are case $(iii)$ points from Definition \ref{chz:setc}.  
Therefore, these points are added correctly in line \Ref{chz:bot:riftadd} to $C_{\text{alg}}$ and $(1)$ still holds 
after that update. One can now verify very easily that both $(2)$ and $(3)$ remain valid at the end of the iteration. 

\noindent 
\textit{Case 2.2: } This iteration does not reach line \Ref{chz:bot:riftadd} \\ 
With the invariants $(2)$ and $(3)$ it follows that this occurs if and only if the interval $[x(c_k), x(h_{\text{supp}}) + w)$ 
does not contain $x(\text{rise}(c_k))$ and $\{h_{\text{supp}},d_{\text{supp}}\} \neq \text{rift-start}(c_k)$~(as illustrated in Figure~\ref{chz:fig:norisenorift}). In this case, 
the support is set to the right vertex of a horizontal edge $a$ with $x_{\min}(a) \in (x(h_{\text{supp}}), x(h_{\text{supp}}) + w)$ 
and $y(a) = y(c_k)$. It is now very straightforward to verify that invariants $(1)-(3)$ still hold at the end of the 
iteration as the set $C_{alg}$ does not change. 

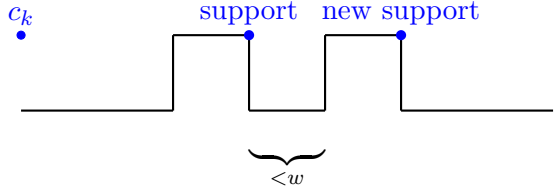
\begin{figure}[!t]
    \centering
    \begin{tikzpicture}
    \draw[thick] (0,0) -- (2,0);
    \draw[thick] (2,0) -- (2,1);
    \draw[thick] (2,1) -- (3,1);
    \draw[thick] (3,1) -- (3,0);
    \draw[thick] (3,0) -- (4,0);
    \draw[thick] (4,0) -- (4,1);
    \draw[thick] (4,1) -- (5,1);
    \draw[thick] (5,1) -- (5,0);
    \draw[thick] (5,0) -- (7,0);

    \filldraw[blue] (0,1) circle (1.5pt) node[above] {$c_k$};
    \filldraw[blue, thick] (3,1) circle (1.5pt) node[above] {support};
    \filldraw[blue, thick] (5,1) circle (1.5pt) node[above] {new support};
    \path (3,0) -- (4,0)
    node[midway, below=10pt]
    {\(\underbrace{\hspace{1cm}}_{\Large < w}\)};
       
\end{tikzpicture}
    \caption{Case 2.2 of Theorem \ref{chz:bottom:correct}}
    \label{chz:fig:norisenorift}
\end{figure}

This concludes the proof of the preservation of the invariants. We will now show that the algorithm terminates. 
Observe for that, that in each iteration of the inner while-loop either the variable $r$ is incremented or the algorithm 
terminates. Thus, the inner while-loop can be executed at most $m$ times. Now we reach line \Ref{chz:bot:firstrift} only 
when $\{h_r,d_r\} \geq_x \text{support} + w$. Although $r$ is not incremented in this part, we update the support variable 
to a vertex further to the right. Thus, after a finite number of such iterations, we will enter the inner-while loop again. 
This shows that the algorithm terminates. 

With this in mind, we can finally show that the algorithm computes the set $C$ correctly. For that, consider the iteration 
of the inner while-loop where $r=m$ and let $c_k$ denote the last element in $C_{\text{alg}}$ at that state. By invariants 
$(2)$ and $(3)$ we know that neither $\text{rise}(c_k)$ nor $\text{rift}(c_k)$ lies on $\APath{c_k,d_m}$. As $d_m$ is now 
the lower vertex of $R(S)$, it follows that neither of them exist on $FB(S)$.
This implies, that $c_k$ is the second to last point in $C$. By invariant $(1)$, we have 
$C_{\text{alg}} = \{c_0,c_1, \ldots ,c_{l-1}\}$ at that state. Therefore, when the final point is added to $C_{\text{alg}}$ 
in line \ref{chz:bot:lastel}, the algorithm returns the complete and correct set $C$ (by the definition of $C$ it is trivial 
that this line computes the last point correctly).

So now it only remains to analyze the runtime complexity. As explained earlier, the inner while-loop can be executed at most $O(nv(S))$ times (as $m\leq nv(S)$). Each iteration of 
this inner loop can be performed in constant time, so the total runtime of this inner loop is $O(nv(S))$. 

Lines \Ref{chz:bot:firstrift} to \Ref{chz:bot:lastrift} can also be executed at most $O(nv(S))$ times, since in each such 
iteration the support variable is updated to a vertex of $FB(S)$ which lies strictly further to the right. Besides the 
$SETUP$\&$MERGE$-calls, all operations in this segment can be performed in constant time. Now observe, that each $SETUP$-call 
processes disjoint ranges of horizontal edges, as after calling $SETUP(\text{start},r)$ we set start to $r$ and never decrement it. 
Thus, each horizontal edge is considered in at most one $SETUP$-call and this bounds the total runtime of the $SETUP$-calls by $O(nv(S))$. 
Now consider the $MERGE$-calls. Observe, that each $MERGE$-call runs in $O(t+s)$ time, where $t$ is a constant and $s$ denotes the number of 
edges this call removes from the first deque parameter. Since any edge that has been removed by a $MERGE$-call is never 
added again to a deque in the algorithm, this also bounds the total runtime for all $MERGE$ calls by $O(nv(S))$. This 
completes the runtime analysis and thus concludes the proof of the theorem. \qedhere 

\end{proof}

\subsubsection{The Top-Function}

We now turn to examining the motion of the horizontal bar across the top boundary of the nice hole. 
The goal will be again to compute all positions where such a bar can be placed in the nice hole without intersecting the 
top boundary. One could of course solve this problem with simply reformulating the Bottom-function to work for the top 
boundary. However, for nice holes the top boundary exhibits a much more constrained structure than the bottom boundary. 
This is because a nice hole may contain lower notches, but no top notches. As a result, the top boundary $FT(S)$ has a 
staircase-like shape, with the only potential exception of the edges associated with the unique falling corner. This 
will simplify this computation significantly, which will be described after the necessary definitions, which are the 
counterparts of the definitions from Section~3.2.2. This simplification of the computation was also 
described in \cite{BLCHZ}, but lacked again formal precision as the return value was not properly defined. Therefore, 
while the corresponding pseudocode is again based on Chazelle's description in \cite{BLCHZ} --- with again only 
minor modifications for improved readability and correctness, particularly to address edge cases which were not handled 
correctly --- the accompanying definitions, lemmas and correctness proof provide an alternative and more rigorous 
description than that given in \cite{BLCHZ}.

\begin{definition}
    Let $S$ be a nice hole. Then we define $S_{top}$ as the nice hole given by the boundary 
    \begin{equation*}
         FT(S)\cup \text{seg}\Big(l_o, (x(L(S), y_{\min}(S)), (x(R(S)), y_{\min}(S)), r_o\Big)
    \end{equation*} 
    where $l_o,r_o$ are the upper vertices of $L(S)$ and $R(S)$ respectively
\end{definition}

\begin{definition}\label{chz:top:g}
    Let $S$ be a nice hole and let $p_1 \in \{x(L(S)),x(L(S))+1, \ldots ,x(R(S)) -w\}$ be arbitrary. We define 
    \begin{equation*}
        g(p_1) := \max\{p_2 \in \mathbb{R} | \text{seg}((p_1,p_2),(p_1 + w , p_2)) \subseteq S_{top}\}
    \end{equation*}
\end{definition}

\begin{definition}\label{chz:setd}
    Let $S$ be a nice hole. Then we define $D$ as the set which contains for every $p_1 \in \{x(L(S)),  \ldots , x(R(S)) - w\}$ exactly the following points 
    \begin{enumerate}[label=(\roman*)]
        \item $(p_1,g(p_1))$ for $p_1 \in \{x(L(S)), x(R(S)) - w\}$
        \item $\{(p_1,g(p_1)), (p_1,z)\}$ if $g(p_1 + 1) = z < g(p_1), p_1 < x(R(S)) - w$
        \item $\{(p_1, z), (p_1,g(p_1))\}$ if $g(p_1 - 1) = z < g(p_1), p_1 > x(L(S))$ 
    \end{enumerate}
\end{definition}

\begin{remark}
    As seen for the set $C$, we will require the set $D$ to remain ordered at all times. The ordering, as well as the 
    convention of duplicates, is defined completely analogously to that of $C$. We can also again assume, without 
    loss of generality, that the first edge of $FT(S)$ has a width of at least $w$.
\end{remark}

Now, the previously mentioned restricted structure of $FT(S)$ will enable us to express $g$ explicitly beforehand. 
For this, let us first define the following notational convention: 

For a nice hole $S$ and a $p_1 \in [x(L(S)), x(R(S))]$,  let
\begin{equation*}
    y_{\max}(S(p_1)) := \max\{p_2 \in \mathbb{R} | (p_1,p_2) \in FT(S)\}
\end{equation*}
Additionally, let us denote by $c_f$ the unique falling corner $c_f(S)$ of $S$ if such exists. 

Now we can explicitly express the function $g$

\begin{lemma}\label{chz:top:explicit:g}
    Let $S$ be a nice hole. Then we have for $p_1 \in \{x(L(S)), \ldots ,x(R(S))-l\}$
    \begin{equation*}
        g(p_1) = \begin{cases}
            y_{\max}(S(p_1)) \ \ \ \textnormal{if } c_f \textnormal{ does not exist or if } x(p_1) \leq x(c_f) - w \\ 
            \min\{y(c_f), y_{\max}(S(p_1))\}  \ \ \textnormal{if } c_f \textnormal{ exists and } x(p_1) > x(c_f) - w
        \end{cases}
    \end{equation*}
\end{lemma}
\noindent 
\begin{proof}
    This follows immediately from the described structure of $FT(S)$ \qedhere
\end{proof}

With this in mind, we can now describe the Top-function. For that, we will again use the analogous notational 
convention that $\big\{ \{h_1,d_1 \}, \{h_2,d_2\} , \ldots ,\{h_{m-1},d_{m-1}\}\big\}$ denotes the vertical 
edges of $FT(S)$ in clockwise order with $h_i >_y d_i \ \forall i \in \{1, \ldots ,m-1\}$. Similarly, we 
denote $\{h_0,d_0\} := L(S), \{h_m,d_m\} := R(S)$

\SetKwInOut{Input}{Input}
\SetKwInOut{Output}{Output}
\SetKw{Continue}{continue}
\SetKw{Break}{break}
\SetKwIF{If}{ElseIf}{Else}{if}{then}{else if}{else}{end}
\SetKw{Goto}{goto}
\begin{algorithm}[H]
\caption{Top-Function}
\Input{Nice hole $S$ , $w\in \mathbb{N}_{>0}$ with $x_{\max}(e) - x_{\min}(e) \geq w $ for the first edge $e$ of $FT(S)$}
\Output{Set $D$ as described in Definition \ref{chz:setd}}
\vspace{0.5cm}
    $D_{\text{alg}} \gets \{h_0\}$ \;
    $r \gets 1$ \;
    \If{$c_f(S)$ \textnormal{does not exist}}{
        \While{$x(\{h_r,d_r\}) \leq x(R(S)) -w $}{
            $D_{\text{alg}}\gets D_{\text{alg}} \cup\{d_r\} \cup \{h_r\}$ \;
            $r \gets r +1 $ \;
        }
    }
    \Else{
        \While{$x(\{h_r,d_r\}) \leq x(c_f) -w $}{
            $D_{\text{alg}} \gets D_{\text{alg}} \cup\{d_r\} \cup \{h_r\}$ \; \label{chz:top:fc_trivial_adds}
            $r \gets r + 1$ \;
        }
        \If{$y(d_r) > y(c_f)$}{
            $D_{\text{alg}} \gets D_{\text{alg}} \cup \{(x(c_f) - w, y(d_r)\} \cup \{(x(c_f) - w, y(c_f))\}$\; \label{chz:top:linecase1}
            \Goto line \Ref{chz:top:lastadd} \;
        }
        \ElseIf{$y(d_r) = y(c_f)$}{
            \If{$d_r = c_f$}{
                $D_{\text{alg}} \gets D_{\text{alg}} \cup \{(x(c_f) - w, y(h_{r-1})\} \cup \{(x(c_f) - w, y(c_f))\}$\; \label{chz:top:algo:weird_edge_case}
            }
            \Goto line \Ref{chz:top:lastadd} \;
        }
        \While{$x(\{h_r,d_r\}) \leq x(R(S)) -w $}{
            \If{$y(h_r) < y(c_f)$ \label{chz:top:last_if}}{
                $D_{\text{alg}} \gets D_{\text{alg}} \cup\{d_r\} \cup \{h_r\}$ \;
                $r \gets r + 1$ \;
            }
            \Else{
                $D_{\text{alg}} \gets D_{\text{alg}} \cup \{d_r\} \cup \{(x(d_r),y(c_f))\}$ \;
                \Break \;
            }
        }
    }
    $D_{\text{alg}} \gets D_{\text{alg}} \cup \{(x(R(S))- w, y(\text{back}(D_{\text{alg}}))\}$ \; \label{chz:top:lastadd}
    \Return $D_{\text{alg}}$ \;
\end{algorithm}

\begin{theorem}\label{chz:top:correct}
    The algorithm Top-Function computes the set $D$ for the specified input correctly and operates in $O(nv(S))$ time.
\end{theorem}

\begin{proof}
    If $S$ does not contain a falling corner, then the correctness is trivial by Lemma \ref{chz:top:explicit:g}. 
    In the case where $S$ does contain such a $c_f(S)$, it also follows directly from that Lemma that the points 
    added in line \Ref{chz:top:fc_trivial_adds} are correct in the sense that the corresponding while loop yields a 
    subset of $D$ with correct ordering and without having skipped a point. Now, let $\{h_r,d_r\}$ be the first 
    vertical edge such that $x(\{h_r,d_r\})> x(c_f) - w$. Three possible cases can now occur:  \\
    \noindent 
    \textit{Case 1: } $y(d_r) > y(c_f)$  (Figure \ref{fig:chz:top:theorem:case1}) \\
    In this case, it is straightforward to verify that line \Ref{chz:top:linecase1} adds the third- and second to last point to $D$ correctly.
    Thus, after the last point is added in line \Ref{chz:top:lastadd} to $D_{alg}$, the algorithm returns the complete and correct set $D$. 

    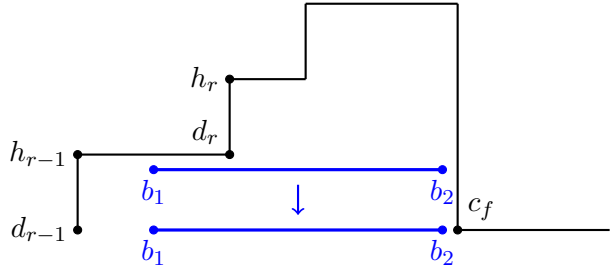
\begin{figure}[H]
        \centering
        \begin{tikzpicture}
    \draw[thick] (0,0) -- (0,1);
    \draw[thick] (0,1) -- (2,1);
    \draw[thick] (2,1) -- (2,2);
    \draw[thick] (2,2) -- (3,2);
    \draw[thick] (3,2) -- (3,3);
    \draw[thick] (3,3) -- (5,3);
    \draw[thick] (5,3) -- (5,0);
    \draw[thick] (5,0) -- (7,0);

    \filldraw[black] (0,0) circle (1.5pt) node[left] {$d_{r-1}$};
    \filldraw[black] (0,1) circle (1.5pt) node[left] {$h_{r-1}$};
    \filldraw[black] (2,1) circle (1.5pt) node[above left] {$d_r$};
    \filldraw[black] (2,2) circle (1.5pt) node[left] {$h_r$};
    \filldraw[black] (5,0) circle (1.5pt) node[above right] {$c_f$};

    \draw[blue, very thick] (1,0.8) -- (4.8, 0.8);
    \draw[blue, very thick] (1, 0) -- (4.8,0);
    \filldraw[blue] (1,0.8) circle (1.5pt) node[below] {$b_1$};
    \filldraw[blue] (4.8,0.8) circle (1.5pt) node[below] {$b_2$};
    \draw[->, blue, thick ] (2.9,0.6) -- (2.9,0.2);
    \filldraw[blue] (1,0) circle (1.5pt) node[below] {$b_1$};
    \filldraw[blue] (4.8,0) circle (1.5pt) node[below] {$b_2$};
    
\end{tikzpicture}
        \caption{Case 1 of Theorem \ref{chz:top:correct}}
        \label{fig:chz:top:theorem:case1}
    \end{figure}

    \noindent 
    \textit{Case 2:} $y(d_r) = y(c_f)$ \\  
    Assume first that we have $x(d_r) \neq x(c_f)$ and thus $x(d_r) < x(c_f)$.
    In this case, $h_{r-1}$ is the second to last point of $D$ and was already added to $D_{\text{alg}}$ in the 
    while-loop in line \Ref{chz:top:fc_trivial_adds}. Thus, we have again that after the last point is added in 
    line \Ref{chz:top:lastadd} to $D_{alg}$, the algorithm returns the complete and correct set $D$. 

    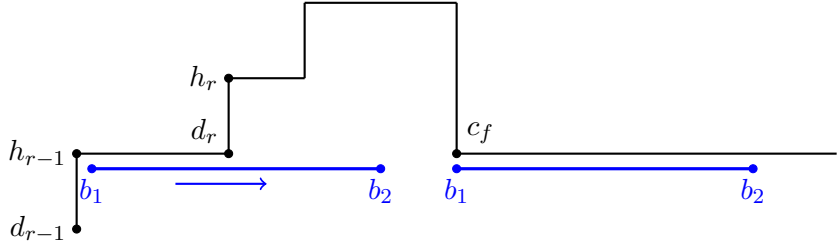
\begin{figure}[H]
        \centering
        \begin{tikzpicture}
    \draw[thick] (0,0) -- (0,1);
    \draw[thick] (0,1) -- (2,1);
    \draw[thick] (2,1) -- (2,2);
    \draw[thick] (2,2) -- (3,2);
    \draw[thick] (3,2) -- (3,3);
    \draw[thick] (3,3) -- (5,3);
    \draw[thick] (5,3) -- (5,1);
    \draw[thick] (5,1) -- (10,1);

    \filldraw[black] (0,0) circle (1.5pt) node[left] {$d_{r-1}$};
    \filldraw[black] (0,1) circle (1.5pt) node[left] {$h_{r-1}$};
    \filldraw[black] (2,1) circle (1.5pt) node[above left] {$d_r$};
    \filldraw[black] (2,2) circle (1.5pt) node[left] {$h_r$};
    \filldraw[black] (5,1) circle (1.5pt) node[above right] {$c_f$};

    \draw[blue, very thick] (0.2,0.8) -- (4, 0.8);
    \draw[blue, very thick] (5, 0.8) -- (8.9,0.8);
    \filldraw[blue] (0.2,0.8) circle (1.5pt) node[below] {$b_1$};
    \filldraw[blue] (4, 0.8) circle (1.5pt) node[below] {$b_2$};
    \draw[->, blue, thick] (1.3,0.6) -- (2.5,0.6);
    \filldraw[blue] (5, 0.8) circle (1.5pt) node[below] {$b_1$};
    \filldraw[blue] (8.9,0.8) circle (1.5pt) node[below] {$b_2$};
    
\end{tikzpicture}
        \caption{Case 2.1 of Theorem \ref{chz:top:correct}}
        \label{fig:chz:top:theorem:case2}
    \end{figure}

    \noindent 
    Now assume, that $d_r = c_f$. This corresponds exactly to the case, where after the last edge 
    $\{h_{r-1},d_{r-1}\}$ with $x(\{h_{r-1},d_{r-1}\}) \leq x(c_f) - w$ the vertical edge corresponding 
    to the falling corner follows directly. Also in this case it is straightforward to verify that line 
    \Ref{chz:top:algo:weird_edge_case} adds the third- and second to last point to $D$ correctly.

    \begin{figure}[H]
        \centering
        \begin{tikzpicture}
    \draw[thick] (0,0) -- (0,1);
    \draw[thick] (0,1) -- (2,1);
    \draw[thick] (2,1) -- (2,3);
    \draw[thick] (2,3) -- (7,3);
    \draw[thick] (7,3) -- (7,1);
    \draw[thick] (7,1) -- (10,1);

    \filldraw[black] (0,0) circle (1.5pt) node[left] {$d_{r-2}$};
    \filldraw[black] (0,1) circle (1.5pt) node[left] {$h_{r-2}$};
    \filldraw[black] (2,1) circle (1.5pt) node[above left] {$d_{r-1}$};
    \filldraw[black] (2,3) circle (1.5pt) node[left] {$h_{r-1}$};
    \filldraw[black] (7,1) circle (1.5pt) node[above right] {$c_f$};

    \draw[blue, very thick] (3,2.8) -- (6.8, 2.8);
    \draw[blue, very thick] (3, 1) -- (6.8,1);
    \filldraw[blue] (3,2.8) circle (1.5pt) node[below] {$b_1$};
    \filldraw[blue] (6.8, 2.8) circle (1.5pt) node[below] {$b_2$};
    \draw[->, blue, thick] (4.9,2.4) -- (4.9,1.4);
    \filldraw[blue] (3, 1) circle (1.5pt) node[below] {$b_1$};
    \filldraw[blue] (6.8,1) circle (1.5pt) node[below] {$b_2$};
    
\end{tikzpicture}
        \caption{Case 2.2 of Theorem \ref{chz:top:correct}}
        \label{fig:chz:top:theorem:case2_2}
    \end{figure}
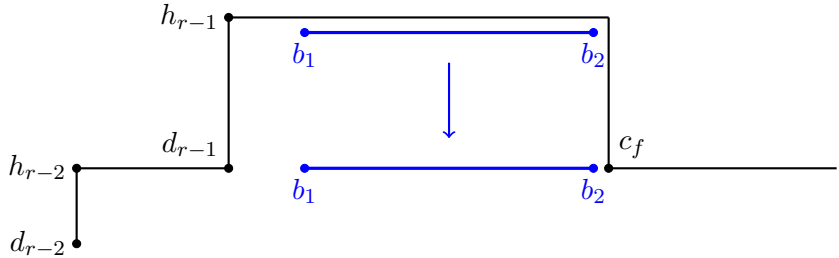

    \noindent 
    \textit{Case 3:} $y(d_r) < y(c_f)$ \\ 
    In this case, we will first encounter a vertical edge $\{h_s,d_s\}$ on our traversal with 
    $y(d_s) < y(c_f), y(h_s) \geq y(c_f)$, before we find an edge $\{h_{s'}, d_{s'}\}$ with $y(d_{s'}) \geq y(c_f)$. 
    It is now again straightforward to verify through a simple case analysis based on the if-condition in line 
    \Ref{chz:top:last_if} that, in this case as well, the algorithm computes the set $D$ correctly up to the last point, 
    which is again added in line \Ref{chz:top:lastadd}.

    \begin{figure}[H]
        \centering
        \begin{tikzpicture}
    \draw[thick] (0,0) -- (0,1);
    \draw[thick] (0,1) -- (2,1);
    \draw[thick] (2,1) -- (2,2);
    \draw[thick] (2,2) -- (3,2);
    \draw[thick] (3,2) -- (3,4);
    \draw[thick] (3,4) -- (5,4);
    \draw[thick] (5,4) -- (5,3);
    \draw[thick] (5,3) -- (10,3);

    \filldraw[black] (0,0) circle (1.5pt) node[left] {$d_{r-1}$};
    \filldraw[black] (0,1) circle (1.5pt) node[left] {$h_{r-1}$};
    \filldraw[black] (2,1) circle (1.5pt) node[above left] {$d_r$};
    \filldraw[black] (2,2) circle (1.5pt) node[left] {$h_r$};
    \filldraw[black] (5,3) circle (1.5pt) node[above right] {$c_f$};

    \draw[blue, very thick] (0.2,0.8) -- (4, 0.8);
    \draw[blue, very thick] (2.2, 1.8) -- (6,1.8);
    \filldraw[blue] (0.2,0.8) circle (1.5pt) node[below] {$b_1$};
    \filldraw[blue] (4, 0.8) circle (1.5pt) node[below] {$b_2$};
    \draw[->, blue, thick] (1.3,0.6) -- (2.5,0.6);
    \filldraw[blue] (2.2, 1.8) circle (1.5pt) node[below] {$b_1$};
    \filldraw[blue] (6,1.8) circle (1.5pt) node[below] {$b_2$};
    
\end{tikzpicture}
        \caption{Case 3 of Theorem \ref{chz:top:correct}}
        \label{fig:chz:top:theorem:case2}
    \end{figure}
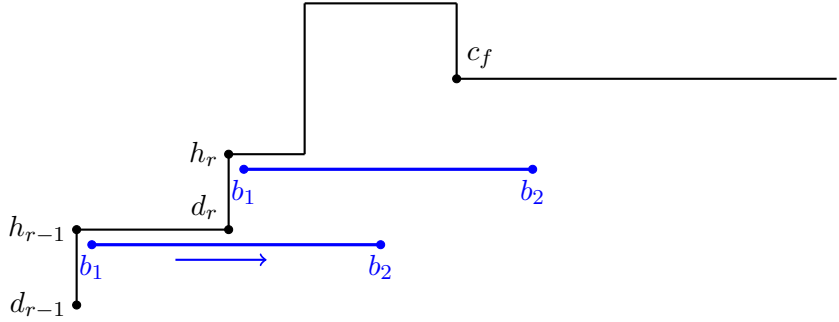

    This concludes the proof of the correctness. The runtime is now clear as no edge is ever considered twice \qedhere. 
    
\end{proof}

\subsubsection{The Placing-Function}

After having examined for a nice hole $S$ how to compute all feasible placements for a horizontal bar within $S_{bot}$ 
and $S_{top}$, we now combine these two results to obtain all feasible BL-stable locations 
for placing the new rectangle $r_{new}$ within $S$. The approach for solving this problem, as well as the corresponding 
pseudocode, is again based on Chazelle's description in \cite{BLCHZ}, with some modifications for readability and 
correctness. However, in a manner similar to Section~3.2.2 and Section~3.2.3, we will present an alternative and more 
rigorous proof of the correctness.

To begin, we derive a criterion that characterizes such feasible BL-stable locations within a nice hole, using the 
previously developed methods. A similar criterion was also implicitly used in Chazelle's description 
\cite{BLCHZ}, but it was neither explicitly stated nor formally proven. 

\begin{lemma}\label{chz:placing_lemma}
    Let $S$ be a nice hole and $p = (p_1,p_2) \in S$. Furthermore, let $f$ and $g$ be as defined in Definition 
    \ref{chz:bot:f} and Definition \ref{chz:top:g} respectively. Then we have that $p$ is a feasible BL-stable location 
    for placing the rectangle $r_{new} = (w,h)$ if and only if the following conditions hold
    \begin{enumerate}[label=(\arabic*)]
        \item $g(p_1) -f(p_1) \geq h$
        \item $f(p_1) = p_2$ 
        \item $p_1 = x(L(S))$ or $f(p_1 - 1) > f(p_1)$ or $g(p_1-1) - f(p_1) < h$
    \end{enumerate}
\end{lemma}

\begin{proof}
    Observe first that with our work in Section~3.2.2 and Section~3.2.3 it is straightforward to verify that for a 
    $p_1^* \in \{x(L(S)), \ldots ,x(R(S))-w\}$ not only the segment $\text{seg}\big((p_1^*,f(p_1^*)), (p_1^*+w,f(p_1^*)\big)$ 
    lies in $S_{bot}$ but also the segment $\text{seg}\big((p_1^*,z), (p_1^*+w,z)\big)$ for each $z \in [f(p_1^*), y_{max}(S)]$. 
    The analogous statement now holds for $S_{top}$. Now as $S = S_{bot} \cap S_{top}$ we get that a location $(p_1^*,f(p_1^*))$ 
    for placing $r_{new}$ is feasible if and only if $g(p_1^*) - f(p_1^*) \geq h$. With this in mind, it is straightforward 
    to show the equivalence of $(1) - (3)$ to the feasibility and BL-stability of $p$, as $(1)$ captures the feasibility 
    of an $x$-coordinate and $(2) + (3)$ capture the BL-stability.
\end{proof}

\begin{remark}\label{chz:ahhh_hilfe_zeit}
From this Lemma it follows, that for any such feasible BL-stable location $p=(p_1,p_2)$ we either have $p_1 = x(L(S))$ 
or that one of the following holds : $g(p_1 - 1) \neq g(p_1)$, $f(p_1-1) \neq f(p_1)$. This implies that for every such 
$p_1$, there exists a point $q\in C \cup D$ such that $x(q) = p_1$. 
Hence, to compute all feasible BL-stable locations it suffices to check conditions $(1) - (3)$ only for those points 
$(p_1, f(p_1))$, where $p_1$ appears as the $x$-coordinate of a point in $C \cup D$.
\end{remark}

This observation will now be formalized by the Placing-Function described in the pseudocode on the next page.

\begin{theorem}\label{chz:placing:correct}
    The Placing-Function computes the set $M$ correctly. It operates in $O(|C|+|D|)$ time.
\end{theorem}

\SetKwInOut{Input}{Input}
\SetKwInOut{Output}{Output}
\SetKw{Continue}{continue}
\SetKw{Break}{break}
\SetKwIF{If}{ElseIf}{Else}{if}{then}{else if}{else}{end}
\SetKw{Goto}{goto}
\begin{algorithm}[!tp]
\caption{Placing-Function}
\Input{Nice hole $S$ , $r_{new} = (w,h) \in \mathbb{N} \times \mathbb{N}$, sets $C =\{c_0, \ldots ,c_{l}\}, D = \{d_0, \ldots ,d_s\}$ 
        from Defintions \ref{chz:setc} and \ref{chz:setd} respectively}
\Output{The set $M$ of all feasible BL-stable locations for placing $r_{new}$ within $S$}
\vspace{0.5cm}
    $M \gets \emptyset$ \;
    \lIf{$d_0 - c_0 \geq h$}{$M \gets M \cup \{c_0\}$}
    $\text{Declare }  g,f \in \mathbb{Z}$ \;
    $i \gets 0, j \gets 0$ \;
    \While{$i < l-1$ \textnormal{or} $j < s - 1$ \label{chz:placing:while}}{
        \If{$x(c_{i+1}) < x(d_{j+1})$}{
            \If{$x(d_j) < x(c_{i+1})$ \textnormal{or} $j = 0$ \label{chz:placing:first:g:comp:start}}{
                $g \gets y(d_j)$\;
            }
            \Else{
                $g\gets \max\{y(d_{j-1}), y(d_j)\}$ \ \label{chz:placing:first:g:comp:end};
            }
            \If{$y(c_{i+1}) > y(c_{i+2})$ \textnormal{and} $g - y(c_{i+2}) \geq h$ \label{chz:placing:first:m:add:start}}{
                $M \gets M \cup \{c_{i+2}\}$ \; \label{chz:placing:first:m:add:end}
            }
            $i \gets i + 2$ \; \label{chz:placing:i:increment:alone}
        }
        \ElseIf{$x(d_{j+1}) < x(c_{i+1})$}{
            \If{$x(c_i) < x(d_{j+1})$ \textnormal{or} $i = 0$}{
                $f \gets y(c_i)$\;
            }
            \Else{
                $f\gets \min\{y(c_{i-1}), y(c_i)\}$ \;
            }
            
            \If{$y(d_{j+1})< y(d_{j+2})$ \textnormal{and} $y(d_{j+1}) - f< h$ \textnormal{and} $y(d_{j+2}) - f\geq h$}{
                $M \gets M\cup \{(x(d_{j+2}), f)\}$ \;
            }
            $j \gets j+2$ \;
        }
        \Else{
            \If{$i= l - 1$ \label{chz:placing:early:return:1}}{
                \If{$y(d_{j+1}) - y(c_l) < h$ \textnormal{and} $y(d_s) - y(c_l) \geq h$}{
                    $M \gets M\cup \{c_l\}$ \;
                }
                \Return $M$ \;
            }
            \If{$j = s - 1$\label{chz:placing:early:return:2}}{
                \If{$y(c_{i+1}) > y(c_{i+2})$ \textnormal{and} $y(d_s) - y(c_l) \geq h$}{
                    $M \gets M \cup \{c_{l}\}$ \;
                }
                \Return $M$ \;
            }
            
            $f \gets \min\{y(c_{i+1}),y(c_{i+2})\}$ \; \label{chz:placing:last:f:comp}
            $g \gets \max\{y(d_{j+1}), y(d_{j+2})\}$ \; \label{chz:placing:last:g:comp}
            \If{$y(c_{i+1}) > y(c_{i+2})$  \textnormal{and} $g - f \geq h$}{
                $M \gets M \cup \{c_{i+2}\}$ \;
            }
            \ElseIf{$y(d_{j+1}) < y(d_{j+2})$ \textnormal{and} $g - f \geq h$ \textnormal{and} $y(d_{j+1}) -f < h $}{
                $M \gets M \cup \{x(d_{j+1}), f\}$ \;
            }
            $i \gets i +2, j \gets j +2$ \; \label{chz:placing:increment:together}
        }
    }
    \Return $M$ \;
\end{algorithm}

\begin{proof} First observe, that throughout the algorithm, the indices $i$ and $j$ are always even. Therefore, 
$c_i$ and $d_j$ are always either the first point of set $C$ or $D$ respectively , or the second point of a point-pair 
as defined in Definition \ref{chz:setc} and \ref{chz:setd} respectively. 
We will begin the proof by showing that the following invariants hold at the beginning of each iteration of the 
while-loop in Line \Ref{chz:placing:while}:
\begin{enumerate}[label=(\arabic*)]
    \item $M$ is the set of all such feasible BL-stable locations $p$ with $x(p) \leq \max\{x(c_i),x(d_j)\}$
    \item $\max\{x(c_i),x(d_j)\} \leq \min\{x(c_{i+1}), x(d_{j+1})\}$
\end{enumerate}
These invariants trivially hold at the start of the first iteration. Now assume they hold at the start of an arbitrary 
iteration. Our goal is now to show that they also hold at the end of this iteration as this would thus prove this preservation inductively.

Now given that invariants $(1)$ and $(2)$ hold at the start of this iteration, we can follow with Remark 
\ref{chz:ahhh_hilfe_zeit} that the next point $p$ for $M$ must have $x(p) \geq \min\{x(c_{i+1}), x(d_{j+1})\}$. \\ 
\textit{Case 1: } $x(c_{i+1}) < x({d_{j+1}})$ \\
Since we have by $(2)$ that $x(d_j) \leq x(c_{i+1}) < x(d_{j+1})$, it follows that lines \Ref{chz:placing:first:g:comp:start} 
to \Ref{chz:placing:first:g:comp:end} set the variable $g$ to $g(x(c_{i+1}))$. Note, that we have $x(c_{i+1}) = x(c_{i+2})$.
Now if we have that $y(c_{i+1}) > y(c_{i+2})$, then these two points define a rift-pair by Definition \ref{chz:setc}. 
Hence by definition we have \begin{equation*}
    y(c_{i+2}) = f(x(c_{i+2})) \text{ and } f(x(c_{i+2}) -1 ) > f(x(c_{i+2}))
\end{equation*}
Thus we have by Lemma \ref{chz:placing_lemma} that $c_{i+2}$ is only added to $M$ in line \ref{chz:placing:first:m:add:end} 
if and only if it corresponds to such a feasible BL-stable location. Therefore, invariant $(1)$ is preserved after the 
increment of $i$ by 2 in line \Ref{chz:placing:i:increment:alone}. This increment now also does not break invariant $(2)$, 
as before the increment we had $x(c_{i+2}) = x(c_{i+1}) < x(d_{j+1})$. \\
\textit{Case 2: }$x(d_{j+1}) < x(c_{i+1})$ \\ 
This case can be shown analogously to Case 1. \\ 
\textit{Case 3: }$x(c_{i+1}) = x(d_{j+1})$ \\ 
If we enter the if-branch at line \Ref{chz:placing:early:return:1}, then we have that 
\begin{equation*}
    x(d_{j+1}) = x(d_{j+2}) = x(c_l) = x(d_s).
\end{equation*} 
Similarly, if we enter the if-branch in line \Ref{chz:placing:early:return:2} then it holds that
\begin{equation*}
    x(c_{i+1}) = x(c_{i+2}) = x(d_{s}) = x(c_l)
\end{equation*}
As by definition we have $y(c_l) = f(x(c_l))$ and $y(d_s) = g(x(d_s))$, it is straightforward to verify that $c_l$ is 
added to $M$ in these lines if and only if it corresponds to such a feasible BL-stable location.
If neither if-branch is entered, then lines \Ref{chz:placing:last:f:comp} and \Ref{chz:placing:last:g:comp} set the 
variable $f$ to $f(x(c_{i+2}))$ and $g$ to $g(x(c_{i+2}))$. Following similar arguments as in the prior cases, one can 
now show that the subsequent lines add a point to $M$ if and only if this point defines a feasible BL-stable location. 
Therefore, invariant $(1)$ is preserved after the increment of both $i,j$ in line \Ref{chz:placing:increment:together}. 
And since $x(c_{i+2}) = x(d_{j+2})$ held before the increment, we have that also invariant $(2)$ is preserved after both increments.

After having thus established the preservation of these invariants, we can now easily prove the correctness of the algorithm: 
If the algorithm terminates as we have both $i=l-1$ and $j=s-1$, then we have the correctness by invariant $(1)$. This is because 
the points $c_l$ and $d_l$ cannot individually define such a feasible BL-stable location, but only if they are not distinct 
from the second point of the final point-pair in $C$ or $D$ respectively.
Now consider the case where the algorithm terminates after entering the if-statement in line \Ref{chz:placing:early:return:1}. 
As argued earlier, we have in this case that $x(d_{j+1}) = x(d_s) = x(c_l)$. Thus, by invariant $(1)$, we have computed the set 
$M$ correctly for all $p$ with $x(p) < x(d_s)$. Furthermore, we have also shown in the Case $3$ analysis that we add the point 
$c_l$ to $M$ if and only if it defines a feasible BL-stable location. Thus, we return the correct $M$ also in this case. Since 
the case where the algorithm terminates after entering the if-statement in line \Ref{chz:placing:early:return:2} is analogous, 
this concludes the proof of the correctness. 

The runtime is now clear, as each iteration of the while-loop can be performed in constant time, and in every iteration either 
$i$ or $j$ is incremented,  or the algorithm returns directly. \qedhere 

\end{proof}

\medskip

With this we can formulate the main theorem of Section~3.2 .
\begin{theorem}\label{chz:nice_hole_main_theorem}
    Let $S$ be a nice hole of a partial packing and $r_{new}$ a rectangle. Then we can compute all feasible BL-stable 
    locations for placing $r_{new}$ within $S$ in $O(nv(S))$ time.
\end{theorem}
\begin{proof}
    Follows directly from Theorem \ref{chz:bottom:correct}, \ref{chz:top:correct} and \ref{chz:placing:correct}. 
\end{proof}

\subsection{Nice Hole Partitioning}

After examining the computation of feasible BLS-stable locations within a nice hole in the previous subsection, 
we now turn to the more general problem of computing them for arbitrary BLS-holes. To achieve this, 
we will partition a BLS-hole into a collection of nice holes, thereby reducing this general case to 
the previously solved one.

\subsubsection{\texorpdfstring{$\bm{QW_i}$- and $\bm{QN_i}$-Partitioning}{QWi- and QNi-Partitioning}}

For this partitioning, we will need the special points $QN_i$ and $QW_i$, which will be defined below. This definition is 
taken from \cite{BLCHZ}.

\begin{definition}
    Let $H$ be a hole with canonical ordering  $L_1,N_2, \ldots ,L_k, R$. Then we define for $i\in \{2, \ldots ,k\}$: 
    \begin{itemize}
        \item $Q_i$ as the upper vertex of $N_i$
        \item $QN_i$ as the first intersection of the boundary of $S$ with the vertical line \\ $\{(x(Q_i), p_2)\mid p_2 > y(Q_i)\}$
        \item $QW_i$ as the first intersection of the boundary of $S$ with the horizontal line \\  $\{(p_1, y(Q_i)) \mid p_1 > x(Q_i)\}$
    \end{itemize}
\end{definition}

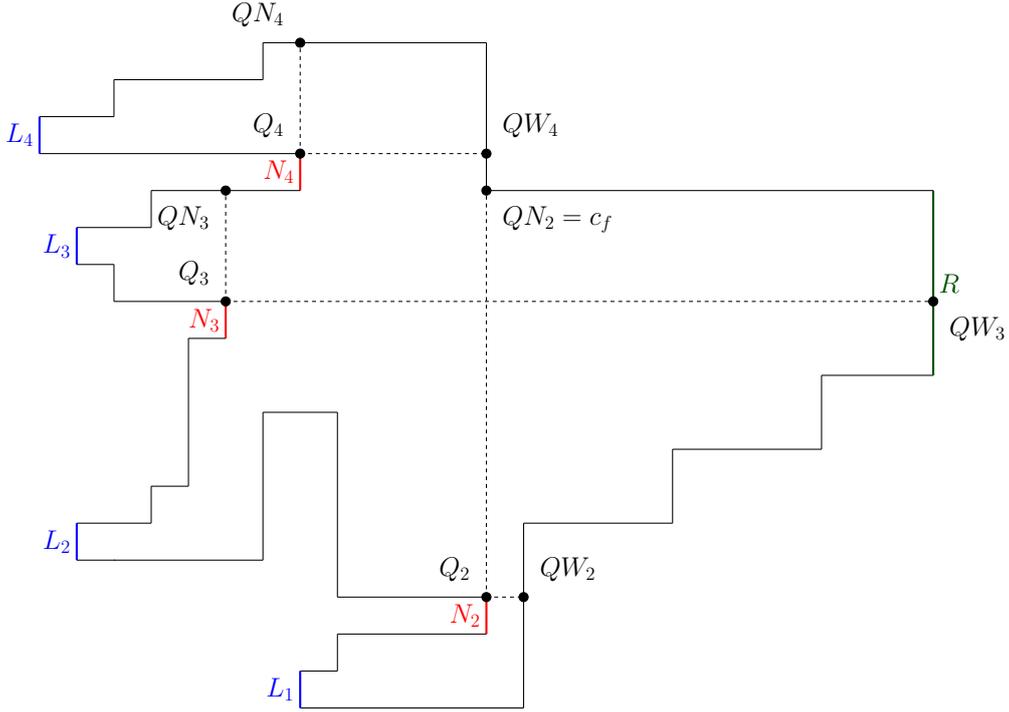
\begin{figure}[H]
    \centering
    \scalebox{0.49}{\begin{tikzpicture}
    \coordinate (A) at (7,1);
    \coordinate (B) at (13,1);
    \coordinate (C) at (13,6);
    \coordinate (D) at (17,6);
    \coordinate (E) at (17,8);
    \coordinate (F) at (21,8);
    \coordinate (G) at (21,10);
    \coordinate (H) at (24,10);
    \coordinate (I) at (24,15);
    \coordinate (J) at (12,15);
    \coordinate (K) at (12,19);
    \coordinate (L) at (6,19);
    \coordinate (M) at (6,18);
    \coordinate (N) at (2,18);
    \coordinate (O) at (2,17);
    \coordinate (P) at (0,17);
    \coordinate (Q) at (0,16);
    \coordinate (R) at (7,16);
    \coordinate (S) at (7,15);
    \coordinate (T) at (3,15);
    \coordinate (U) at (3,14);
    \coordinate (V) at (1,14);
    \coordinate (W) at (1,13);
    \coordinate (X) at (2,13);
    \coordinate (Y) at (2,12);
    \coordinate (Z) at (5,12);
    \coordinate (Z') at (5,11);
    \coordinate (Y') at (4,11);
    \coordinate (X') at (4,7);
    \coordinate (W') at (3,7);
    \coordinate (V') at (3,6);
    \coordinate (U') at (1,6);
    \coordinate (T') at (1,5);
    \coordinate (S') at (2,5);
    \coordinate (R') at (2,5);
    \coordinate (Q') at (6,5);
    \coordinate (P') at (6,9);
    \coordinate (O') at (8,9);
    \coordinate (N') at (8,4);
    \coordinate (M') at (12,4);
    \coordinate (L') at (12,3);
    \coordinate (K') at (8,3);
    \coordinate (J') at (8,2);
    \coordinate (I') at (7,2);

    \draw[thick] (A) -- (B);
    \draw[thick] (B) -- (C);
    \draw[thick] (C) -- (D);
    \draw[thick] (D) -- (E);
    \draw[thick] (E) -- (F);
    \draw[thick] (F) -- (G);
    \draw[thick] (G) -- (H);
    \draw[ultra thick, darkgreen] (H) -- (I) node[midway, right] {\huge $R$};
    \draw[thick] (I) -- (J);
    \draw[thick] (J) -- (K);
    \draw[thick] (K) -- (L);
    \draw[thick] (L) -- (M);
    \draw[thick] (M) -- (N);
    \draw[thick](N) -- (O);
    \draw[thick] (O) -- (P);
    \draw[ultra thick, blue] (P) -- (Q) node[midway, left] {\huge $L_4$};
    \draw[thick] (Q) -- (R);
     \draw[ultra thick, red] (R) -- (S) node[midway, left] {\huge $N_4$};
    \draw[thick] (S) -- (T);
    \draw[thick] (T) -- (U);
    \draw[thick] (U) -- (V);
    \draw[ultra thick, blue] (V) -- (W) node[midway, left] {\huge $L_3$};
    \draw[thick] (W) -- (X);
    \draw[thick] (X) -- (Y);
    \draw[thick] (Y) -- (Z);
    \draw[ultra thick, red] (Z) -- (Z') node[midway, left] {\huge $N_3$};
    \draw[thick] (Z') -- (Y');
    \draw[thick] (Y') -- (X');
    \draw[thick] (X') -- (W');
    \draw[thick] (W') -- (V');
    \draw[thick] (V') -- (U');
    \draw[ultra thick, blue] (U') -- (T') node[midway, left] {\huge $L_2$};
    \draw[thick] (T') -- (S');
    \draw[thick] (S') -- (R');
    \draw[thick] (R') -- (Q');
    \draw[thick] (Q') -- (P');
    \draw[thick] (P') -- (O');
    \draw[thick] (O') -- (N');
    \draw[thick] (N') -- (M');
    \draw[ultra thick, red] (M') -- (L') node[midway, left] {\huge $N_2$};
    \draw[thick] (L') -- (K');
    \draw[thick] (K') -- (J');
    \draw[thick] (J') -- (I');
    \draw[ultra thick, blue] (I') -- (A) node[midway, left] {\huge $L_1$};

    \coordinate (Q2) at (M');
    \coordinate (Q3) at (Z);
    \coordinate (Q4) at (R);

    \filldraw[black] (J) circle (3.5pt) node[below right = 0.3cm ] {\huge $QN_2 = c_f$};

    \filldraw[black] (M') circle (3.5pt) node[above left = 0.3cm] {\huge $Q_2$};

    \filldraw[black] (Z) circle (3.5pt) node[above left = 0.3cm] {\huge $Q_3$};

    \filldraw[black] (R) circle (3.5pt) node[above left = 0.3cm] {\huge $Q_4$};

    \coordinate (QW2) at ($(M') + (1,0)$);

    \filldraw[black] (QW2) circle (3.5pt) node[above right = 0.3cm] {\huge $QW_2$};

    \coordinate (QW3) at ($(Z) + (19,0)$);

    \filldraw[black] (QW3) circle (3.5pt) node[below right = 0.3cm] {\huge $QW_3$};

    \coordinate (QW4) at ($(R) + (5,0)$);

    \filldraw[black] (QW4) circle (3.5pt) node[above right = 0.3cm] {\huge $QW_4$};

    \coordinate (QN2) at ($(M') + (0,11)$);

    \coordinate (QN3) at ($(Z) + (0,3)$);

    \filldraw[black] (QN3) circle (3.5pt) node[below left = 0.3cm] {\huge $QN_3$};

    \coordinate (QN4) at ($(R) + (0,3)$);

    \filldraw[black] (QN4) circle (3.5pt) node[above left = 0.3cm] {\huge $QN_4$};

    \draw[thick, dashed] (Q2) -- (QW2) ;
    \draw[thick,dashed] (Q2) -- (QN2) ;

    \draw[thick,dashed] (Q3) -- (QW3) ;
    \draw[thick,dashed] (Q3) -- (QN3) ;

    \draw[thick,dashed] (Q4) -- (QW4) ;
    \draw[thick,dashed] (Q4) -- (QN4);

\end{tikzpicture}}
    \caption{Special Points $QN_i \& QW_i$}
    \label{fig:enter-label}
\end{figure}

The idea is now to use the $\text{seg}(Q_i,QN_i)$ segments and the $\text{seg}(Q_i,QW_i)$ segments to partition a BLS-hole. 
This method was also described by Chazelle in \cite{BLCHZ}. However, Chazelle did neither address how to compute all such 
special points efficiently, nor did he provide a rigorous proof for the correctness of the partitioning. Section~3.3 will 
now fill that gap by presenting both an efficient computation method for the special points and a formal proof for the partitioning approach.

Moreover, Chazelle's approach was limited to partitions using only the $Q_i-QN_i$ segments. This limitation is now exactly 
where the flaw in his runtime analysis arises, as such partitions alone are not sufficient enough to guarantee a linear runtime. 
This will be discussed later in greater detail together with the presentation of a counter-example. This paper now 
corrects this flaw by introducing an alternative partitioning approach using a combination of $Q_i-QN_i$- and 
$Q_i-QW_i$-segments. This combined partitioning will then enable the computation of BL-stable locations for a BLS-hole in linear time. 

We start by introducing a characterization of the $QN_i$ and $QW_i$ points, which will enable us to compute those special 
points efficiently for a nice hole.

For this characterization we will first prove an auxiliary lemma
\begin{lemma}\label{chz:auxiliary:rme}
    Let $H$ be a BLS-hole and let $q,p$ be two points on the boundary of $H$ such that $x(q) = x(p), \ y(q) < y(p)$. If 
    the last edge on the path $\Path{p,q}$ is not a (clockwise) downward edge of $H$, then $\Path{p,q}$ contains $R(H)$
\end{lemma}
\begin{proof}
   Let $p^*$ be the point on the path $\Path{p,q}$ with $y(p^*)$ maximal among those with $x(p^*) = x(q)$. As the point 
   $p$ itself lies on that path, that implies $y(p^*) > y(q)$. 
   Now consider the path $\Path{p^*, q}$ to define $z$ as follows: if $\Path{p^*, q}$ begins with a downward edge, then 
   $z$ is the lower endpoint of that edge, otherwise $z = p^*$. Observe, that with our assumption we have $z \neq q$ in 
   both cases. Next, consider the path $\Path{z,q}$ and let $z'$ be the first point on that path distinct from $z$ with 
   $x(z') = x(z) = x(q)$. By the maximality of $p^*$, we have $y(z') < y(z)$. By the minimality of $z'$, we have 
   $x_{\min}(\Path{z,z'}) \geq x(z)$ or $x_{\max}(\Path{z,z'}) \leq x(z)$. And as $\Path{z,z'}$ starts with a horizontal 
   edge, we also have that there exists a vertex $w$ on that path with $x(w) \neq x(z)$. Thus by 
   Lemma~\ref{chz:special_edges_equiv}, we have that $\Path{z,z'}$ contains a right notch or a rightmost edge. As the 
   former is impossible, the path has to contain a rightmost edge and as this path is a subpath of $\Path{p,q}$ our statement is shown. \qedhere

\end{proof}

With this auxiliary lemma, we can now present the characterization for $QN_i$ and $QW_i$

\begin{lemma}\label{chz:qni_char}
    Let $H$ be a BLS-hole with canonical ordering $L_1,N_2, \ldots ,L_k, R$ and let $i\in \{2, \ldots ,k\}$ be arbitrary. 
    Furthermore, let $p^*$ be the first point encountered on a clockwise traversal of the boundary of $H$ starting at $Q_i$, 
    which satisfies $x(p^*) = x(Q_i), p^* \neq Q_i$. Then the following holds : \\ If $p^*$ is the upper vertex of a 
    downward edge $e$ of $H$, then $QN_i$ is the lower vertex of $e$. Otherwise, $QN_i$ is the point $p^*$ itself. 
    Additionally, we have that $R$ does not lie on $\Path{Q_i,QN_i}$
\end{lemma}

\begin{proof}
    Let $p$ be the first such point encountered, or, if this point is the upper endpoint of a downward edge $e$, 
    let $p$ instead be the lower endpoint of $e$. Since the first edge on the path $\Path{Q_i,p}$ is a leftward edge, 
    we have $x_{\min}(\Path{Q_i,p}) < x(Q_i)$. Moreover, by the minimality of $p$ we thus have $x_{\max}(\Path{Q_i,p})\leq x(Q_i)$. 
    Therefore, we must have $y(p) > y(Q_i)$, as otherwise this path would contain a right notch by Lemma \ref{chz::special_existence}. 
    Now assume that $p$ is not $QN_i$. This would imply that there exists a $\tilde{p}$ with $x(\tilde{p}) = x(p)$, $y(Q_i) < y(\tilde{p}) < y(p)$ 
    that lies on $\Path{p,Q_i}$. Now assume that $\tilde{p}$ is the first such point on this path.

    Observe that by the minimality of $\tilde{p}$ and by the definition of $p$ we have that the last edge on the path 
    $\Path{p,\tilde{p}}$ is not a downward edge. Therefore, by Lemma \ref{chz:auxiliary:rme} we have that $\Path{p,\tilde{p}}$ 
    contains a rightmost edge of $H$. Now, since $Q_i$ is the upper point of a left notch, the last edge on the path 
    $\Path{\tilde{p},Q_i}$ can also not be a downward edge. Thus, by the same lemma, also $\Path{\tilde{p},Q_i}$ contains a 
    rightmost edge of $H$. However, these paths are edge-disjoint by the definition of $\tilde{p}$. This implies that $H$ 
    contains 2 distinct rightmost edges, contradicting Corollary \ref{chz:exactly:one:rme:and:tme} $\lightning$. 

    Therefore, we conclude that $p=QN_i$. Furthermore, this proof shows that the path $\Path{QNi,Q_i}$ contains $R$, which 
    shows the second part of the lemma \qedhere
\end{proof}

\begin{remark}\label{chz:flip:remark}
    Observe that the proof relies only on the fact that $\pred(Q_i)$ is an upward edge and $\succv(Q_i)$ is a leftward edge. 
    Now these conditions are also fulfilled for the right vertex of a bottom notch. This implies that the analogous 
    statement holds for such vertices $B_j$ and their corresponding first intersection point $BN_j$ of a vertical line 
    extending upward from $B_j$ with the boundary of $H$.
\end{remark}

\begin{lemma}\label{chz:qwi:char}
    Let $H$ be a BLS-hole with canonical ordering $L_1,N_2, \ldots ,L_k, R$ and let $i\in \{2, \ldots ,k\}$ be arbitrary. 
    Furthermore, let $p^*$ be the first point encountered on an anticlockwise traversal of the boundary of $H$ starting 
    at $Q_i$, which satisfies $y(p^*) = y(Q_i), p^* \neq Q_i$. Then the following holds : \\ If $p^*$ is the right 
    vertex of an anticlockwise-leftward edge $e$ of $H$, then $QW_i$ is the left vertex of $e$. Otherwise, $QW_i$ is 
    the point $p^*$ itself. Additionally, we have that the unique upmost edge of $H$ does not lie on $\APath{Q_i,QW_i}$.
\end{lemma}

\begin{proof}
    Observe that the hole $H'$, which arises from $H$ by first rotating the hole 90 degrees anticlockwise and then 
    mirroring it across the vertical axis, remains a BLS-hole. Under this transformation, left notches from $H$ 
    correspond to bottom notches of $H'$ and the top vertex $Q_i$ of a left notch of $H$ corresponds to the right 
    vertex $B_i$ of a bottom notch of $H'$. Moreover, we have that the points $QW_i$ of $H$ correspond to the points 
    $BN_i$ of $H'$ as described in Remark \ref{chz:flip:remark}. The statement now directly follows from this remark 
    together with Lemma \ref{chz:qni_char}, since an anticlockwise traversal of $H$ corresponds to a clockwise traversal 
    of $H'$ due to the mirroring.
\end{proof} 

Lemma \ref{chz:qni_char} now enables us to compute all these points $QN_i$ with a single clockwise traversal of $H$. 
To do so, we start the clockwise traversal at the unique rightmost edge of $H$. During this traversal, we keep track of 
the $x-$coordinates of each vertex $Q_i$ that has already been encountered but for which the corresponding $QN_i$ was not 
identified yet. For such a $Q_i$, the first point that we will now encounter with that same $x-$coordinate, will be 
exactly $QN_i$ by Lemma \ref{chz:qni_char}, except for the special case described in that lemma, which can be handled 
accordingly. Furthermore, according to the same lemma, we will identify all such points before returning to the rightmost edge.

Similarly, Lemma \ref{chz:qwi:char} enables us to analogously compute all the points $QW_i$ with a single anticlockwise 
traversal of $H$, starting at the unique upmost edge.

After establishing these characterizations which allow us to compute these special points efficiently, we will now 
present a rigorous proof of the structure of the partitioning that arises when all segments $\text{seg}(Q_i,QN_i)$ are 
added to a BLS-hole. To carry out this proof, we will introduce new notational conventions: 

For a hole $H$, we denote by $\text{LME}(H)$ the set of all leftmost edges of H. Similarly, $FC(H)$ denotes the set of 
all falling corner,  $LN(H)$ denotes the set of all left notches and $Q(H)$ denotes the set of all upper vertices of the 
left notches. $QN(H)$ and $QW(H)$ are now defined analogously. 

Furthermore, when working with multiple holes in a proof simultaneously, we will explicitly indicate by indexing the 
hole to which a predecessor, successor, or a path refers.

\begin{theorem}\label{chz:partitioning:theorem}
    Let $H$ be a BLS-hole with canonical ordering $L_1(H), N_2(H), \ldots ,L_k(H),\\ R(H)$. Then, after adding the segments 
    $\text{seg}(Q_i,QN_i)$ for all $i \in \{2, \ldots ,k\}$, the following conditions hold:
    \begin{enumerate}[label=(\arabic*)]
        \item The hole $H$ is partitioned in $k$ nice holes $S_1, \ldots ,S_k$
        \item Each leftmost edge $L_i(H)$ is contained in exactly one such nice hole and every such nice hole contains exactly one leftmost edge 
        \item Let $S'$ be a nice hole of the partitioning. Then we have 
            \begin{equation*}
                R(S') = 
                    \begin{cases}
                        R(H) & \text{if } L(S') = L_1(H) \\ 
                        \{{QN_i}^*, Q_i\} & \text{if } L(S') = L_i(H) \text{ for } i \in \{2, \ldots ,k\}
                    \end{cases}
            \end{equation*}
            Here $QN_i^*$ is defined as follows : If $QN_i$ is the lower vertex of a downward edge $e$, then $QN_i^*$ is 
            the upper vertex of $e$. Otherwise, $QN_i^*$ is just $QN_i$.

        \item We have $O(\sum_{i=1}^k nv(S_i)) = O(nv(H))$
        \end{enumerate}
        
\end{theorem}

\begin{proof}
   We will show this with induction over $k = |\text{LME}(H)|$. Furthermore, we will assume in the proof that 
   $QN_i = QN_i^*$ for all $i \in \{2, \ldots ,k\}$, as the proof can be easily adapted to work for the general case 
   where this does not necessarily hold, but doing so would impact readability without yielding additional insights. 
   \\ 
   \textit{Induction start: } $k = 1$ \\ 
   In this case, $H$ is already a nice hole and thus $(1) - (4)$ are trivially fulfilled. \\
   \textit{Induction step: } $k-1 \to k$ \\ 
   So let $k\geq 2$. We first add $\text{seg}(Q_2,QN_2)$. That partitions $H$ in two holes $H_1$ and $H_2$, where 
   the boundary of $H_1$ is given by 
   \begin{equation}\label{chz:eq:borderh1}
        \Path{QN_2,Q_2}(H) \cup \text{seg}(Q_2,QN_2)
   \end{equation} and the boundary of $H_2$ is given by 
   \begin{equation}\label{chz:eq:borderh2}
       \Path{Q_2,QN_2}(H) \cup \text{seg}(QN_2,Q_2)
   \end{equation}\\
   We first want to show that the following conditions hold: 
   \begin{enumerate}[label=(\roman*)]
       \item $H_1$ and $H_2$ are BLS-holes again and $|FC(H)| \geq |FC(H_1)| + |FC(H_2)|$
       \item $\text{LME}(H) = \text{LME}(H_1) \sqcup \text{LME}(H_2)$ ("$\sqcup$" denotes disjoint union) 
       \item $R(H_1) = R(H), L_1(H_1) = L_1(H), R(H_2) = \{QN_2,Q_2\}, L_1(H_2) = L_2(H)$
       \item $Q(H) \setminus \{Q_2\} = Q(H_1) \sqcup Q(H_2)$
       \item $QN(H)\setminus \{QN_2\} = QN(H_1) \sqcup QN(H_2)$
       \item Let $k'$ be maximal such that $L_{k'}(H)$ is contained in $H_2$. Then the canonical ordering of $H_2$ is 
       \begin{equation*}
           L_2(H),\tilde{N}_3,L_3(H), \ldots ,\tilde{N}_{k'},L_{k'}(H),R(H_2)
       \end{equation*}
       and that of $H_1$ is
       \begin{equation*}
            L_1(H),\tilde{N}_{k'+1},L_{k'+1}(H), \ldots ,\tilde{N}_{k},L_{k}(H),R(H_2)
       \end{equation*}
       where $\tilde{N_i}$ is a left notch with the same upper vertex as $N_i(H)$ but with a potentially different lower vertex.
   \end{enumerate}

    $(i):$ By the specification of the boundaries of $H_1$ and $H_2$ in equations (\ref{chz:eq:borderh1}) and 
    (\ref{chz:eq:borderh2}), it is clear that no right or top notches can occur along their boundaries. To show 
    that $|FC(H)| \geq |FC(H_1)| + |FC(H_2)|$, recall that a falling corner is defined as a vertex with a preceding 
    downward edge and a succeeding rightward edge. Thus, if $H_1$ or $H_2$ contains a falling corner, it must be an 
    inner vertex of $\Path{QN_2,Q_2}(H)$ or $\Path{Q_2,QN_2}(H)$ respectively. This is because $\{Q_2,QN_2\}$ is a 
    downward edge with a succeeding leftward edge in $H_2$, and part of an upward edge in $H_1$. Hence, neither $Q_2$ 
    nor $QN_2$ can correspond to falling corners in either subhole. 

    Therefore, a falling corner in $H_1$ or $H_2$ must already be a falling corner of $H$ as the segments defined by 
    $\Path{QN_2,Q_2}(H)$ or $\Path{Q_2,QN_2}(H)$ are identical in $H$ and $H_1$ or $H_2$ respectively. Since the only 
    shared vertices of $H_1$ and $H_2$ are now $Q_2$ and $QN_2$, this also implies that there is no double counting and 
    we have $|FC(H)| \geq |FC(H_1)| + |FC(H_2)|$. This now also proves that $H_1$ and $H_2$ have both at most one falling 
    corner, making them both BLS-holes by definition.

    $(ii):$ Analogous to the argument for falling corners, one can show that any leftmost edge $L\in \text{LME}(H)$ must 
    be an inner edge of either $\Path{QN_2,Q_2}(H)$ or the path $\Path{Q_2,QN_2}(H)$ (as a leftmost edge is defined by 
    three subsequent edges by Lemma~\ref{chz:special_edges_equiv}). Therefore, we have $\text{LME}(H) \subseteq \text{LME}(H_1) \cup \text{LME}(H_2)$. 
    The reverse inclusion can be shown analogously. The disjointness now follows directly from the fact that $Q_2$ and $QN_2$ 
    are the only shared vertices of both subholes, which by the same reasoning cannot be vertices of leftmost edges.

    $(iii):$ One can directly deduce by the boundary definition of $H_1$ that both $R(H)$ and $L_1(H)$ lie on $H_1$. From 
    this it follows by $(i)$, that $R(H)$ is the unique rightmost edge of $H_1$ and by $(ii)$ that $L_1(H_1) = L_1(H)$, as 
    we thus cannot encounter another leftmost edge beforehand. 

    Furthermore, it follows from Lemma \ref{chz:qni_char} that $\{QN_2,Q_2\}$ is the vertical edge with maximum 
    $x$-coordinate of $H_2$, which makes it the unique rightmost edge of $H_2$ by the proof of Lemma \ref{chz:geq_one_most_edge}.
     And $L_1(H_2) = L_2(H)$ now follows directly from the canonical ordering of $H$, as the first leftmost edge on the 
     path $\Path{Q_2,QN_2}(H)$ is thus precisely $L_2(H)$.

   $(iv): $ We first show the disjointness. For that, observe that neither $Q_2$ nor $QN_2$ can be in $Q(H_1)$ or $Q(H_2)$. 
   This is because $Q_2$ is not even a vertex of $H_1$ and for $H_2$ it is a vertex with a preceding downward edge, thus 
   disqualifying it from being the upper vertex of a left notch. Similarly, for $QN_2$ its succeeding edge in $H_1$ cannot 
   be a leftward edge and its succeeding edge in $H_2$ is a downward edge, thus proving the disjointness.

    For equality, we first prove $Q(H) \setminus \{Q_2\} \subseteq  Q(H_1) \sqcup Q(H_2)$. For this, let 
    $\{p_i,Q_i\} \in LN(H) \setminus \{N_2\}$. We trivially have that $\{p_i,Q_i\} \cap \{Q_2\} = \emptyset$ and 
    $Q_i \neq QN_2$. Thus, if we would also have $p_i \neq QN_2$, then this edge would be an inner edge of either 
    $\Path{QN_2,Q_2}(H)$ \text{or} $\Path{Q_2,QN_2}(H)$, which would then prove the inclusion as shown in the previous 
    arguments. However, in this case $p_i = QN_2$ could actually occur, which would lead to the left notch $\{p_i,Q_i\}$ 
    neither appearing in $H_1$ nor in $H_2$. But in this situation, the edge $\{\pred_H(Q_i), Q_i\}$ would be a left notch 
    of $H_1$. Although this notch now differs from $\{p_i,QN_i\}$, it shares the same upper vertex $Q_i$ which proves the 
    inclusion also in this case. The reverse inclusion can now be shown analogously. 

    \begin{figure}[H]
        \centering
        \begin{tikzpicture}
    \draw[thick] (0,0) -- (5,0);
    \draw[thick] (0,0) -- (0,1);
    \draw[thick] (0,1) -- (2,1);
    \draw[thick] (2,1) -- (2,2);
    \draw[thick] (2,2) -- (0,2);
    \draw[thick] (0,2) -- (0,4);
    \draw[thick] (0,4) -- (2,4);
    \draw[thick] (2,4) -- (2,5);
    \draw[thick] (2,5) -- (0,5);
    \draw[thick] (0,5) -- (0,6);
    \draw[thick] (0,6) -- (5,6);

    \filldraw[black] (2,1) circle(1.5pt) node[above right] {$\pred_H(Q_2)$};
    \filldraw[black] (2,2) circle (1.5pt) node[above right] {$Q_2$};
    \filldraw[black] (2,4) circle (1.5pt) node[above right] {$QN_2 = p_i$};
    \filldraw[black] (2,5) circle (1.5pt) node[above right] {$Q_i$};
    \draw[dashed] (2,2) -- (2,4);
    \node at (1, 3) {$H_2$};
    \node at (4,3) {$H_1$};
\end{tikzpicture}
        \caption{Proof for constraint $(iv)$}
        \label{fig:enter-label}
    \end{figure}
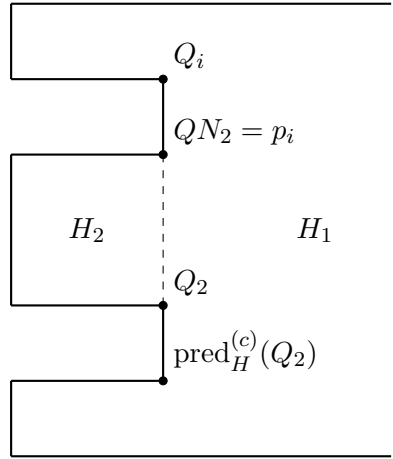

    \noindent
    $(v): $ Recall, that we have shown in $(iv)$ that every $Q_i \in Q(H) \setminus \{Q_2\}$ is either an inner vertex of the
     path $\Path{QN_2,Q_2}(H)$ or of the path $ \Path{Q_2,QN_2}(H)$. Suppose first, that $Q_i$ lies on $\Path{QN_2,Q_2}(H)$. 
     Then by Lemma \ref{chz:qni_char}, the corresponding point $QN_i$ for $H$ must lie on $\Path{Q_i, R(H)}(H)$ which is a 
     subpath of $\Path{QN_2,Q_2}(H)$. Since the boundary of $H_1$ is identical to that of $H$ on this segment, this means that 
     the point $QN_i$ is also identical for both holes (again by Lemma \ref{chz:qni_char}). 

    Now consider the case, where $Q_i$ lies on $\Path{Q_2,QN_2}(H)$. Then the $QN_i$ point for $H_2$ must also lie on this 
    path, as $\{QN_2,Q_2\}$ is the rightmost edge of $H_2$ by $(iii)$. Since the boundary of $H_2$ is now again identical 
    to that of $H$ on this segment, this means that the point $QN_i$ is also identical for both holes.  

    $(vi): $ This follows now directly with $(i)-(v)$, as this is exactly the order of the edges in that they appear on 
    the boundary of $H_1$ and $H_2$.

    \medskip 

    After showing the validity of these conditions, we now proceed to proving the main statement.

    For this, we first add all remaining segments $\text{seg}(Q_i,QN_i)$ for $i \in \{3, \ldots ,k\}$. By $(iv)$ and $(v)$ we 
    have therefore also added all $Q_j(H_1)-QN_j(H_1)$ and all $Q_j(H_2)-QN_j(H_2)$ segments. With this now in place, we 
    can apply the induction hypothesis separately to $H_1$ and $H_2$, as by $(ii)$ both holes contain at least one leftmost 
    edge less than $H$. 

    Now by induction hypothesis, $H_1$ is partitioned into $|\text{LME}(H_1)|$ nice holes and $H_2$ is partitioned into 
    $|\text{LME}(H_2)|$ nice holes. This implies by $(ii)$ that this partitions $H$ into $|\text{LME}(H)|$ nice holes, proving $(1)$ for $H$.

    We also have by $(2)$ for $H_1$, that every nice hole $S$ in the partitioning of $H_1$ contains exactly one leftmost 
    edge of $H_1$ and that every such leftmost edge appears in exactly one such nice hole. Since the analogous statement 
    also holds for $H_2$, we have with $(ii)$ that $(2)$ now also holds for $H$.

    Now to prove $(3)$ for $H$, one can observe that since $L_1(H_1) = L_1(H)$ and $R(H_1) = R(H)$, it follows from $(3)$ 
    applied to $H_1$ that the nice hole $S$ of the partitioning of $H$ that contains $L_1(H)$ satisfies $R(S) = R(H)$. 
    Similarly, since $L_1(H_2) = L_2(H)$ and $R(H_2) = \{QN_2,Q_2\}$, it follows that for the nice hole $S'$ of the 
    partitioning of $H$ that contains $L_2(H)$ we have $R(S') = \{QN_2,Q_2\}$. By now applying the same reasoning for 
    all remaining nice holes of the partitioning, we get with $(vi)$ that $(3)$ now also holds for $H$.

    Finally, $(4)$ now follows from the fact that $H_1$ and $H_2$ share at most two vertices and either of them contains 
    at most only one vertex that is not present in $H$. Therefore, we have $O(nv(H_1)+nv(H_2)) = O(nv(H))$ and $(4)$ for 
    $H$ now follows directly from $(4)$ applied to $H_1$ and $H_2$. This concludes the proof. \qedhere

\end{proof}

An analogous statement could be shown for the partitioning that arises from adding all $\text{seg}(Q_i,QW_i)$ segments.
 For our purposes however, a weaker statement will suffice for this type of partitioning: 

\begin{lemma}\label{chz:lemma:qwi:part}
    Let $H$ be a BLS-hole with canonical ordering $L_1(H),N_2(H), \ldots ,L_k(H),\\ R(H)$. Furthermore, let the index set $I 
    \subseteq \{2, \ldots ,k\}$ be arbitrary. Then after adding the segments $\text{seg}(Q_i,QW_i)$ for all $i \in I$ the 
    following conditions hold:
    \begin{enumerate}[label=(\arabic*)]
        \item The BLS-hole H is partitioned into $|I|+1$ BLS-holes $H_1, \ldots ,H_{|I|+1}$
        \item $Q(H)\setminus\{\bigsqcup_{j=1}^{|I|+1}Q_j(H)\}
        =\bigsqcup_{j=1}^{|I|+1} Q_i(H_j)$
        \item $QW(H)\setminus\{\bigsqcup_{j=1}^{|I|+1}QW_j(H)\}
        =\bigsqcup_{j=1}^{|I|+1} QW_i(H_j)$
        \item We have $O(\sum_{j=1}^{|I|+1}nv(H_i)) = O(nv(H))$

    \end{enumerate}
    
\end{lemma}

\begin{proof}
    This can be shown completely analogously to the proof of the previous theorem \qedhere.
\end{proof}

So in other words, this lemma states that after adding an arbitrary subset of segments $\text{seg}(Q_i,QW_i)$, 
the remaining $Q_j$ and $QW_j$ points remain unchanged, and the total number of vertices in the resulting partitioning is 
again linear in the total number of vertices of the original hole.

\subsubsection[Computing BL-stable locations for BLS-Holes]{Computing BL-stable locations for BLS-Holes}

After presenting the different types of partitioning methods, we will now examine in detail how to use them 
to reduce the problem of finding all feasible BL-stable locations within a BLS-hole to the previously solved problem of 
finding them within a nice hole. Observe that using the partitioning induced by adding all segments $\text{seg}(Q_i,QN_i)$ 
for computing the BL-stable locations for each nice hole of the partitioning separately is not sufficient. This is because 
it could occur, that a rectangle can only be placed in such a way that it lies partially in one such nice hole and 
partially in another. 

One could then try the following approach to solve this problem : For each nice hole $S$ in the partitioning with 
$R(S) = \{QN_i^*,Q_i\}$, store not only the point $QN_i$ but also the corresponding point $QW_i$. Now, when invoking 
the bottom function on $S$, do not stop as soon as the right endpoint of the horizontal bar hits the rightmost edge of 
$S$. Instead, continue sliding the segment. By definition, $QW_i$ will now be the first point encountered by the right 
point of the horizontal bar. Up to the hit of $QW_i$, no other edge needs to be considered and the algorithm can then just 
simply continue from $QW_i$ as long as the left part of the horizontal bar still lies in $S$. An illustration of this 
approach is shown below in figure \ref{fig:chz:counterexample}.

\begin{figure}[H]
    \centering
    \scalebox{0.7}{\begin{tikzpicture}
    \coordinate (A) at (7,0);
    \coordinate (B) at (13,0);
    \coordinate (C) at (13,1);
    \coordinate (D) at (14,1);
    \coordinate (E) at (14,2);
    \coordinate (F) at (15,2);
    \coordinate (G) at (15,4);
    \coordinate (H) at (16,4);
    \coordinate (I) at (16,5);
    \coordinate (J) at (17,5);
    \coordinate (K) at (17,6);
    \coordinate (L) at (18,6);
    \coordinate (M) at (18,7);
    \coordinate (N) at (19,7);
    \coordinate (O) at (19,8);
    \coordinate (P) at (2,8);
    \coordinate (Q) at (2,6);
    \coordinate (R) at (10,6);
    \coordinate (S) at (10,5);
    \coordinate (T) at (2,5);
    \coordinate (U) at (2,4);
    \coordinate (V) at (5,4);
    \coordinate (W) at (5,3);
    \coordinate (X) at (11,3);
    \coordinate (Y) at (11,2);
    \coordinate (Z) at (7,2);

    \draw[thick] (A) -- (B);
    \draw[thick] (B) -- (C);
    \draw[thick] (C) -- (D);
    \draw[thick] (D) -- (E);
    \draw[thick] (E) -- (F);
    \draw[thick] (F) -- (G);
    \draw[thick] (G) -- (H) node[midway, below] {\Large $e$};
    \draw[thick] (H) -- (I);
    \draw[thick] (I) -- (J);
    \draw[thick] (J) -- (K);
    \draw[thick] (K) -- (L);
    \draw[thick] (L) -- (M);
    \draw[thick] (M) -- (N);
    \draw[thick] (N) -- (O);
    \draw[thick] (O) -- (P);
    \draw[thick] (P) -- (Q);
    \draw[thick] (Q) -- (R);
    \draw[thick] (R) -- (S);
    \draw[thick] (S) -- (T);
    \draw[thick] (T) -- (U);
    \draw[thick] (U) -- (V);
    \draw[thick] (V) -- (W);
    \draw[thick] (W) -- (X);
    \draw[thick] (X) -- (Y);
    \draw[thick] (Y) -- (Z);
    \draw[thick] (Z) -- (A);

    \coordinate (Q2) at (X);
    \coordinate (QW2) at (15,3);
    \coordinate (QN2) at (11,8);

    \coordinate (Q3) at (R);
    \coordinate (QW3) at (K);
    \coordinate (QN3) at (10,8);

    \filldraw[black] (Q2) circle (2.5pt) node[below left] {\Large $Q_2$};
    \filldraw[black] (QW2) circle (2.5pt) node[below right] {\Large $QW_2$};
    \filldraw[black] (QN2) circle (2.5pt) node[above right] {\Large $QN_2$};

    \filldraw[black] (Q3) circle (2.5pt) node[above right] {\Large $Q_3$};
    \filldraw[black] (QW3) circle (2.5pt) node[above left] {\Large $QW_3$};
    \filldraw[black] (QN3) circle (2.5pt) node[above left] {\Large $QN_3$};

    \draw[dashed] (Q2) -- (QW2);
    \draw[dashed] (Q2) -- (QN2);
    \draw[dashed] (Q3) -- (QW3);
    \draw[dashed] (Q3) -- (QN3);

    \node at (9,1) {\huge $S_1$};
    \node at (6,3.5) {\huge $S_2$};
    \node at (7,7) {\huge $S_3$};

    \draw[blue, very thick] (2.2,4.2) -- (8.2,4.2);

    \draw[blue, very thick] (8.8,3.2) -- (14.8, 3.2);

    \filldraw[blue] (2.2,4.2) circle (2pt) node[above right] {\Large $b_1$};
    \filldraw[blue] (8.2,4.2) circle (2pt) node[above left] {\Large $b_2$};

    \filldraw[blue] (8.8,3.2) circle (2pt) node[above left] {\Large $b_1$};
    \filldraw[blue] (14.8,3.2) circle (2pt) node[above left] {\Large $b_2$};

    \draw[->,blue, very thick] (4.2,4.5) -- (6.2,4.5);

\end{tikzpicture}}
    \caption{Illustration of a counter-example}
    \label{fig:chz:counterexample}
\end{figure}
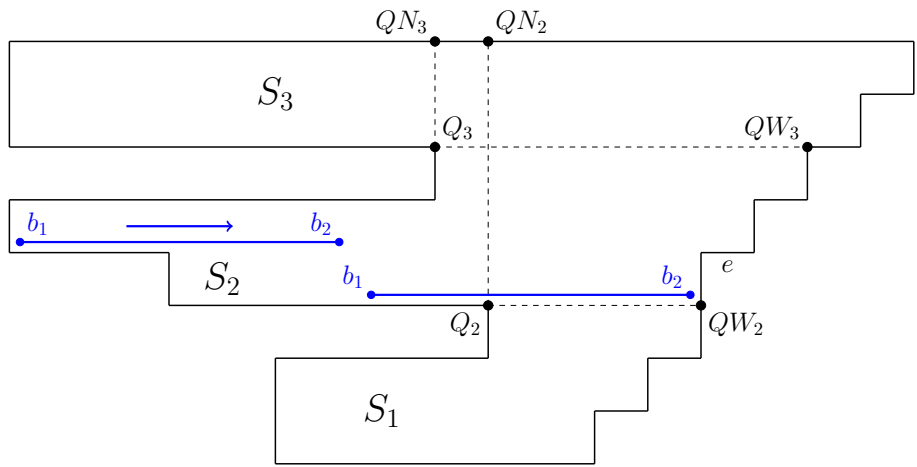

This is now exactly the type of approach described by Chazelle in \cite{BLCHZ}. However, Chazelle did not explicitly 
state whether the sliding should stop upon reaching $QW_i$, or whether it should continue beyond that point. As 
illustrated in Figure \ref{fig:chz:counterexample}, stopping at $QW_2$ is clearly insufficient. That is because even 
after raising the bar so that it is supported by the edge $e$, the bar would still be partially contained in $S_2$.

However, continuing beyond that point could potentially break the linear runtime guarantee. The reason for this is also 
illustrated in Figure \ref{fig:chz:counterexample}, as continuing the slide after reaching $QW_2$ would cause the right 
staircase-like structure (starting from $QW_2$) to be processed multiple times --- once during the \texttt{Bottom}-call 
for $S_2$ and once during the \texttt{Bottom}-call for $S_1$. In addition, depending on the width of the bar, it could 
also happen that part of this staircase-like structure is also considered in the \texttt{Bottom}-call for $S_3$. Observe 
that this issue cannot simply be resolved by just stopping the traversal of the boundary of $S_1$ once $QW_2$ is 
reached --- although the left endpoint of the horizontal bar may then lie within $S_2$ for a while, continuing further 
up the staircase could eventually result in the bar being fully placed inside $S_1$ again. 

More generally, one can now easily construct in a similar manner an example where such a staircase-like structure 
consists of $O(p)$ edges - where $p$ is the number of vertices of the entire BLS hole - and is processed additionally 
in $O(p)$ separate \texttt{Bottom} calls. This clearly breaks the linear runtime and constitutes the core flaw in 
Chazelle's runtime analysis.

To address this issue, we will now present an alternative partitioning approach.
This approach is motivated by the observation that, if in our example from Figure \ref{fig:chz:counterexample} we 
had $x(QW_2) - x(Q_2) \geq w$ --- where $w$ is the width of the horizontal bar --- then this repeated consideration 
of the staircase-like structure would not occur. This is because in this case the sliding could safely stop upon 
reaching $QW_2$, since any further continuation would cause it to lie entirely outside of the nice hole $S_2$.

More generally, we claim that if this condition holds for all such $Q_j,QW_j$ points of a BLS-hole, then a slightly 
modified variant of the partitioning method from Theorem \ref{chz:partitioning:theorem} can be applied to enable the 
computation of the feasible BL-stable locations in linear time. This claim will now be formalized in the following 
theorem, after which we will show how to reduce the general case to that special setting.

\begin{theorem}\label{chz:qwi_dist_big_theorem}
    Let $H$ be a BLS-hole with canonical ordering $L_1,N_2, \ldots ,L_k,R$ and $r = (w,h)$ a rectangle. If we have 
    $x(QW_i) - x(Q_i) \geq w$ for all $i \in \{2, \ldots ,k\}$, then we can compute the set of all feasible BL-stable 
    locations for placing $r$ within $H$ in $O(nv(H))$ time. 
\end{theorem}

\begin{proof}
   Consider again the partitioning $S_1, \ldots ,S_k$ of $H$ as described in Theorem \ref{chz:partitioning:theorem}, 
   where $R(S_1) = R(H)$ and $R(S_i) = \{QN_i^*,Q_i\}$ for all $i \in \{2, \ldots ,k\}$. As argued earlier, it is not 
   sufficient to simply compute the feasible BL-stable locations for each such nice hole $S_i$ separately, since it 
   is possible that the bottom-left corner of the rectangle $r$ lies in $S_i$, but part of $r$ lies in another nice 
   hole of the partitioning. So instead, we want to identify for each such nice hole $S_i$ the set of all such feasible 
   BL-stable locations, where the bottom left corner of $r$ lies in $S_i$, while $r$ itself is not necessarily fully 
   contained in $S_i$, but is fully contained in the original BLS-hole $H$. Observe that for the nice hole $S_1$, this 
   approach does not differ from the first one --- since $R(S_1) = R(H)$, it is not possible for the bottom-left corner 
   of $r$ to lie in $S_1$ while being partially contained in another nice hole. 

   For the other nice holes $S_i$ with $i \in \{2, \ldots ,k\}$, our approach can now be reduced to computing the set of 
   such feasible BL-stable locations not within $S_i$, but within the expanded hole 
   \begin{equation*}
       S_i^* = (S_i \cup B_i) \cap H = S_i \cup (B_i \cap H)
   \end{equation*} where $B_i$ is just the rectilinear set defined as
   \begin{equation*}
       B_i := [x(Q_i),x(QW_i)]\times [y(Q_i),y(QN_i)]
   \end{equation*}Observe that this reduction is only possible since we have $x(QW_i)-x(Q_i) \geq w$ as assumption. 

   However, it is not immediately clear that $S_i^*$ is itself a nice hole again, and even if it is, it is not obvious 
   that $O(nv(S_i^*)) = O(nv(S_i))$. To resolve this, we will first prove a structural claim about $S_i^*$ from which 
   this property will follow. 

   \newpage 
   \textit{Claim: }If the hole $H$ contains no falling corner or if its unique falling corner $c_f$ is not contained 
   in $B_i$, then we have 
   \begin{equation}\label{chz:eq:my_boy:eq1}
       B_i \cap H = B_i
   \end{equation}
   Otherwise, if instead we have $c_f \in B_i$, then we have 
   \begin{equation}\label{chz:eq:my_boy:eq2}
       B_i \cap H = ([x(Q_i),x(QW_i)] \times [y(Q_i),y(c_f)]) \cup ([x(Q_i),x(c_f)] \times [y(c_f),y(QN_i)])
   \end{equation}

    \begin{figure}[H]
        \centering
        \scalebox{0.7}{\begin{tikzpicture}
    \coordinate (A) at (0,0);
    \coordinate (B) at (0,2);
    \coordinate (C) at (2,2);
    \coordinate (Q2) at (2,3);
    \coordinate (E) at (0,3);
    \coordinate (F) at (0,6);
    \coordinate (QN2) at (2,6);
    \coordinate (H) at (2,7);
    \coordinate (I) at (6,7);
    \coordinate (cf) at (6,5);
    \coordinate (J) at (9,5);
    \coordinate (K) at (9,3);
    \coordinate (QW2) at (7,3);
    \coordinate (L) at (7,0);

    \draw[thick] (A) -- (B);
    \draw[thick] (B) -- (C);
    \draw[thick] (C) -- (Q2);
    \draw[thick] (Q2) -- (E);
    \draw[thick] (E) -- (F);
    \draw[thick] (F) -- (QN2);
    \draw[thick] (QN2) -- (H);
    \draw[thick] (H) -- (I);
    \draw[thick] (I) -- (cf);
    \draw[thick] (cf) -- (J);
    \draw[thick] (J) -- (K);
    \draw[thick] (K) -- (QW2);
    \draw[thick] (QW2) -- (L);
    \draw[thick] (L) -- (A);

    \filldraw[black] (Q2) circle (2pt) node[above left] {\large $Q_2$};
    \filldraw[black] (QN2) circle (2pt) node[below left] {\large $QN_2$};
    \filldraw[black] (QW2) circle (2pt) node [above right] {\large $QW_2$};
    \filldraw[black] (cf) circle (2pt) node[above right] {\large $c_f$};

    \draw[dashed] (Q2) -- (QN2);
    \draw[dashed] (Q2) -- (QW2);

    \fill[pattern = north east lines, pattern color = darkgreen] (Q2) -- (QN2) -- (6,6) -- (cf) -- (7,5) -- (QW2) -- (Q2);

    \node at (4,4.5) {\Large $B_2 \cap H$};
    \node at (1,4.5) {\Large $S_2$};

\end{tikzpicture}}
        \caption{Case 2 of the claim}
        \label{fig:enter-labe}
    \end{figure}
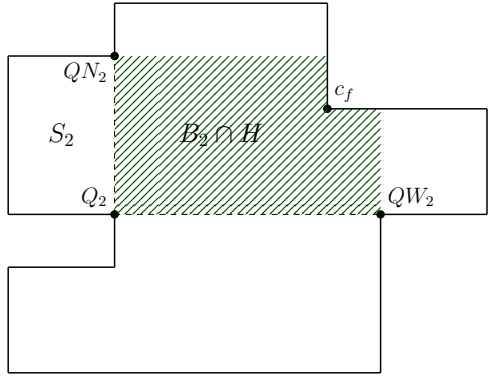

   \noindent 
   \textit{Proof of the claim: }Observe, that the horizontal bar $\text{seg}(b_1,b_2) = \text{seg}(Q_i,QW_i)$ is 
   contained in $H$ by definition of $QW_i$. We now lift this bar vertically upwards as far as possible, while ensuring 
   that it remains entirely within $H$. More precisely, we define 
   \begin{equation*}
       y^* = \max\{ y'\in \mathbb{R} | \ \text{seg}(Q_i +_y \tilde{y}, QW_i +_y \tilde{y}) \subseteq H \ \forall \tilde{y} \in [0,y']\}
   \end{equation*}
   If $y^* \geq y(QN_i) - y(Q_i)$, then it follows immediately that $B_i \cap H = B_i$. So suppose instead that we 
   have $y^* < y(QN_i) - y(Q_i)$. 

   In this case, there must exist a horizontal edge $e$ of $H$ with $y(e) = y(Q_i) + y^*$ such that 
   \begin{equation}
       (x_{\min}(e), x_{\max}(e)) \cap (x(Q_i), x(QW_i)) \neq \emptyset 
   \end{equation} and such that directly above $e$ the outer region of $H$ lies. Therefore, $e$ is a rightward edge by 
   Lemma \ref{chz::horizontal_interior}.
   Since no edge of $H$ can now intersect the segment $\text{seg}(Q_i,QN_i)$, our edge $e$ must fulfill 
   $x_{\min}(e) > x(Q_i)$. Now consider the vertical edge $e' = \pred(e)$, i.e the edge that ends at the left vertex 
   of $e$. By the minimality of $y^*$, it now follows that $y_{\min}(e') = y(e)$. Since $e'$ now ends in the left 
   vertex of a rightward edge and begins above it, this means that $e'$ is a downward edge. This however now implies 
   that this left endpoint of $e$ is the unique falling corner of $H$ by definition. Consequently, this case can thus 
   only occur if $H$ contains such a falling corner, which is also contained in $B_i$. This now  proves 
   (\ref{chz:eq:my_boy:eq1}). To show (\ref{chz:eq:my_boy:eq2}), observe that in this case we have thus already proven that 
   \begin{equation*}
       [x(Q_i),x(QW_i)] \times [y(Q_i),y(c_f)] \subseteq B_i \cap H
   \end{equation*}
   Therefore, to complete the proof of (\ref{chz:eq:my_boy:eq2}) it only remains to show that 
   \begin{equation}\label{chz:eq:my_boy:eq3}
       [x(Q_i),x(c_f)] \times [y(Q_i),y(QN_i)] \subseteq B_i \cap H
   \end{equation} since 
   \begin{equation*}
       [x(c_f),x(QW_i)] \times [y(c_f), y(QN_i)] \subseteq B_i \setminus H
   \end{equation*}
   is straightforward to verify. Now (\ref{chz:eq:my_boy:eq3}) can be argued in a fully analogous manner as before --- 
   by using the same horizontal bar lifting procedure. This time, however, we lift the segment $\text{seg}(Q_i, (x(c_f),y(Q_i))$. 
   In this case, no horizontal edge can obstruct the bar from reaching height $y(QN_i)$, as this would imply the existence 
   of a falling corner that is distinct from $c_f$, which cannot happen for a BLS-hole $H$. Therefore, 
   (\ref{chz:eq:my_boy:eq3}) is shown and this concludes the proof of the claim.

   \vspace{0.5cm}

   This claim now directly implies that $S_i^*$ is a nice hole and $O(nv(S_i^*)) = O(nv(S_i))$. With this in mind, 
   we can now complete the proof of the theorem.
   
    To accomplish that, we first compute the $QN_i$ vertices. As described in Section~3.3.1, this can be done 
    with just a single clockwise traversal of the boundary of $H$. 
    
    Afterwards, one computes the holes $S_1, \ldots ,S_k$ of the partitioning, which is straightforward by Theorem 
    \ref{chz:partitioning:theorem}. We then construct the corresponding expanded holes $S_1^*,S_2^* \ldots ,S_k^*$, 
    using the structural result established in the claim, where $S_1^*$ is just $S_1$ itself.
    
    For each such nice hole $S_i^*$ of them, we apply Theorem \ref{chz:nice_hole_main_theorem} to compute the set of 
    all feasible BL-stable locations for placing $r$ within $S_i^*$ in $O(nv(S_i^*))$ time. The union of all these 
    sets yields exactly the set of all feasible BL-stable location for placing $r$ within $H$, as described earlier. 
    Finally, since we have \begin{equation*}
        O(\sum_{i=1}^k nv(S_i^*)) =  O(\sum_{i=1}^k nv(S_i)) = O(nv(H))
    \end{equation*} where the last equality follows from Theorem \ref{chz:partitioning:theorem}, this now concludes 
    the proof. \qedhere 
   
\end{proof}

Having hereby established how to compute the feasible BL-stable locations in this special case, we now proceed to 
reduce the general case to this setting and thus prove the main theorem of this Section~3.3. 

\begin{theorem}\label{chz:qwi_dist_arb_theorem}
    Let $H$ be a BLS-hole with canonical ordering $L_1,N_2, \ldots ,L_k,R$ and $r = (w,h)$ a rectangle. We can compute 
    the set of all feasible BL-stable locations for placing $r$ within $H$ in $O(nv(H))$ time. 
\end{theorem}

\begin{proof}
    If for every $i\in \{2, \ldots ,k\} $ we have $x(QW_i) - x(Q_i) \geq w$, then this follows directly from Theorem 
    \ref{chz:qwi_dist_big_theorem}. So suppose this condition does not hold for every such $Q_i$ and let 
    $I \subseteq \{2,..,k\}$ be the  set of all indices $i \in I$ such that $x(QW_i) - x(Q_i) < w$.
    
    Now insert the segments $\text{seg}(Q_i,QW_i)$ for all $i \in I$ into $H$. By Lemma 
    \ref{chz:lemma:qwi:part}, this partitions $H$ into $l := |I|+1$ BLS-holes $H_1, \ldots ,H_{l}$. Observe that it 
    is not possible to place the rectangle $r$ in such a way that it lies partially in one such BLS-hole $H_i$ and 
    partially in such another BLS-hole $H_j$. This is because their separating segments $\text{seg}(Q_i,QW_i)$ for 
    every $i\in I$ have horizontal length strictly smaller than $w$, so any rectangle of width $w$ cannot cross this 
    segment from one side to another. 

    Therefore we have, that if we can compute for every such $H_i$ separately the set of all feasible BL-stable 
    locations for placing $r$ within $H_i$, then the union of all those sets will give us exactly the set of all 
    feasible BL-stable locations for placing $r$ within $H$.

    Now, for this separate computation, observe that by Lemma \ref{chz:lemma:qwi:part} the $Q_j$ and $QW_j$ points 
    for $j \notin I$ remain unchanged in these subholes $H_1, \ldots ,H_l$. This implies that by our choice of $I$, the 
    remaining $Q_j$ vertices in every such subhole now satisfy $x(QW_j) - x(Q_j) \geq w$ in their respective subholes. 
    This now enables us to use Theorem \ref{chz:qwi_dist_big_theorem} on each such $H_i$ and thus perform this 
    computation in $O(nv(H_i))$ time. Since we have now by Lemma \ref{chz:lemma:qwi:part} that 
    \begin{equation*}
        O(\sum_{i=1}^l nv(H_i) ) = O(nv(H))
    \end{equation*} it follows that the total computation can be done in $O(nv(H))$ time.

    Moreover, the partitioning itself can also be carried out in $O(nv(H))$ time. This is done by first computing 
    all the $QW_i$ vertices, again using only a single anticlockwise traversal, as described earlier. It is then 
    straightforward to verify that the partitioning step for all $i\in I$ can be also performed in a total of $O(nv(H))$ 
    time. This concludes the proof \qedhere 
\end{proof}

\subsection{Maintaining the Data Structure}

After having established in the previous subsections how to compute the set of all feasible BL-stable locations 
within a single BLS-hole, Section~3.4 concludes the description of Chazelle's implementation by explicitly describing 
how to maintain the data structure storing the holes throughout the whole algorithm. Furthermore, we will prove that 
this data structure requires only linear space with respect to the number of already placed rectangles. We then show 
that the data structure can also be updated in linear time after placing a new rectangle, thereby establishing the overall 
linear runtime of the implementation.

The data structure for storing a single hole was already described in Section~3.1, but for clarity reasons, we briefly 
recall it here --- a single hole is stored as a doubly-linked list of its vertices, in the order in which they appear 
on a traversal on the boundary of the hole. In addition, we store special pointers to the special vertical edges of the 
hole --- namely to all edges involved in the canonical ordering of the hole, along with a pointer to the unique falling 
corner if such exists. To store multiple holes, we use an array in which each entry holds the data structure representing 
one such hole. 

It is clear, that the size of the data structure is linear in the total number of vertices across all stored holes. 
We now aim to show that this total number is, in turn, linear in the number of already placed rectangles. While this 
result was also proven in Chazelle's description \cite{BLCHZ}, we will present here an alternative proof. Although our 
proof will yield a much more crude bound, it is simpler and still sufficient for establishing the asymptotic bound.

\begin{lemma}\label{chz:hole_rectangle_num_lemma_bound}
    Let $sp$ be a feasible partial packing of a SPI \  \SPI and let $H_1, \ldots ,H_k$ denote the holes of $sp$. Then we have
    \begin{equation*}
        \sum_{i=1}^k nv(H_i) \leq 8|R_{sp}| +4 
    \end{equation*}
    and thus 
    \begin{equation*}
        O(\sum_{i=1}^k nv(H_i)) = O(|R_{sp}| + 4)  
    \end{equation*}
    
\end{lemma}

\begin{proof}
    Observe that every vertex of a hole $H_i$ coincides either with a corner of a placed rectangle, or with a corner of 
    the strip. Since two distinct holes can intersect only at vertices, but not along edges, it follows that a corner of 
    a rectangle can coincide with vertices of at most two distinct holes. Furthermore, by the same argument, a corner of 
    the strip can coincide with the vertex of at most one hole. Since each placed rectangle and the strip itself have 
    four corners, this proves the claimed bound. \qedhere
\end{proof}

Thus, by Theorem \ref{chz:qwi_dist_arb_theorem}, if we are given the described data structure which stores all holes of 
our feasible partial packing $sp$, then we can find the set of all feasible BL-stable locations for placing the next 
rectangle $r$ in $O(|R_{sp}|+4)$ time. Therefore, it only remains to show for our runtime analysis that we can update 
our data structure in $O(|R_{sp}|+4)$ time after placing such a rectangle, since initializing the data structure at the 
beginning where no rectangle is placed yet is a trivial manner. This updating procedure was also described by Chazelle 
in \cite{BLCHZ}, and our proof of that statement is inspired by that same description.

\begin{lemma}\label{chz:hole_updating_lemma}
    Let $sp$ be a feasible partial packing of a SPI \  \SPI \  and let $H_1, \ldots ,H_k$ denote the holes of $sp$. 
    Furthermore, let $r = (w,h)$ be a rectangle and $(x,y)$ a feasible BL-stable location for placing $r$ within $H_i$ 
    for some $i\in \{1, \ldots ,k\}$. Then we can update the data structure for the feasible extension of $sp$ resulting from 
    placing $r$ at this position in $O(|R_{sp}|+4)$ time.
\end{lemma}

\begin{proof}
    It is clear, that we only have to update the corresponding entry of our array which corresponds to the data structure 
    storing $H_i$. It is now geometrically evident, that we can update the corresponding doubly linked-list(s) of the new 
    hole(s) $H_{i_1}, \ldots ,H_{i_l}$ arising from $H_i$ in time $O(nv(H_i))$, by just performing a traversal of the boundary 
    of $H_i$ while checking the traversed edges against those of the newly placed rectangle.
    A formal proof for this that considers all cases would be very lengthy and technical, without providing much 
    additional insight. Since the result is evident, we thus omit the details here.
    
    Once these doubly linked-lists for the new holes $H_{i_j}$ have been computed, we iterate over each of them to 
    initialize the special links to the special vertical edges in their canonical ordering, as well as to the unique 
    falling corner, if it exists. This can be done by just a single clockwise traversal starting from the rightmost edge. 
    By Lemma \ref{chz:hole_rectangle_num_lemma_bound} we have $O(\sum_{j=1}^l nv(H_{i_j})) = O((|R_{sp}|+1)+4)$, which thus 
    completes the runtime argument. \qedhere
    
\end{proof}

\subsection{Overview of the full Implementation}

Building on the results examined in the previous subsections --- specifically Theorem~\ref{chz:qwi_dist_arb_theorem}, 
Lemma \ref{chz:hole_rectangle_num_lemma_bound} and Lemma \ref{chz:hole_updating_lemma} --- we have essentially already 
established both the correctness as well as the runtime complexity of Chazelle's implementation. Nevertheless, 
for the sake of clarity and better understanding, this subsection provides a summary of all relevant steps of this 
implementation, outlined in the pseudocode below.

\SetKwInOut{Input}{Input}
\SetKwInOut{Output}{Output}
\SetKw{Continue}{continue}
\SetKw{Break}{break}
\SetKwIF{If}{ElseIf}{Else}{if}{then}{else if}{else}{end}
\SetKw{Goto}{goto}
\begin{algorithm}[!t]
\caption{Chazelle's Implementation for the BL-heuristic}
\Input{\SPI , where $R = \{r_1, \ldots ,r_n\}$}
\Output{Corresponding feasible packing produced by the BL-heuristic}
\vspace{0.5cm}
    $sp(r_i) \gets (\infty, \infty)  \ \forall i \in [n]$ \;
    Initialize the hole data structure $\mathcal{H}$ \;
    \For{$i= 1$ \KwTo $n$}{
        $M \gets \emptyset$ ; \tcp{Set of all feasible BL-stable locations for placing $r_i$}

        \For{BLS-hole $H \in \mathcal{H}$ \label{chz:calH:loop}}{
            Let $L_1,N_2, \ldots ,L_k,R$ be the canonical ordering of $H$ ; \tcp{Definition \ref{chz:canonical:ordering}}
            $I \gets \{j \in \{2, \ldots ,n\}\mid  x(QW_j) - x(Q_j) < w(r_i)\}$ \;
            Let $H_1, \ldots ,H_l$ be the holes of the partitioning of $H$ that arises by inserting the segments $\text{seg}(Q_j,QW_j)$ into $H$ \ $\forall j \in I$ ; \tcp{See proof of Theorem \ref{chz:qwi_dist_arb_theorem}}
            $M_H \gets \emptyset$ ; \tcp{Set of all feasible BL-stable locations for placing $r_i$ within $H$}
            \For{$H_j \in \{H_1, \ldots ,H_l\}$}{
                Let $S_1, \ldots ,S_{k'}$ be the nice hole partitioning of $H_j$ described in Theorem \ref{chz:partitioning:theorem} \;
                \For{$S_p \in \{S_1, \ldots ,S_{k'}\} $}{
                    $M_H \gets M_H \cup \texttt{All-BL-Locs}(S_p^*,r_i)$ ; \tcp{$S_p^*$ as defined in the proof of Theorem \ref{chz:qwi_dist_big_theorem} }
                }
            }
            $M \gets M \cup M_H$ \;
        }
        Let $(x,y) \in M$ be the location in $M$ that minimizes $(y,x)$ lexicographically \;
        $sp(r_i) \gets (x,y)$ \;
        Update $\mathcal{H}$ accordingly ; \label{chz:update_step_line} \tcp{Lemma \ref{chz:hole_updating_lemma}} 
    }
    \Return $sp$ \;

    \vspace{0.5cm}

    \SetKwBlock{Fn}{Function \texttt{All-BL-Locs}($S, r$)}{end}
    \tcc{Computes all feasible BL-stable locations for placing $r$ within the nice hole $S$.}
    \Fn{
    $C \gets \texttt{BOTTOM-Function}(S,w(r))$; \tcp{see Theorem \ref{chz:bottom:correct}}
    $D \gets \texttt{TOP-Function}(S,w(r))$; \tcp{see Theorem \ref{chz:top:correct}}
    \Return $\texttt{PLACING-Function}(C,D,r)$; \tcp{see Theorem \ref{chz:placing:correct}}
    }
\end{algorithm}

\begin{theorem}
    For a SPI \ \SPI, Chazelle's implementation computes the packing produced by the BL-heuristic in $O(n^2)$ time, where $n := |R|$.
\end{theorem}

\begin{proof}
    The details of the implementation are described in the listed pseudocode below. 
    The correctness of the partitioning approach follows from the proofs of proofs of Theorem~\ref{chz:qwi_dist_big_theorem} and Theorem \ref{chz:qwi_dist_arb_theorem}. The correctness for 
    the computation of feasible locations within nice holes follows from Theorem~\ref{chz:bottom:correct}, 
    Theorem~\ref{chz:top:correct} and Theorem~\ref{chz:placing:correct}. The linear runtime of the inner for-loop in 
    line~\ref{chz:calH:loop} follows from Lemma~\ref{chz:lemma:qwi:part}, Theorem~\ref{chz:partitioning:theorem} and 
    Lemma~\ref{chz:hole_rectangle_num_lemma_bound}. The update step in line~\ref{chz:update_step_line} can also be 
    performed in linear time by Lemma~\ref{chz:hole_updating_lemma}, which thus shows that each iteration of the outer 
    for-loop has linear runtime with regard to $n$. This implies, apart from the correctness, the total runtime of $O(n^2)$.
\end{proof}
\newpage 

\section{Conclusion}

We have thus successfully revisited and analyzed Chazelle's classical implementation of the BL-heuristic.
Our investigation identified and corrected a critical flaw in the original runtime analysis and 
yielded a corrected and fully rigorous description of the algorithm. 

This refined description confirms that Chazelle's implementation --- after modifications in the partitioning step --- indeed 
runs in $O(n^2)$ time and requires only $O(n)$ space. This establishes its asymptotic optimality among known implementations 
of the BL-heuristic. 

Although this refined implementation is primarily of theoretical interest --- as it is rather intricate and therefore unlikely to 
be used in practice --- it provides valuable insights into the structure of packings produced by the BL-heuristic. We hope 
that the insights gained from this description will inspire simpler and more practical implementations of the 
heuristic in the future.

\section*{Acknowledgments}

I would like to thank Stefan Hougardy for introducing me to the research topic of the BL-heuristic
and for his valuable guidance throughout my research.

\printbibliography    

\end{document}